\documentclass[twocolumn,appendixfloats,numberedappendix]{aastex7}

\usepackage{apjfonts}
\usepackage{times}
\usepackage{CJK}
\usepackage{amsmath}
\usepackage{hyperref}
\usepackage{cleveref}
\usepackage{natbib}
\usepackage{amssymb}
\usepackage{lineno}
\usepackage{rotating}
\usepackage{lipsum} 
\usepackage{academicons}
\usepackage{graphicx}
\usepackage{booktabs}

\usepackage{color}

\shorttitle{}
\shortauthors{Le et al.}

\newcommand{\Hb}{H{$\beta$}}
\newcommand{\Ha}{H{$\alpha$}}

\newcommand{\OIII}{[\ion{O}{3}]}
\newcommand{\OII}{[\ion{O}{2}]}
\newcommand{\NII}{[\ion{N}{2}]}

\newcommand{\m}{$M_{\rm BH}$}
\newcommand{\msun}{$\rm M_{\odot}$}

\def\gsim{\mathrel{\rlap{\lower4pt\hbox{\hskip1pt$\sim$}}
    \raise1pt\hbox{$>$}}}         

\def\lsim{\mathrel{\rlap{\lower4pt\hbox{\hskip1pt$\sim$}}
    \raise1pt\hbox{$<$}}}         

\newcommand{\ustc}{\affil{Department of Astronomy, University of Science and Technology of China, Hefei 230026, China; \href{mailto: lha@ustc.edu.cn}{lha@ustc.edu.cn}; \href{mailto: xuey@ustc.edu.cn}{xuey@ustc.edu.cn}}}
\newcommand{\sustc}{\affil{School of Astronomy and Space Science, University of Science and Technology of China, Hefei 230026, China}}
\newcommand{\snu}{\affil{Astronomy Program, Department of Physics \& Astronomy, Seoul National University, Seoul 08826, Republic of Korea}}

\begin{document}

\begin{CJK*}{UTF8}{gbsn}

\title{A Comparison of Star Formation Rates by Different Tracers in Nearby Galaxies}

\author[0000-0003-1270-9802]{Huynh Anh N. Le} \ustc \sustc
\email{lha@ustc.edu.cn} 
\author[0000-0002-8055-5465]{Jong-Hak Woo} \snu 
\email{woo@astro.snu.ac.kr} 
\author[0000-0002-1935-8104]{Yongquan Xue}  \ustc \sustc 
\email{xuey@ustc.edu.cn}
\author[0009-0002-9526-5820]{Ashraf Ayubinia} \snu
\email{a.ayubinia@gmail.com} 
\author[0000-0002-2156-4994]{Changseok Kim} \snu
\email{kcs1996kcs@snu.ac.kr} 
\author[0000-0002-4926-1362]{Xiaozhi Lin} \ustc \sustc
\email{xzlin@ustc.edu.cn} 

\correspondingauthor{Yongquan Xue (\href{mailto: xuey@ustc.edu.cn}{xuey@ustc.edu.cn})}


\begin{abstract} 

We utilize a large sample of $\sim$113,000 galaxies ($z < 0.3$) from the Sloan Digital Sky Survey with high-quality data to compare star formation rates (SFRs) across multiple diagnostic methods and examine their connection to Active Galactic Nuclei (AGNs) strength, indicated by Eddington ratio. Our sample encompassed star-forming (SF), composite, Seyfert, and LINER galaxies. Our analysis utilizes various SFRs indicators, including observed infrared flux ($\rm SFR_{FIR}$) from AKARI/Herschel ($\sim$4,100 sources), the MPA-JHU catalog ($\rm SFR_{Dn4000}$), the ANN catalog ($\rm SFR_{ANN}$), the GSWLC catalog ($\rm SFR_{SED}$ and $\rm SFR_{MIR}$), as well as \OII\ and \Ha\ emission lines ($\rm SFR_{[OII]}$ and $\rm SFR_{H\alpha}$).  Within SF galaxies, SFRs measurements from different tracers exhibited differences, with offsets and scatter below 0.26 dex and 0.29 dex, respectively. Moreover, non-SF galaxies (composite, Seyfert, and LINER) displayed discrepancies among SFR tracers, particularly for LINER galaxies, with offsets below 0.86 dex and a scatter of 0.57 dex. Additionally, our findings revealed robust correlations between SFRs and specific SFRs (sSFRs) with Eddington ratios. Eddington ratio exhibited gradual transitions in the (s)SFRs-stellar mass diagrams. Galaxies with high Eddington ratios displayed high star formation activity, similar to blue SF galaxies. Furthermore, we observed decreasing sSFR trends from SF galaxies to composite, Seyfert, and LINER galaxies. Our results may provide insight into our understanding of (s)SFRs traced by different approaches and their connection to AGN activities.

\end{abstract}


\section{INTRODUCTION}\label{section:intro}

Many studies have been conducted to study the relationship between AGNs and the star formation (SF) activities of their hosts. However, the current results are still controversial. Studies of AGN feedback have yielded diverse results. There is a positive correlation (positive feedback, e.g., \citealp{Zhuang+20}), a negative correlation (negative feedback, e.g., \citealp{Page+12}), or no/weak correlation (e.g., \citealp{Shimizu+15}) between AGN strength and SF rates (SFRs) of their hosts. Additionally, both negative and positive feedback have been observed in some individual targets using Integral Field Spectroscopy \citep[e.g.,][]{Luo+19, Shin+19}. Notably, strong correlations between AGN activity and SFRs are significant for the local redshift sample, which consists of the most luminous AGNs. However, for low-luminosity AGNs or higher redshift sources, there are no correlations between AGN activity and SFRs of their host galaxies \citep[e.g.,][]{Rosario+12}. As discussed by \citet{Hickox+14}, when using a high-stellar mass sample to study the relationship between AGN-SFR, a strong correlation between AGN-SFR is expected because both are initially correlated with high stellar mass.

One of the challenges in understanding the physical nature of black holes and host galaxy coevolution lies in the difficulty of accurately estimating SFRs. Estimating SFRs is challenging due to the high uncertainties associated with various SFR indicators, and each SFR tracer also has its limitations. In this paper, we primarily focus on the UV/optical and infrared (IR) bands. The rest-frame UV/optical emission directly traces the SFRs by the young stellar population because O and B star emission, characteristic of young stars, dominates the UV band compared to the longer wavelength range. However, the measurement of SFRs using the UV band may lack information about SF in dust-obscured regions and is strongly influenced by dust attenuation corrections \citep[e.g.,][]{Salim+07}. The UV/optical emission lines such as \OII\ and \Ha\ provide direct measurements of the young stellar population in low-redshift AGN sources. However, the emission from these lines may have uncertainties due to contamination from other sources, such as AGN emission and shocks from the interstellar region. Additionally, SFRs traced by the UV/optical emission lines strongly depend on dust attenuation corrections \citep[e.g.,][]{Kennicutt+98, Kennicutt+12, Calzetti+13}.

In contrast to the UV/optical bands, SFRs traced by the IR band indirectly measure the stellar light, but rather measure the dust heated by stellar light and re-radiated in the far-infrared (FIR) band. The FIR diagnostic is considered a robust indicator for measuring SFRs by tracing cold dust emission from the Interstellar Medium (ISM) while having less contribution from ISM shocks or AGN emissions \citep[e.g.,][]{Ellison+16a}. However, there are limitations to SFRs traced by IR emission. For example, mid-infrared (MIR) emission may originate from the hot dusty torus around AGNs \citep[e.g.,][]{Netzer+07}. Additionally, there may be the potential for starlight to be absorbed by dust or for dust heated by evolved stars ($>$100$-$200 Myr) to contribute to the FIR emission. Both of these effects can result in the underestimation or overestimation of SFRs \citep[e.g.,][]{Hirashita+01, Kennicutt+12}. Nevertheless, FIR emission is generally considered a reliable tracer for measuring SFRs in many studies in the literature \citep[e.g.,][]{Rosario+12, Matsuoka+15, Xu+20, Kim+22}. Until now, the sources that have SFR estimation using the IR band are relatively less due to the lower spatial resolution and sensitivity of IR/FIR observational facilities, such as the AKARI spacecraft and the Herschel Space Observatory. It is crucial to have large samples with robust SFR estimations using the FIR diagnostic to study SF activity effectively. Recently, some studies have leveraged artificial neural networks (ANNs) trained on FIR-detected galaxies to accurately predict IR luminosities and SFRs \citep[e.g.,][]{Ellison+16a, Ayubinia+25}. 

Recently, some studies found that using Spectral Energy Distribution (SED) fitting, a careful decomposition of dust heated by AGN is necessary to obtain a reliable measurement of the IR emission peak at $\sim$100 $\mu\mathrm{m}$, allowing for a robust SFR estimation. Therefore, the observations of FIR and/or Sub-millimeter (Sub-mm) are essential for robust estimations of SFR. Sub-mm fluxes will play a key role in the SED fitting models, particularly for targets without observed FIR fluxes. If the FIR/Sub-mm fluxes are not applied in the SED fitting models, the estimated IR luminosity will be exceeded by a factor of 2 \citep[e.g.,][]{Kim+22, Le+24}. Unfortunately, the number of galaxies with observations of Sub-mm data is still limited; a larger sample with Sub-mm observations is crucial. In short, the estimation of SFRs is associated with high uncertainty when using different tracers. Many observational efforts with high-quality spectra should be conducted to probe reliable SFRs in AGNs.

A series of statistical studies on gas outflows in type 2 AGNs have been conducted recently using a large sample of $\sim$113,000 galaxies ($\rm z < 0.3$) with high signal-to-noise (S/N) ratio in both the continuum and emission lines, based on the kinematics of the \OIII\ 5007 \AA\ emission line \citep[e.g.,][]{Bae+14, Woo+16, Woo+17, Woo+20}. By using this large and high-quality data sample, firstly, we aim to properly compare SFRs in the host galaxies using different diagnostic methods and investigate any discrepancies in SFR estimations from various tracers. Secondly, we aim to examine in detail the relationship between AGN strength (e.g., indicated by Eddington ratio) and SF activities. Previous works \citep[e.g.,][]{Woo+17, Woo+20}, found a strong positive correlation between SFRs and Eddington ratio, indicating a lack of negative AGN feedback and suggesting that a dynamical timescale may be required for AGN to quench SF activities. Motivated by this result, in this study, we aim to examine the relationship between Eddington ratio and SFRs or specific SFRs (sSFRs) in more detail by investigating how Eddington ratio varies across the (s)SFR-stellar mass plane. Particularly, we aim to analyze the role of Eddington ratio in distinguishing between active (blue) and quenched (red) (s)SFRs relative to the main-sequence (MS) line on the (s)SFR-stellar mass diagram.

Additionally, in our study, by using the \OII\ and \Ha\ emission lines, we estimate SFRs for this large sample and compare them with SFR determinations from commonly used tracers in the literature. These include IR-based SFRs derived from the IR luminosity in the AKARI/Far-infrared Surveyor (FIS) all-sky survey bright source catalog \citep{Yamamura+10} and the Herschel/PACS (Photodetector Array Camera and Spectrometer) Point Source catalog \citep{Pilbratt+10}. Additionally, we compare our results with SFRs based on the \Ha\ line and $\rm D_{n}4000$ ($\rm SFR_{Dn4000}$) from the Max Planck Institute for Astrophysics and Johns Hopkins University (MPA-JHU) catalog \footnote{http://www.mpa-garching.mpg.de/SDSS/} \citep{Brinchmann+04}, as well as IR-based SFR predictions from the ANN technique \citep{Ellison+16a}. Furthermore, \citet{Salim+16} and \citet{Salim+18} recently used UV-optical-MIR spectral energy distribution (SED) fitting and MIR flux to determine SFRs for $\sim$700,000 galaxies at $z < 0.3$ in the GALEX-SDSS-WISE Legacy Catalog (GSWLC). In this study, we also aim to compare our estimated SFRs with SFRs determined by the UV/optical SED fitting ($\rm SFR_{SED}$) and MIR flux ($\rm SFR_{MIR}$) from the GSWLC-2 catalog.

Comparing various SFR estimators may help us to understand the discrepancies in estimating SFRs through different diagnostic tracers and deepen our understanding of their impacts on the relationship between AGN and SF activities. Moreover, in \citet{Woo+17}, there is a concern about the robustness of both AGN bolometric luminosity and SFRs. Firstly, the AGN bolometric luminosity was estimated using the \OIII\ emission line for type-2 AGNs, but the \OIII\ emission line flux can be contributed to by both SF and AGN fractions. Secondly, SFRs were adopted based on the ANN catalog, necessitating a reliability test by comparing them to the estimated SFRs using other observational tracers. Therefore, in this work, we aim to consider the AGN and SF fractions in the determinations of \OII, \OIII, and H$\alpha$ luminosities to estimate both AGN bolometric luminosity and SFRs robustly.
  
In Section \ref{section:sample}, we describe the sample selection process. Sections \ref{section:meas} and \ref{section:result} provide detailed information on the measurements and present the results, respectively. The discussions are presented in Section \ref{section:discuss}, and the summary is provided in Section \ref{section:sum}. Similar to \citet{Brinchmann+04}, we adopted the \citet{Kroupa01} stellar initial mass function (IMF) between 0.01 to 100 \msun in this work. The following cosmological parameters are used throughout the paper: $H_0 = 70$~km~s$^{-1}$~Mpc$^{-1}$, $\Omega_{\rm m} = 0.30$, and $\Omega_{\Lambda} = 0.70$. 
  
\section{Sample Selection}\label{section:sample}

\begin{figure}
	\includegraphics[width=0.4\textwidth]{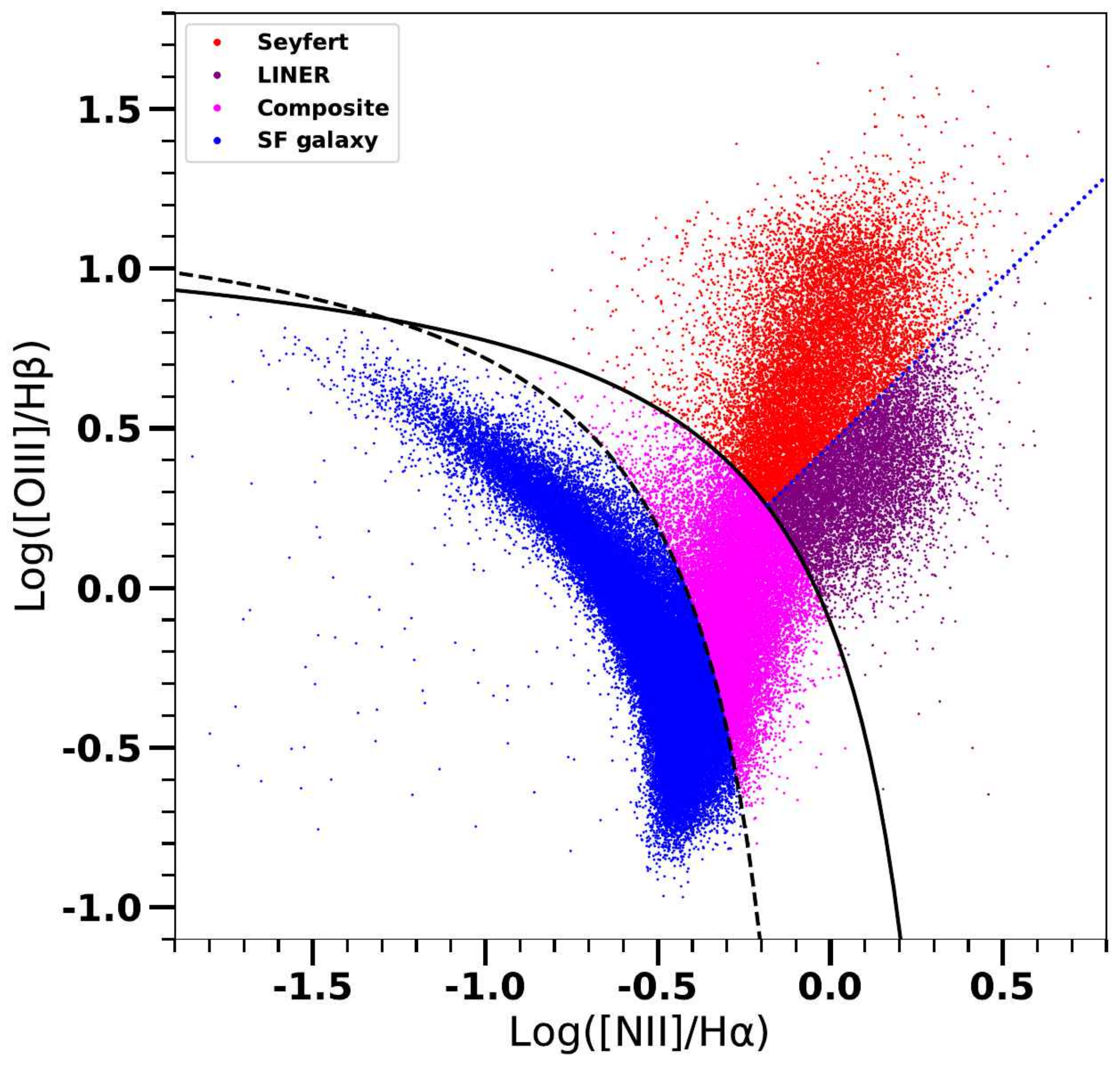}
	\centering
	\caption{BPT diagram of the selected sample, $\sim$113,000 galaxies (z $<$ 0.3) in this work. The classification of galaxies into SF, composite, Seyfert, and LINER categories is based on \citet{Kauffmann+03} (dashed line), \citet{Kewley+06} (solid line), and \citet{Schawinski+07} (dotted line), respectively.  
	\label{fig:sample}}
\end{figure}

We selected the sample from the MPA-JHU catalog, based on the Sloan Digital Sky Survey (SDSS) Data Release 7 \citep{Abazajian+09}. Firstly, the initial sample consisted of 235,922 emission line galaxies (z $<$ 0.3), satisfying the criteria of having a S/N $\geq$ 10 in the continuum and S/N $\geq$ 3 in the four emission lines: \Hb, \OIII\ 5007 \AA, \Ha, and \NII. Secondly, we selected sources with an amplitude-to-noise (A/N) ratio (the peak of an emission line to the spectral noise, e.g., \textbf{\citealp{Sarzi2006}}) larger than five for the \OIII\ and \Ha\ emission lines. Finally, our sample included 112,726 emission line galaxies, classified into star formation (SF), composite, Seyfert, and LINER categories (Figure \ref{fig:sample}) using the Baldwin-Phillips-Terlevich (BPT) diagram \citep{Baldwin+81}. The number of targets in each group is as follows: 69,262 SF, 20,712 composite, 14,067 Seyfert, and 8,685 LINER galaxies. Note that the classification of different galaxy types remains challenging. The BPT diagram has been widely used in the literature. However, the boundary lines between galaxy types are still under debate, especially in transitional or ambiguous regions such as the composite area between SF and AGNs. Recently, many efforts have been made to improve the \textit{true} classification of galaxy types \textbf{\citep[e.g.,][]{deSouza+17, Teimoorinia+18, Zhang+20, Ji+20}}. Depending on the diagnostic methods used, the identification of SF, composite, Seyfert, and LINER galaxies may vary slightly. As a continuation of our previous works \citep[e.g.,][]{Woo+16, Le+25}, we adopt the traditional BPT diagram classification scheme for our sample.

In this study, by matching our large sample with the available SFR catalogs in the literature (e.g., MPA-JHU, ANN, and GSWLC catalogs), we aim to investigate the differences in SFRs in each classification type galaxy estimated using various diagnostic methods, including UV-based SFRs by the \OII\ and \Ha\ emission lines, $\rm SFR_{Dn4000}$, IR-based SFRs, $\rm SFR_{SED}$, and $\rm SFR_{MIR}$. Additionally, we will analyze the relationships between these compared SFR tracers and the strength of nucleus activities, as indicated by Eddington ratio. Table \ref{table:sfr_summary} shows the list of SFR tracers used in this work. Note that \OIII\ luminosity and stellar mass ranges of our selected sample are consistent with those from \citet{Zhuang+19} for calibrating \OII\ SFRs ($\rm \log L_{[OIII]}\ =\ 39.3 - 44.4\ erg\ s^{-1}$, $\rm \log M_{\odot} = 8.5 - 11$).

\begin{figure*}
	\includegraphics[width=0.9\textwidth]{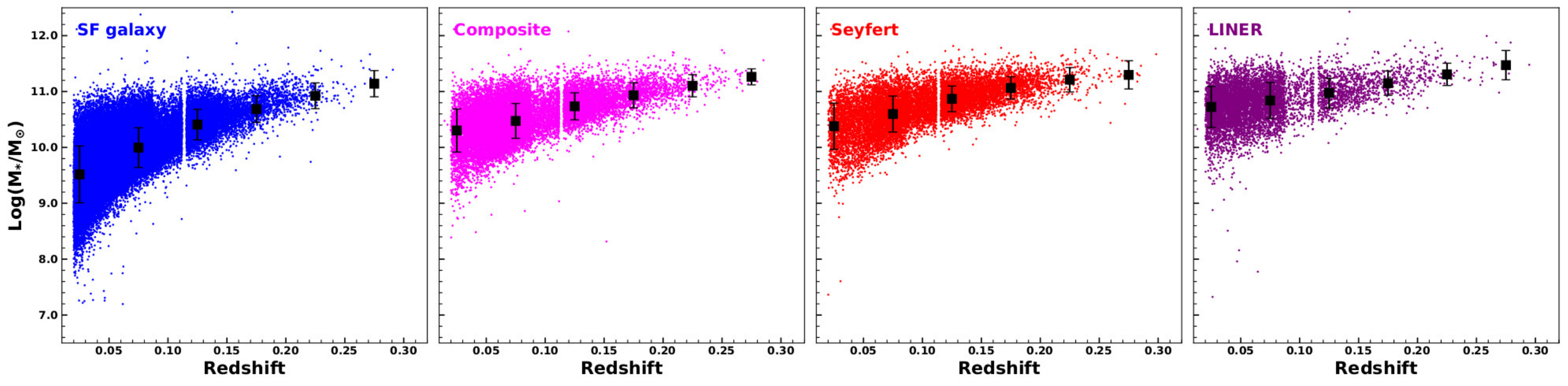}
	\includegraphics[width=0.9\textwidth]{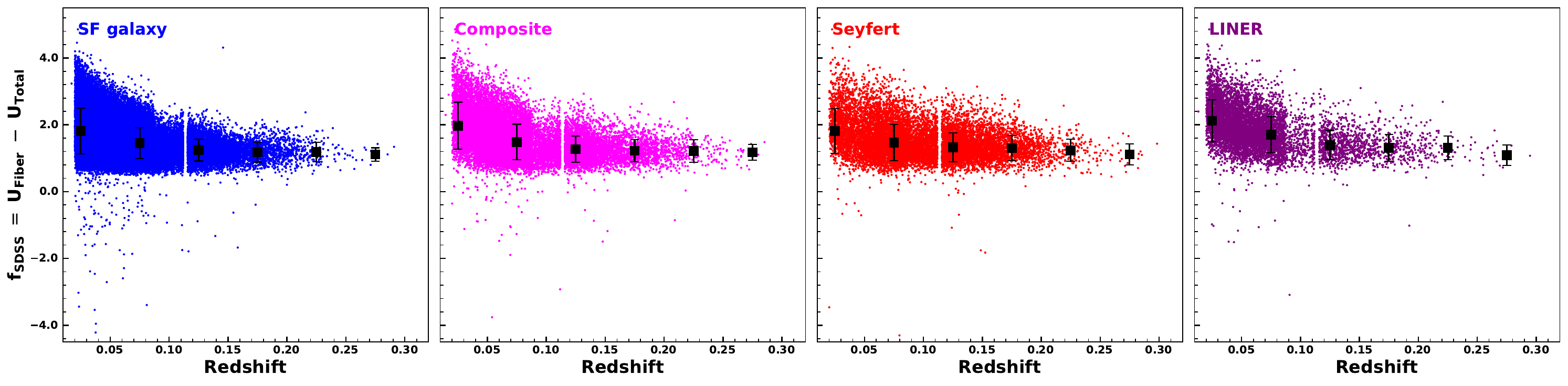}
	\centering
	\caption{Top panels: Stellar masses of the selected sample as a function of redshift. Bottom panels: Fraction of the u-band light captured by the SDSS fiber, defined as the difference in magnitude, $\rm f_{SDSS} = U_{Fiber} - U_{Total}$, shown as a function of redshift. The black squares show the median stellar mass values of each redshift bin $\rm \Delta z = 0.05$. Each panel corresponds to a different galaxy classification: SF, composite, Seyfert, and LINER, respectively.
	\label{fig:stellar_fsdss}}
\end{figure*}

As our selected sample spans a redshift range of $z < 0.3$, it may be affected by aperture size effects due to the fixed SDSS fiber diameter (3$\arcsec$), which corresponds to different physical scales at different redshifts. At higher redshifts, the SDSS fiber can encompass most of the entire galaxy, while at lower redshifts, it may capture the central region. To assess the impact of aperture size effects in our sample, Figure~\ref{fig:stellar_fsdss} presents the stellar mass distribution (adopted from the MPA-JHU catalog) and the fraction of u-band light captured by the SDSS fiber, defined as the difference in magnitude, $\rm f_{SDSS} = U_{Fiber} - U_{Total}$, as a function of redshift. $\rm U_{Fiber}$ and $\rm U_{Total}$ are taken from the SDSS photometry catalog, corresponding to $\rm FIBERMAG$ and $\rm MODELMAG$ in the u-band, respectively. We find that within each redshift bin of $\Delta z = 0.05$, the stellar mass distribution remains relatively consistent across galaxy types, ranging from $\rm \log(M_*/M_\odot) \sim8-11$ for SF galaxies and $\sim9-11$ for non-SF galaxies. In general, higher redshift bins tend to contain galaxies with slightly higher stellar masses, median $\rm \log(M_*/M_\odot) \sim 11$, whereas at lower redshifts, the mass range extends to lower values, median $\rm \log(M_*/M_\odot) \sim 10.5$, though still within the standard deviation of the higher-redshift mass distribution.

The fraction of u-band light captured by the SDSS fiber, $\rm f_{SDSS}$, is shown in the bottom panel of Figure~\ref{fig:stellar_fsdss}. For low-redshift galaxies ($z < 0.05$), the SDSS fiber captures a lower fraction of the total galaxy light compared to those at higher redshifts. In contrast, for galaxies at $z > 0.05$, the fraction of light captured by the SDSS fiber is higher and remains relatively constant with redshift. The loss of light captured by the SDSS fiber for galaxies at $z < 0.05$ may affect the comparison of SFRs derived from \OII\ and \Ha\ emission lines with those obtained from other tracers. We further discuss this issue in Section~\ref{section:discuss}.

\section{Measurements}\label{section:meas}

The analysis routine of the SDSS spectra in our sample has been thoroughly described in Section 2 of \citet{Bae+14}. In this paper, we present the measurements of SFRs using the \OII\ and \Ha\ emission line fluxes in our large sample. Additionally, we cross-matched the sample with the available IR fluxes observed by the AKARI/FIS and Herschel/PACS catalogs to estimate IR-based SFRs in our sample. To expand IR-based SFR sources, we matched our sample with the ANN catalog. Furthermore, we linked our sample with the MPA-JHU catalog, which provides SFRs based on the \Ha\ line and $\rm SFR_{Dn4000}$. We also connected our sample with the GSWLC-2 catalog that used UV-optical-MIR SED fitting and MIR flux to determine SFRs. Using these determined SFRs, we compare the discrepancies among multiple tracers for each galaxy type within our large sample and analyze their relationships with the nucleus strength indicator probed by Eddington ratio.

In this work, we correct the emission line fluxes for dust extinction with observed Balmer decrements and the Milky Way extinction curve of \citet{Cardelli+89} with $\rm R_{V} = A_{V}/E(B-V) = 3.1$. For electron temperatures of $\rm T_{e} = 10^4\ K$ and electron densities ranging from $\rm n_{e} \approx 10^2$ to $\rm 10^4\ cm^{-3}$, the intrinsic values of \Ha/\Hb\ are 2.86 and 3.1 for SF galaxies and AGNs, respectively \citep{Osterbrock+06}. 

\subsection{SFRs Based on \OII\ Emission Line}\label{subsec:oii_sfr}

To estimate SFRs based on the \OII\ emission line, we select sources with an S/N $\geq$ 3 of \OII\ line from our sample of 112,726 sources. Additionally, we excluded sources with \Ha/\Hb\ ratios below 2.86 in SF galaxies and 3.1 in AGNs. Balmer decrements that deviate below the theoretical value may arise from errors in subtracting the stellar continuum or inaccuracies in flux calibration and measurement. Applying these criteria, we find that 98,933 sources meet the requirement, including 62,134 SF, 18,638 composite, 12,089 Seyfert, and 6,072 LINER galaxies.  

\begin{figure}
	\includegraphics[width=0.47\textwidth]{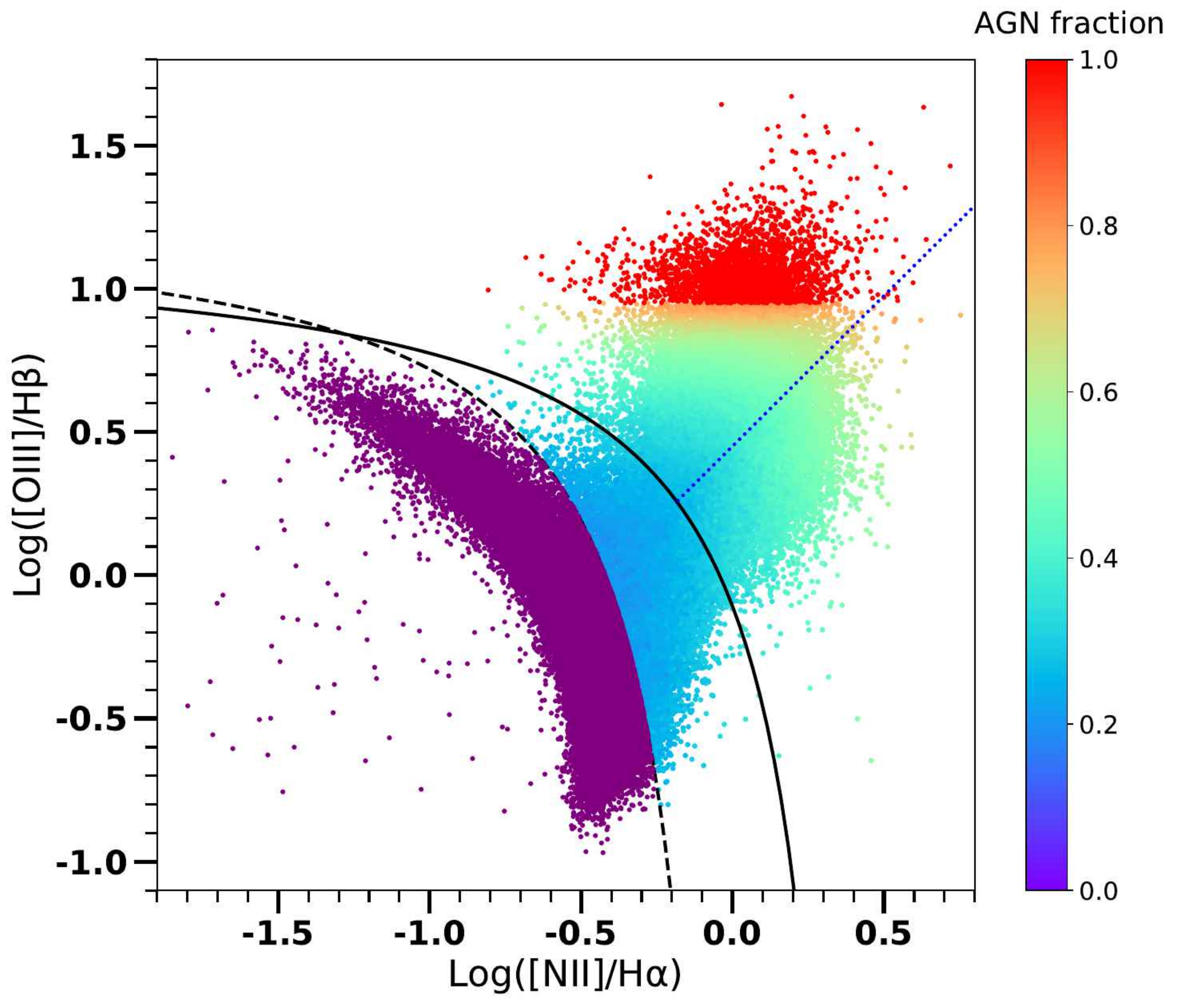}
	\centering
	\caption{AGN fraction estimations obtained using the MCMC Rainbow Python package. The color scale represents the estimated fractions based on large potential referent points of pure SF galaxies (magenta) and AGNs (red).   
	\label{fig:agnfraction}}
\end{figure}

\begin{figure}
	\includegraphics[width=0.40\textwidth]{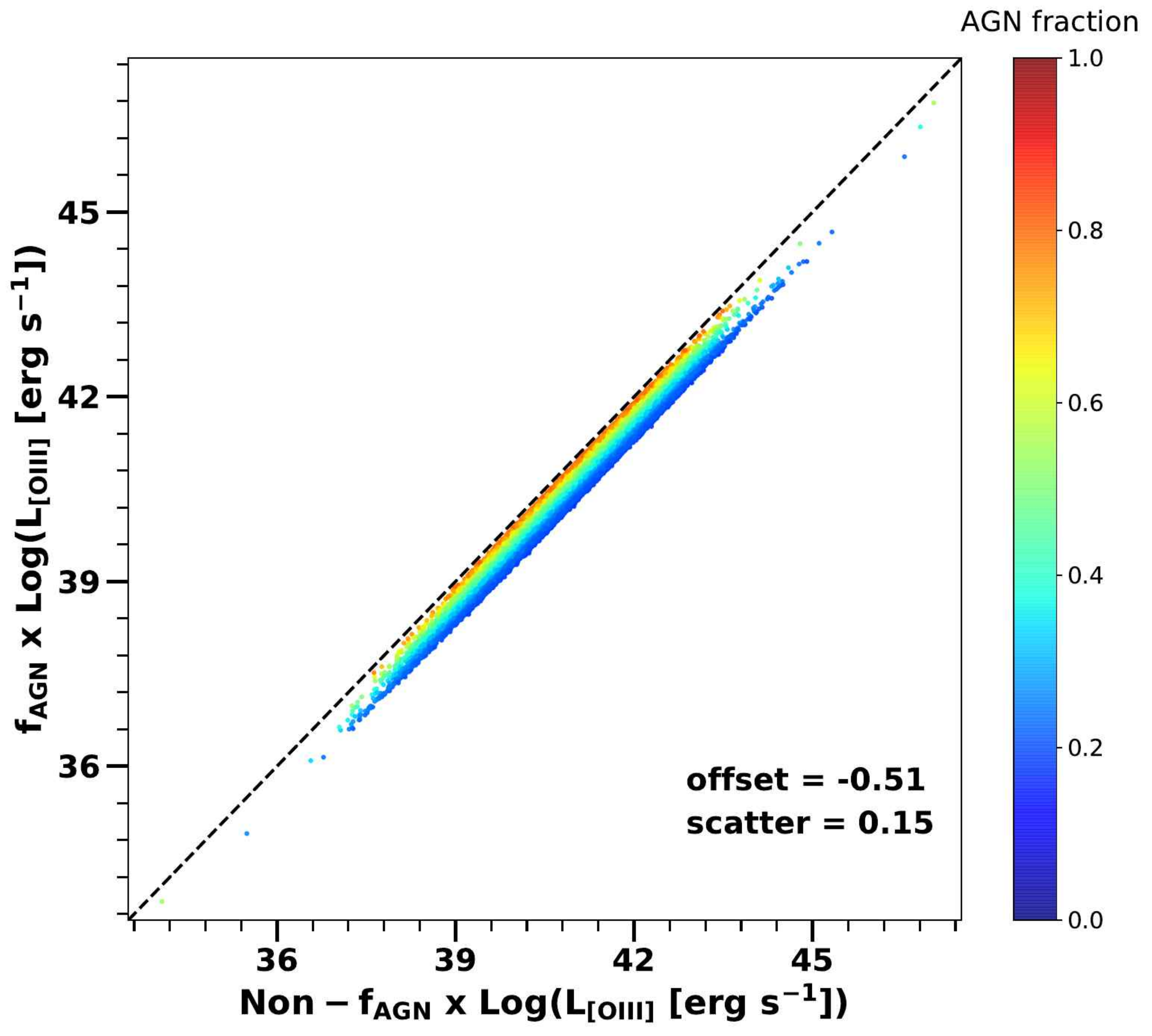}
	\includegraphics[width=0.40\textwidth]{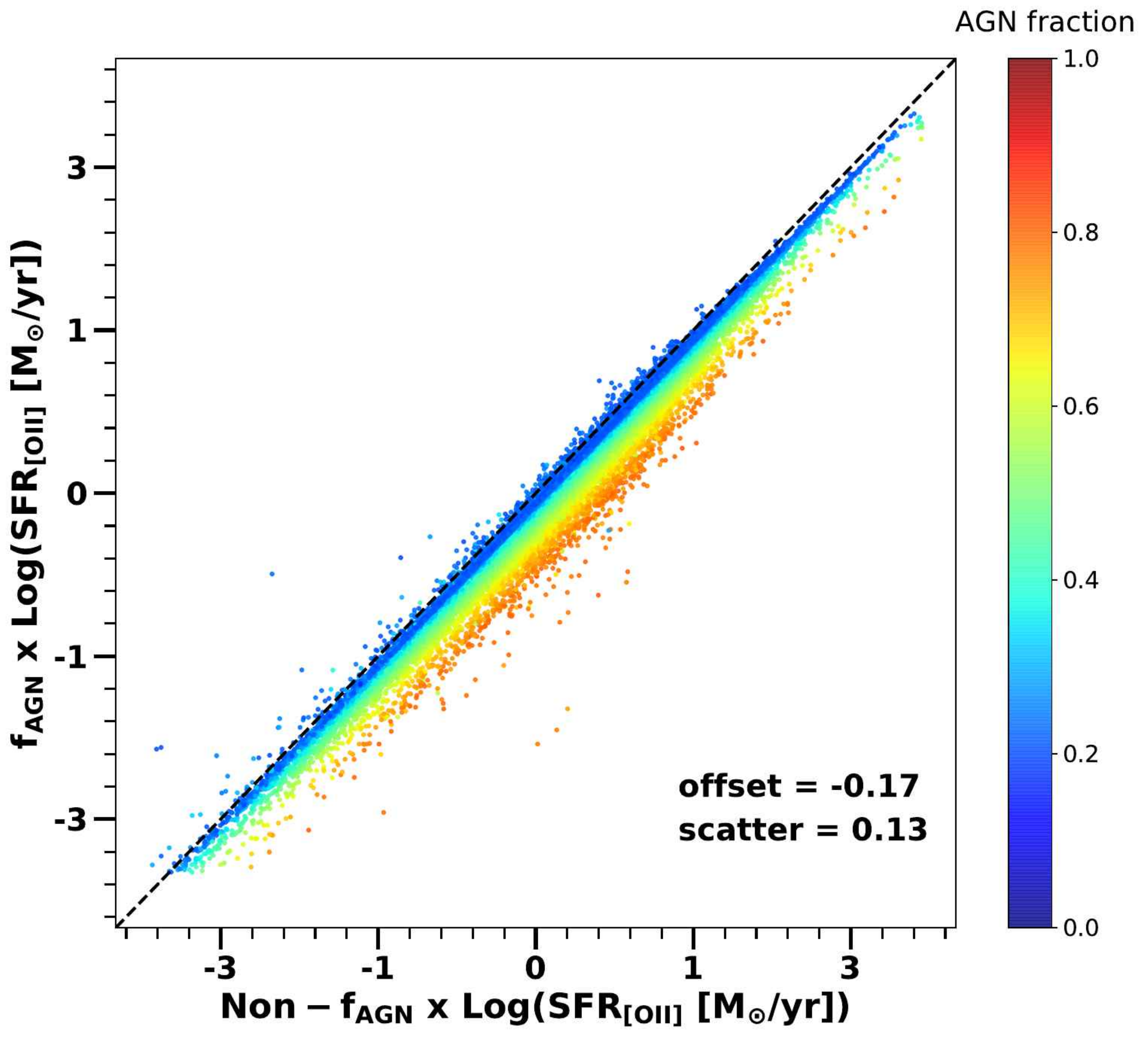}
	\centering
	\caption{Comparison between \OIII\ luminosity (top panel) and $\rm SFR_{[OII]}$ (bottom panel) with and without correction for AGN fraction ($\rm f_{AGN}$). We only show sources where $\rm 0 < f_{AGN} < 1$. 
	\label{fig:agnfraction_com}}
\end{figure}

We utilized the method proposed by \citet{Zhuang+19} to estimate the SFR based on \OII\ in our sample. The approach developed by \citet{Zhuang+19} assumes that all the \OIII\ emission originates from AGN. However, the \OIII\ emission may also include contributions from regions associated with SF (e.g., \citealp{Kauffmann09}, \citealp{Davies+16}). In this study, we estimate the fraction of AGN contribution to the \OIII\ emission line in each source of our sample using a Python package called Rainbow \textbf{\citep{Smirnova19}}. We then applied these fractions to the entire \OIII\ emission when estimating the SFRs based on \OII. The Rainbow package is specifically designed to determine ionization fractions in AGN host galaxies. It employs the concept of the mixing sequence (e.g., \citealp{Davies+16}), where each point in the BPT diagram is considered a linear combination of basic vectors that characterize the ionization fractions of AGN and SF. Rainbow utilizes a Bayesian approach and the Monte Carlo Markov Chain (MCMC) algorithm to estimate the SF and AGN fractions (for more detailed information, see Section 3 of \citealp{Smirnova+22}). Figure \ref{fig:agnfraction} displays the estimated AGN fractions ($\rm f_{AGN}$) of all the sources in our sample. We also examine the discrepancies between \OIII\ luminosity and SFRs based on \OII\ with and without correction for AGN fraction. Figure \ref{fig:agnfraction_com} shows differences when applying AGN fraction correcting for \OIII\ luminosity and SFRs based on \OII\ for sources where $\rm 0 < f_{AGN} < 1$.

We estimate the SFRs for non-SF sources in our sample by using the equation provided by \citet{Zhuang+19},
\begin{equation}\label{eq:sfroii}
\text{SFR}_{\text{[OII]}} (\text{M}_\odot \, \text{yr}^{-1}) = \frac{{5.3 \times 10^{-42} (\rm L_{\text{[OII]}} - 0.109 \rm L_{\text{[OIII]}}) \, \text{(erg s}^{-1})}}{\rm F(x)},
\end{equation}
Here, $\rm SFR_{[OII]}$ is SFR estimated based on \OII\ emission line. $\rm L_{[OII]}$ and $\rm L_{[OIII]}$ represent total extinction-corrected luminosities of \OII\ and \OIII, respectively. The function $\mathrm{F}(x) = -4373.14 + 1463.92x - 163.045x^{2} + 6.04285x^{3}$ is determined from the dependence of the extinction-corrected \NII /\OII\ ratio on the oxygen abundance $\mathrm{x} = \log(\mathrm{O}/\mathrm{H}) + 12 = 28.0974 - 7.23631 (\log \mathrm{M}_*) + 0.850344 (\log \mathrm{M}_*)^2 - 0.0318315 (\log \mathrm{M}_*)^3$ (see Figure 1 in \citealp{Zhuang+19}). For SF sources in our sample, we also use Equation \ref{eq:sfroii} but without the parameter $\rm L_{[OIII]}$, as it is utilized to subtract AGN emission in active galaxies. We adopted the stellar mass from the MPA-JHU catalog for our sample. 

\subsection{SFRs Based on \Ha\ Emission Line}\label{subsec:ha_sfr}

The \Ha\ emission line is commonly regarded as a reliable tracer of ongoing SF activity in galaxies. As discussed in Section \ref{subsec:oii_sfr}, to estimate SFRs based on the \Ha\ emission line, we excluded 2,527 sources with \Ha/\Hb\ ratios below 2.86 in SF galaxies and 3.1 in AGNs, which are below the theoretical values of Balmer decrements. All of our sources have an S/N of \Ha\ $>$ 3, as mentioned in Section \ref{section:sample}. Furthermore, to take into account the contamination of AGN emission in the \Ha\ emission line, we excluded the AGN fractions in the \Ha\ emission line for each source in our sample based on the estimated fractions shown in Figure \ref{fig:agnfraction}. 

We utilized equation 2 in \citet{Murphy+11} to estimate SFRs based on \Ha\ emission line with the adopted Kroupa IMF for our sample: 
\begin{equation}\label{eq:sfrhalpa}
\rm SFR_{\text{H}\alpha}\ (\text{M}_\odot \, \text{yr}^{-1}) = 5.3 \times 10^{-42}\ \rm L_{\text{H}\alpha}\ (\text{erg} \, \text{s}^{-1}),
\end{equation}
Here, $\rm SFR_{H\alpha}$ is SFR estimated based on \Ha\ emission line. $\rm L_{H\alpha}$ is dust-extinction corrected \Ha\ luminosity.  

\subsection{SFRs Based on Observed Infrared Fluxes}\label{subsec:ir_sfr}

To compare the SFRs determined by the \OII\ and \Ha\ emission lines in our sample, we also utilize IR luminosity to estimate the IR-based SFRs for our sample. Firstly, we performed a cross-matching procedure between our sample and the AKARI/FIS catalog \citep{Yamamura+10}. This catalog provided flux measurements in four distinct bands: 65 $\rm \mu$m, 90 $\rm \mu$m, 140 $\rm \mu$m, and 160 $\rm \mu$m. To ensure the reliability of our analysis, we specifically focused on high-quality detection (indicated by the quality flag FQUAL $=$ 3) in the 90 $\rm \mu$m band, as it closely aligns with the peak of the SED fitting within the IR range. Through this process, we successfully identified 2,439 counterpart sources. Secondly, we utilized the Herschel/PACS catalog \citep{Pilbratt+10} to access flux measurements across the 70 $\rm \mu$m, 100 $\rm \mu$m, and 160 $\rm \mu$m bands. Employing a matching radius of 3$\arcsec$ and a criterion of S/N ratio greater than 3, we identified counterparts in the 100 $\rm \mu$m band for a total of 1,748 sources. Finally, based on the provided IR luminosity of a total of 4,187 sources, we determine the SFRs by adopting the equation proposed by \citet{Kennicutt+98} revised for the Kroupa IMF,  
\begin{equation}\label{eq:sfr_ir}
\rm \log SFR_{\text{FIR}}\ (\text{M}_\odot \, \text{yr}^{-1}) = \rm \log L_{\text{FIR}}\ (\text{erg} \, \text{s}^{-1}) - 43.519,
\end{equation}

Here, $\rm SFR_{FIR}$ is estimated based on the observed FIR fluxes of the AKARI/Herschel catalog. $\rm L_{FIR}$ represents the total IR luminosity, encompassing the entire range of near-infrared (NIR) wavelengths from 0.7 to 2.5 $\rm \mu$m, MIR wavelengths from 2.5 to 25 $\rm \mu$m, and FIR wavelengths from 25 to 1000 $\rm \mu$m. Therefore, our determined SFRs using the monochromatic luminosity at 90 and 100 $\rm \mu$m may have associated uncertainties. However, for our statistical comparison, these uncertainties have a relatively minor impact due to the dominance of FIR luminosity across the entire IR spectrum. In addition, the AKARI and Herschel fluxes represent different wavelength regions of the SED at 90 $\rm \mu$m and 100 $\rm \mu$m, respectively. The rest-frame of each source may vary due to redshift, which can introduce uncertainties. These effects may contribute to $\sim$2\% uncertainty in our determined SFRs \citep[e.g.,][]{Matsuoka+15, Woo+20}.

\subsection{SFRs Based on Predicted Infrared Flux by the ANN}\label{subsec:ir_ann}

Due to the limited number of sources in our sample that provide fluxes from the AKARI and Herschel catalogs, we have employed the catalog created by \citet{Ellison+16a}. This catalog utilizes the ANN technique to predict the total IR luminosity. \citet{Ellison+16a} used a training set of 1,136 galaxies from the Herschel Stripe 82 sample to predict the total IR luminosity for $\sim$330,000 galaxies in the SDSS. By cross-matching our sample with the ANN catalog, we identified 100,017 counterparts. Among them, there were 62,902 SF, 18,793 composite, 12,194 Seyfert, and 6,128 LINER galaxies. These counterparts were selected based on the criterion that the uncertainty of the ANN-predicted IR luminosity ($\rm \sigma_{ANN}$) was less than 0.3. Following the equation \ref{eq:sfr_ir}, we determine the SFRs based on the ANN for our sources as follows,
\begin{equation}\label{eq:sfr_ann}
\rm \log SFR_{\text{ANN}}\ (\text{M}_\odot \, \text{yr}^{-1}) = \rm \log L_{\text{AIR}}\ (\text{erg} \, \text{s}^{-1}) - 43.519,
\end{equation}
Here, $\rm SFR_{ANN}$ is estimated based on the predicted IR fluxes of the ANN catalog. $\rm L_{AIR}$ is the predicted total IR luminosity by the ANN catalog with $\rm \sigma_{ANN}$ $<$ 0.3. 

\subsection{SFRs Based on the MPA-JHU and GSWLC catalogs}\label{subsec:mpajhu_sfr}

Using a large sample from the SDSS, the MPA-JHU catalog \citep{Brinchmann+04} utilized the \Ha\ emission line to estimate SFRs for SF galaxies. They also employed the anti-correlation between $\rm D_{n}4000$ and specific SFRs from SF galaxies to determine SFRs for non-SF objects. The SFRs of the MPA-JHU catalog, $\rm SFR_{Dn4000}$, have been widely used in the literature. However, some studies have found that the $\rm SFR_{Dn4000}$ values for non-SF sources are lower ($\sim$0.3 dex) compared to SFRs derived from other diagnostics such as \OII, IR, and ANN \citep[e.g.,][]{Ellison+16a, Ellison+16b, Zhuang+20}. In this study, utilizing our large sample, we aim to use the estimated SFRs based on \OII, \Ha, IR, and ANN to compare them with the $\rm SFR_{Dn4000}$ values for both SF and non-SF sources. We will also discuss the differences observed among these SFR estimates.

Additionally, by cross-matching our sample with the GSWLC-2 catalog, we identified 98,341 matched sources. We aim to compare the estimated SFRs based on \OII, \Ha, $\rm D_{n}4000$, IR, and ANN with those determined by UV/optical SED fitting ($\rm SFR_{SED}$) and MIR flux ($\rm SFR_{MIR}$) from \citet{Salim+16}. Since the SFRs provided by \citet{Salim+16} assume a Chabrier IMF \citep{Chabrier03}, based on the Kroupa IMF, we convert all their SFRs by a factor of 0.94.

\subsection{Nucleus Strength Physical Properties}\label{subsec:others}

To investigate the relationship between SFRs and the physical properties of the nucleus strength, we determined the AGN bolometric luminosity, black hole mass, and Eddington ratio for each source in our sample. First, we used the \OIII\ luminosity as a proxy for bolometric luminosity by applying a bolometric correction to the extinction-corrected \OIII\ luminosity ($\rm L_{Bol} = 600 \times L_{OIII}$, \citealp{Kauffmann09}). The bolometric luminosity of our sample falls within the range of $\rm 10^{40} - 10^{47}\ erg\ s^{-1}$. Second, we determined the black hole mass for our sample based on the black hole mass-stellar mass relation \citep{Marconi03}, using the adopted stellar mass from the MPA-JHU catalog. The determined black hole masses for our sample fall within the range of $\rm 10^{6} - 10^{9}\ M_{\odot}$. Finally, based on the estimated bolometric luminosity and black hole mass, we calculated the Eddington ratio for our sample.

\section{Results}\label{section:result}

In this section, we present comparisons between various SFRs mentioned in Section \ref{section:meas}. In this comparison, we present the SFR ratios as a function of redshift to examine their variation from low-to-high redshift galaxies. The $\rm D_{n}4000$ index is also shown as a color scale to indicate the stellar population age. We separate the comparison into a small sub-sample ($\sim$4,100 sources), for which IR fluxes are obtained from the AKARI/Herschel catalog. We also compare SFRs among a large SDSS sample ($\sim$110,000 sources) using SFRs based on the ANN predictions. Table \ref{table:catalog} lists target properties of our sample.


%

\subsection{SFRs by Observed IR Fluxes versus Other Tracers}\label{subsec:sfr_ir}

\begin{figure*}
\centering
        \includegraphics[width=0.9\textwidth]{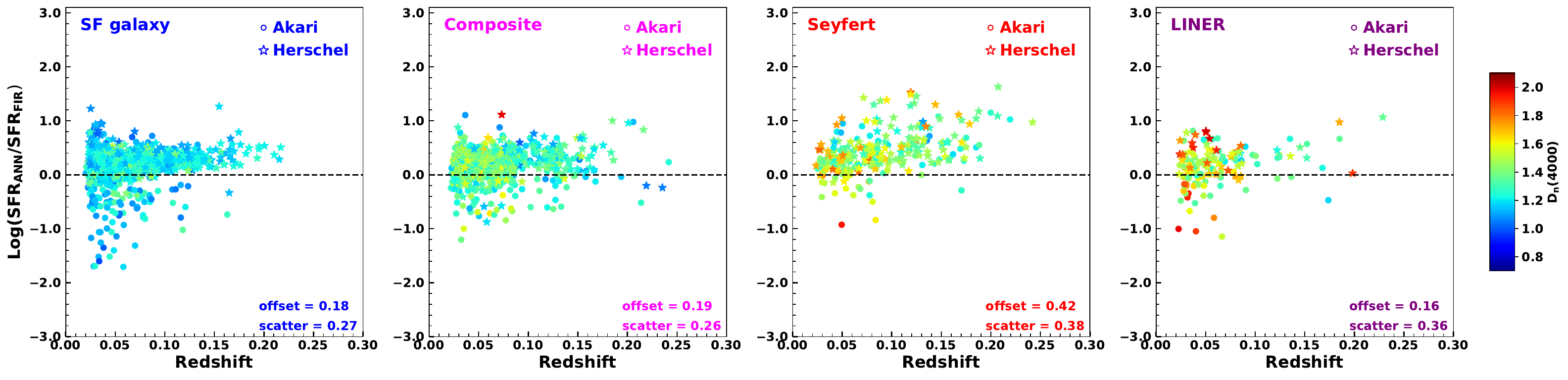}
        \includegraphics[width=0.9\textwidth]{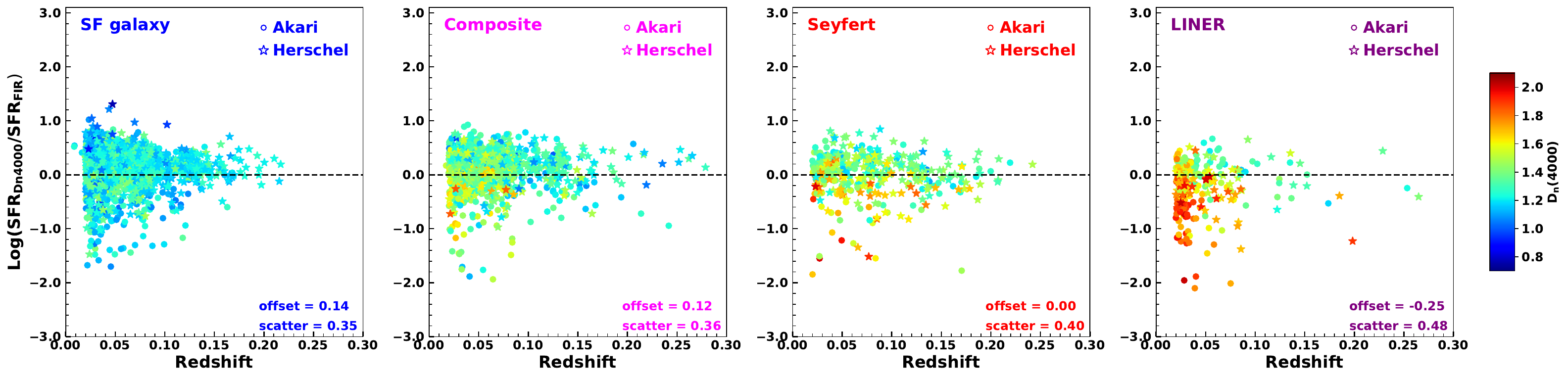}	
        \includegraphics[width=0.9\textwidth]{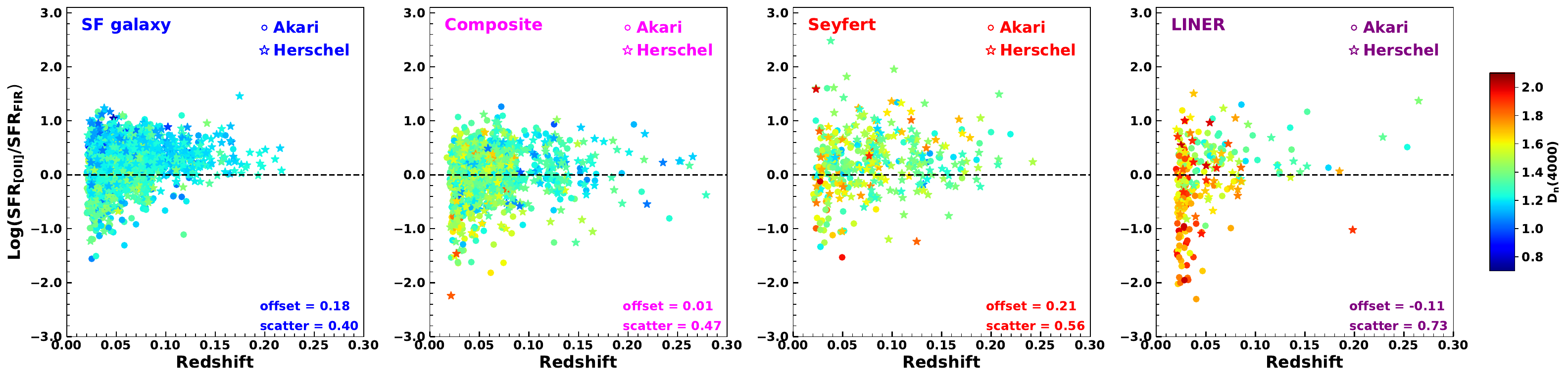}
        \includegraphics[width=0.9\textwidth]{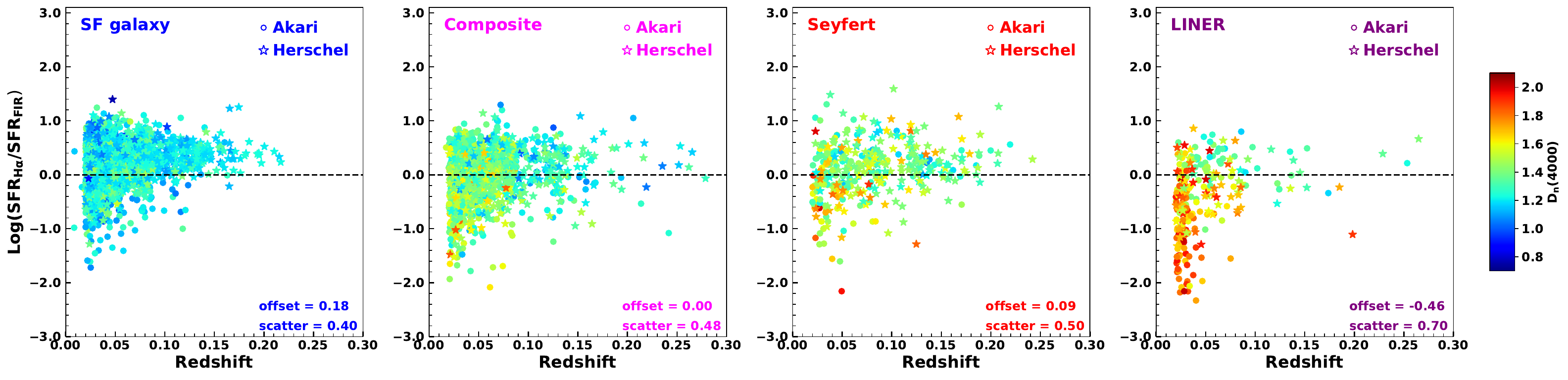}
        \includegraphics[width=0.9\textwidth]{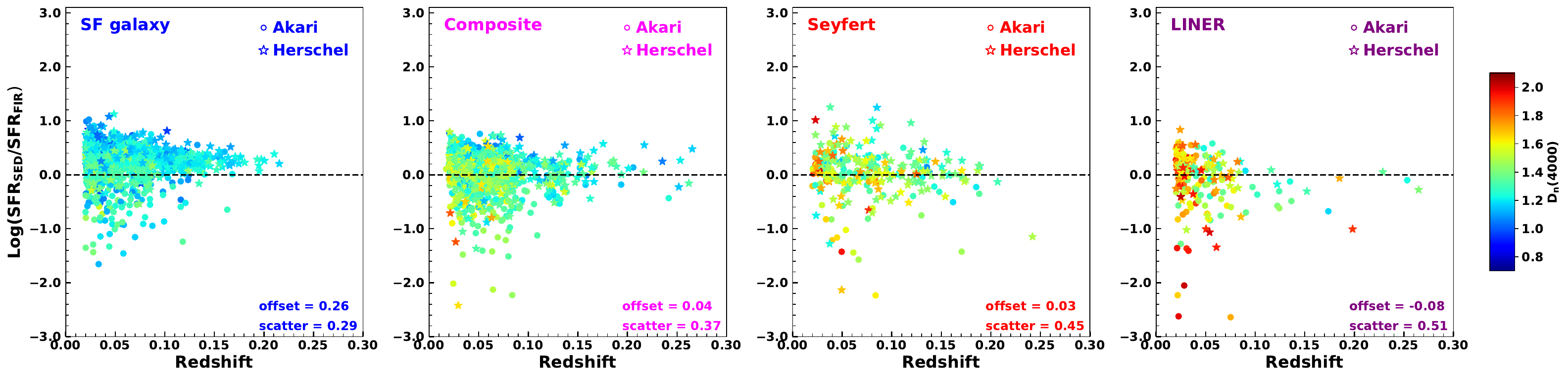}
        \includegraphics[width=0.9\textwidth]{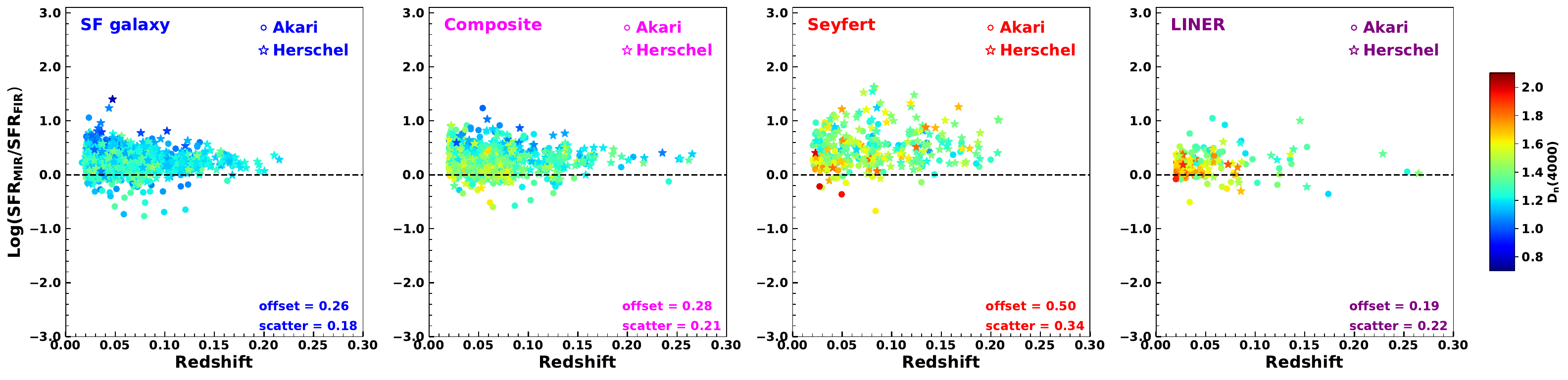}
	\caption{Ratios between SFRs determined by other tracers ($\rm SFR_{ANN}$, $\rm SFR_{Dn4000}$, $\rm SFR_{[OII]}$, $\rm SFR_{H\alpha}$, $\rm SFR_{SED}$ and $\rm SFR_{MIR}$) and $\rm SFR_{FIR}$. The circles and stars represent $\rm SFR_{FIR}$ determined by IR fluxes observed by Akari and Herschel, respectively. Each panel illustrates the classifications among galaxies, such as SF, composite, Seyfert, and LINER, respectively. The color scales represent the $\rm D_{n}(4000)$ values. 
	\label{fig:ir_compare}}
\end{figure*}

In this section, we use SFRs based on IR fluxes as the standard base for comparison with other SFR tracers. Figure \ref{fig:ir_compare} presents the ratios between other SFR tracers ($\rm SFR_{ANN}$, $\rm SFR_{Dn4000}$, $\rm SFR_{[OII]}$, $\rm SFR_{H\alpha}$, $\rm SFR_{SED}$, and $\rm SFR_{MIR}$) and $\rm SFR_{FIR}$ as a function of redshift to examine their difference from low-to-high redshift galaxies. In this comparison, we also utilize the $\mathrm{D_{n}4000}$ to check the stellar population age and to see how the SFR ratios change in relation to their sensitivity in tracing the stellar age of the sample. $\mathrm{D_{n}4000}$ is sensitive to the age of the stellar population, e.g., $\mathrm{D_{n}4000} < 1.8$ indicates blue, active SF galaxies, while higher values represent red, older galaxies \citep[e.g.,][]{Kauffmann+03}. From the compared results, we found that SF galaxies show blue and young stellar population age. The stellar population of composite, Seyfert, and LINER galaxies becomes older and redder. Table \ref{table:result} shows the scatter and offset of the comparison between various SFR tracers. Here we list the result of the comparison for each SFR tracer.


$\rm SFR_{ANN}$ is slightly higher than $\rm SFR_{FIR}$, with an offset of less than 0.19 dex for SF, composite, and LINER galaxies. But for Seyfert galaxies, $\rm SFR_{ANN}$ is higher than $\rm SFR_{FIR}$, with offsets of 0.42 dex and a large scatter of 0.38 dex. Particularly, $\rm SFR_{ANN}$ becomes higher than $\rm SFR_{FIR}$ for higher redshift galaxies.

We find that $\rm SFR_{Dn4000}$ generally shows slightly higher values than $\rm SFR_{FIR}$, with an offset of less than 0.14 dex and a scatter of 0.40 dex for SF, composite, and Seyfert galaxies. However, for LINER galaxies, $\rm SFR_{Dn4000}$ is lower than $\rm SFR_{FIR}$, with an offset of $-$0.25 dex and a scatter of 0.48 dex. We found that $\rm SFR_{Dn4000}$ is quite consistent with $\rm SFR_{FIR}$ for young population galaxies. But, for the older galaxies, $\rm SFR_{Dn4000}$ becomes lower compared to $\rm SFR_{FIR}$. 

$\rm SFR_{[OII]}$ is slightly higher than $\rm SFR_{FIR}$, with an offset of 0.18 dex and a scatter of 0.40 dex for SF galaxies. For composite galaxies, the two indicators are correlated, with $\rm SFR_{[OII]}$ being slightly lower than $\rm SFR_{FIR}$ by 0.01 dex, with a scatter of 0.47 dex. In Seyfert galaxies, $\rm SFR_{[OII]}$ is higher than $\rm SFR_{FIR}$, showing an offset of 0.21 dex and a scatter of 0.56 dex. For LINER galaxies, $\rm SFR_{[OII]}$ is lower than $\rm SFR_{FIR}$, with an offset of $-$0.11 dex and a large scatter of 0.73 dex. We found that targets of older stellar population age at lower redshift galaxies show significantly lower $\rm SFR_{[OII]}$ compared to $\rm SFR_{FIR}$. We will discuss this further in the discussion part.

$\rm SFR_{H\alpha}$ exhibits a comparable correlation with $\rm SFR_{FIR}$. For SF galaxies, $\rm SFR_{H\alpha}$ is slightly higher than $\rm SFR_{FIR}$, with an offset of 0.18 dex and a large scatter of 0.40 dex. In composite galaxies, $\rm SFR_{H\alpha}$ is comparable to $\rm SFR_{FIR}$, with a scatter of 0.48 dex. Seyfert galaxies show a correlation where $\rm SFR_{H\alpha}$ is higher than $\rm SFR_{FIR}$ by 0.09 dex, with a large scatter of 0.50 dex. For LINER galaxies, $\rm SFR_{H\alpha}$ is significantly lower than $\rm SFR_{FIR}$, with an offset of $-$0.46 dex and a scatter of 0.70 dex. Similar to $\rm SFR_{[OII]}$, targets with older and lower redshift show significant low $\rm SFR_{H\alpha}$ compared to $\rm SFR_{FIR}$.


For SF galaxies, $\rm SFR_{SED}$ is higher than $\rm SFR_{FIR}$, with an offset of 0.26 dex and a scatter of 0.29 dex. In non-SF galaxies, there is a strong correlation between the two tracers. Among composite and Seyfert galaxies, the offsets are 0.04 dex and 0.03 dex, with scatters of 0.37 dex and 0.45 dex, respectively. In LINER galaxies, $\rm SFR_{SED}$ is slightly lower than $\rm SFR_{FIR}$, with an offset of $-$0.08 dex and a scatter of 0.51 dex. We also found that targets with older stellar populations tend to show significantly lower $\rm SFR_{SED}$ compared to $\rm SFR_{FIR}$.

$\rm SFR_{MIR}$ is higher than $\rm SFR_{FIR}$ in SF galaxies, showing an offset of 0.26 dex and a scatter of 0.18 dex. Composite galaxies exhibit higher values, with an offset of 0.28 dex and a scatter of 0.21 dex. In Seyfert galaxies, $\rm SFR_{MIR}$ is significantly higher than $\rm SFR_{FIR}$, with an offset of 0.50 dex and a scatter of 0.34 dex. For LINER galaxies, $\rm SFR_{MIR}$ also exceeds $\rm SFR_{FIR}$, with an offset of 0.19 dex and a scatter of 0.22 dex.

\subsection{SFRs by the ANN versus Other Tracers}\label{subsec:sfr_annbase}

Using SFRs based on the ANN prediction, we show the SFR ratios between other tracers and the IR-based (ANN) estimation with a larger sample ($\sim$110,000 sources). With $\mathrm{D_{n}4000}$ as color schemes, we can check the trends between young and old stellar populations in the SFR comparison plots.

For $\rm SFR_{ANN}$ and $\rm SFR_{Dn4000}$, these two SFR indicators show a comparable correlation, with a scatter (0.19 dex) for SF galaxies. In the case of composite galaxies, $\rm SFR_{Dn4000}$ is lower than $\rm SFR_{ANN}$, exhibiting an offset of $-$0.19 dex and a scatter of 0.27 dex. For Seyfert and LINER galaxies, the offsets become significantly larger, with $\rm SFR_{Dn4000}$ being lower than $\rm SFR_{ANN}$ by a scatter of 0.41 dex and offsets of $-$0.59 dex and $-$0.70 dex, respectively. Targets of older stellar population show significantly lower $\rm SFR_{Dn4000}$ compared to $\rm SFR_{ANN}$. Also, as the redshift becomes higher, $\rm SFR_{ANN}$ shows higher values compared to $\rm SFR_{Dn4000}$. 

$\rm SFR_{ANN}$ and $\rm SFR_{[OII]}$ exhibit a comparable correlation for SF galaxies. However, $\rm SFR_{[OII]}$ is slightly smaller than $\rm SFR_{ANN}$, with an offset of $-$0.02 dex and a scatter of 0.31 dex. For non-SF galaxies, the offsets become more pronounced: $-$0.36 dex for composite, $-$0.29 dex for Seyfert, and $-$0.52 dex for LINER galaxies. Furthermore, the correlations display relatively large scatters of 0.42 dex, 0.56 dex, and 0.65 dex for composite, Seyfert, and LINER galaxies, respectively. Similarly to $\rm SFR_{[OII]}$, both $\rm SFR_{ANN}$ and $\rm SFR_{H\alpha}$ show a comparable correlation for SF galaxies. $\rm SFR_{H\alpha}$ is slightly smaller than $\rm SFR_{ANN}$, with an offset of $-$0.11 dex and a scatter of 0.33 dex. For non-SF galaxies, the offsets are $-$0.39 dex (composite), $-$0.51 dex (Seyfert), and $-$0.86 dex (LINER), with scatters of 0.40 dex, 0.46 dex, and 0.57 dex, respectively. We found that targets with older stellar populations tend to have lower $\rm SFR_{[OII]}$ and $\rm SFR_{H\alpha}$ compared to $\rm SFR_{ANN}$, particularly for lower redshift sources.

For SF galaxies, $\rm SFR_{SED}$ is slightly higher than $\rm SFR_{ANN}$, with a small offset of 0.06 dex and a scatter of 0.19 dex. For non-SF galaxies, the offsets become larger, with values of $-$0.21 dex (composite), $-$0.52 dex (Seyfert), and $-$0.65 dex (LINER), and corresponding scatters of 0.32 dex, 0.49 dex, and 0.62 dex for composite, Seyfert, and LINER galaxies, respectively. Targets with older stellar populations tend to have lower $\rm SFR_{SED}$ compared to $\rm SFR_{ANN}$.

$\rm SFR_{ANN}$ and $\rm SFR_{MIR}$ show a good-comparable correlation for both SF and non-SF galaxies. $\rm SFR_{MIR}$ is slightly lower than $\rm SFR_{ANN}$, with an offset of $-$0.01 dex and a scatter of 0.18 dex. Composite galaxies show an offset of $-$0.04 dex and a scatter of 0.22 dex, while Seyfert galaxies exhibit an offset of $-$0.01 dex and a scatter of 0.31 dex. For LINER galaxies, the offset and scatter are $-$0.13 dex and 0.25 dex, respectively.

\begin{figure*}
\centering
        \includegraphics[width=0.95\textwidth]{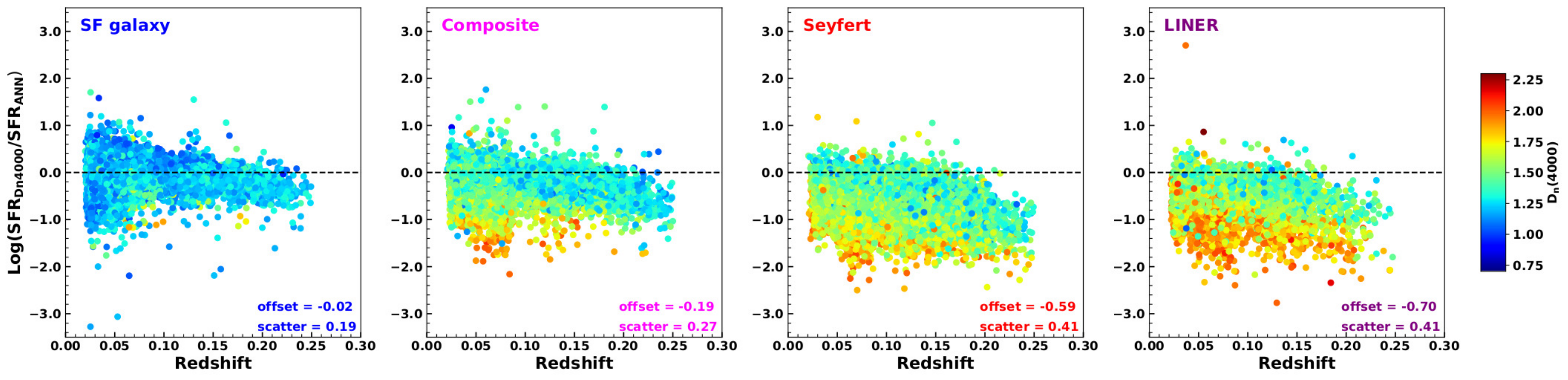}	
        \includegraphics[width=0.95\textwidth]{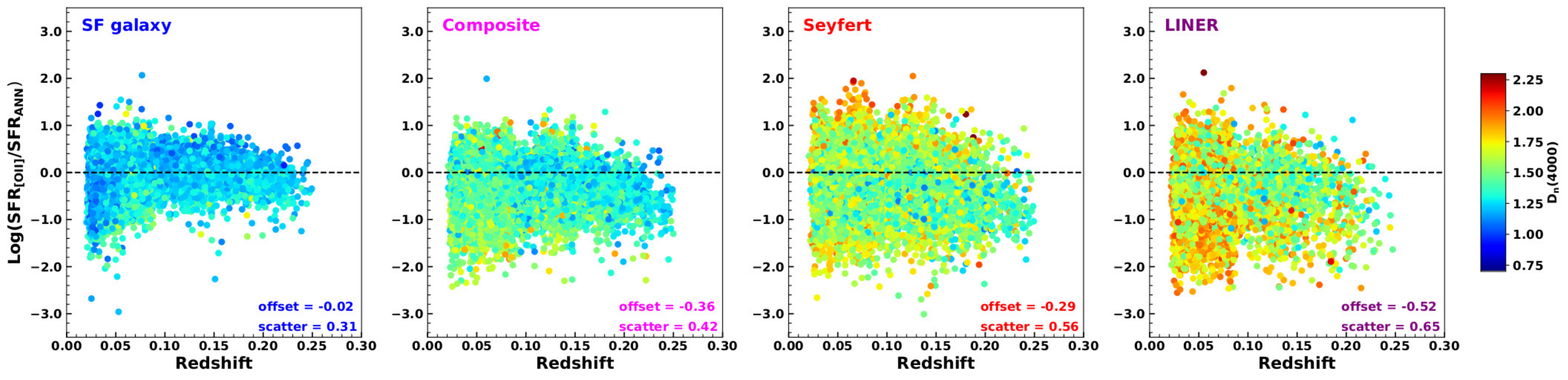}
        \includegraphics[width=0.95\textwidth]{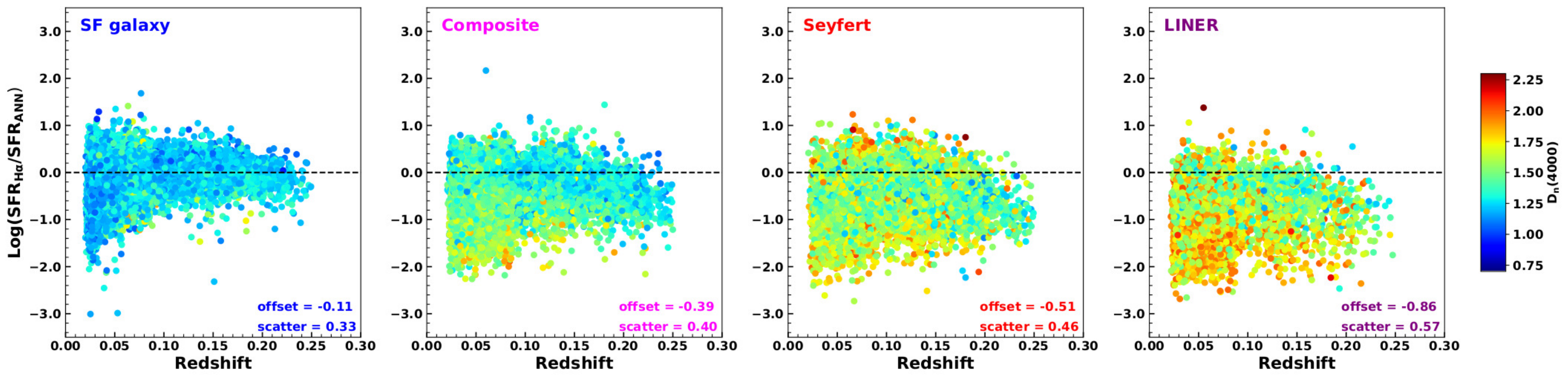}
        \includegraphics[width=0.95\textwidth]{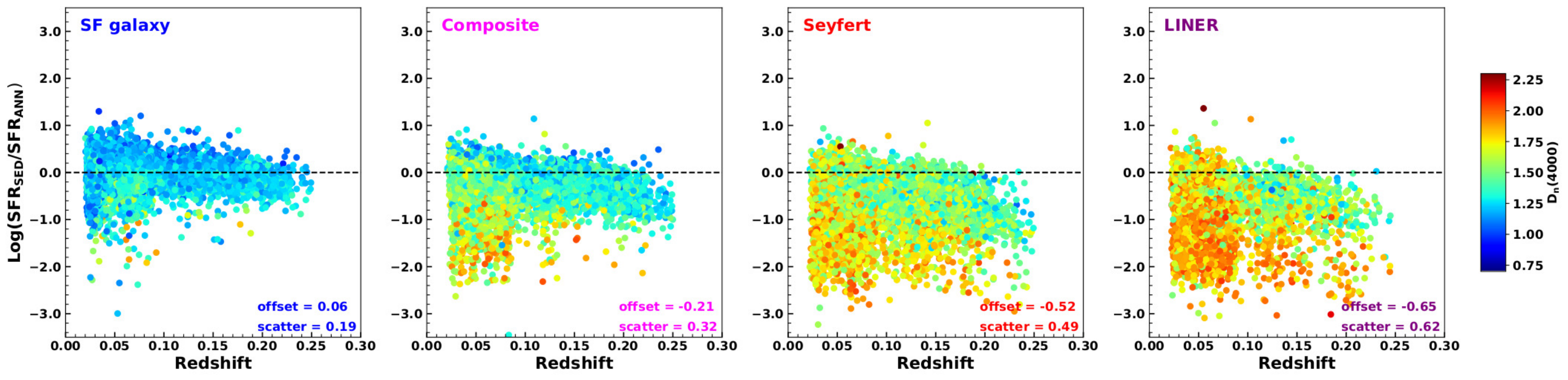}
        \includegraphics[width=0.95\textwidth]{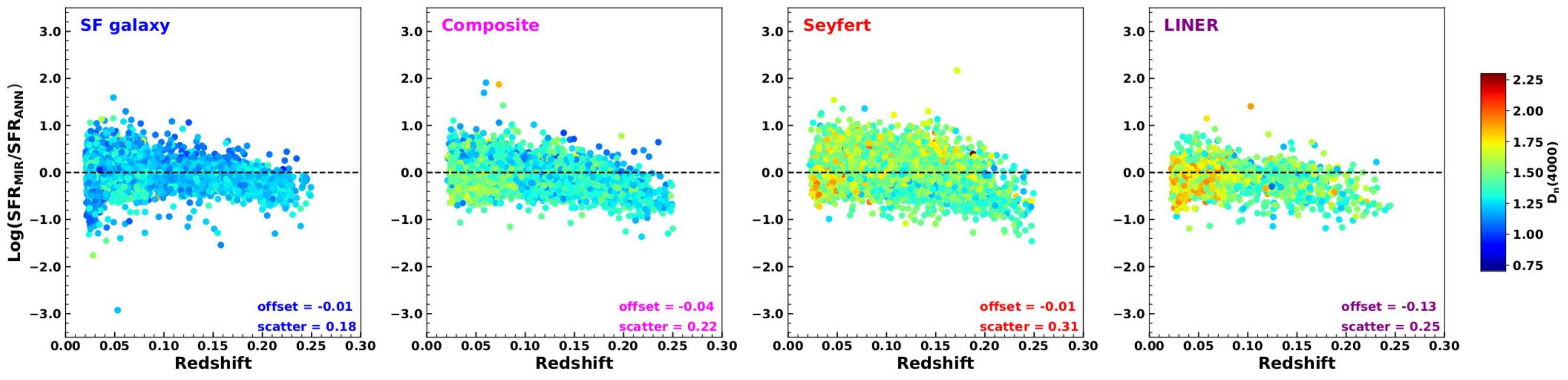}
	\caption{Ratios between SFRs determined by other tracers ($\rm SFR_{Dn4000}$, $\rm SFR_{[OII]}$, $\rm SFR_{H\alpha}$, $\rm SFR_{SED}$ and $\rm SFR_{MIR}$) and $\rm SFR_{ANN}$. $\rm SFR_{ANN}$ determined with $\rm \sigma_{ANN} < 0.3$. Each panel illustrates the classifications among galaxies, such as SF, composite, Seyfert, and LINER, respectively. The color scales represent the $\rm D_{n}(4000)$ values. 
	\label{fig:ann_compare}}
\end{figure*}

\subsection{Additional Comparison of Other SFR Tracers}\label{subsec:sfr_mpabase}

In Table \ref{table:result}, we also compare SFRs determined by $\rm SFR_{Dn4000}$, $\rm SFR_{[OII]}$, $\rm SFR_{H\alpha}$, and $\rm SFR_{SED}$. Among these tracers, we find good correlations with small offset differences for SF galaxies. Notably, there is a strong correlation between $\rm SFR_{Dn4000}$ and $\rm SFR_{SED}$. The largest difference is observed for $\rm SFR_{H\alpha}$, which is lower than $\rm SFR_{SED}$, with an offset of $-$0.18 dex and a scatter of 0.36 dex. For composite galaxies, the largest difference occurs between $\rm SFR_{H\alpha}$ and $\rm SFR_{Dn4000}$, with $\rm SFR_{H\alpha}$ being smaller, exhibiting an offset of $-$0.21 dex and a scatter of 0.41 dex. Additionally, $\rm SFR_{H\alpha}$ is also smaller than $\rm SFR_{SED}$, with a similar offset of $-$0.20 dex and scatter of 0.48 dex. In Seyfert galaxies, the largest difference is observed between $\rm SFR_{[OII]}$ and $\rm SFR_{Dn4000}$, with $\rm SFR_{[OII]}$ showing larger values, an offset of 0.31 dex, and a scatter of 0.67 dex. For LINER galaxies, the largest difference is between $\rm SFR_{[OII]}$ and $\rm SFR_{H\alpha}$, with $\rm SFR_{[OII]}$ being larger, exhibiting an offset of 0.34 dex and a scatter of 0.25 dex.

\subsection{SFRs versus Eddington ratio} \label{subsec:sfr_edd}

In this section, we select one SFR tracer to investigate the relationship between AGN strength and SFRs in different galaxy types. FIR-based SFRs are widely regarded as a robust measure of star formation in the literature. From our comparison of various tracers, we find that $\rm SFR_{SED}$ correlates most strongly with $\rm SFR_{FIR}$ and shows the smallest offsets for non-SF galaxies (Figure~\ref{fig:ir_compare}). We therefore adopt $\rm SFR_{SED}$ to examine the connection between AGN and star formation activities in our sample.

\begin{figure*}
\centering

\includegraphics[width=0.223\textwidth]{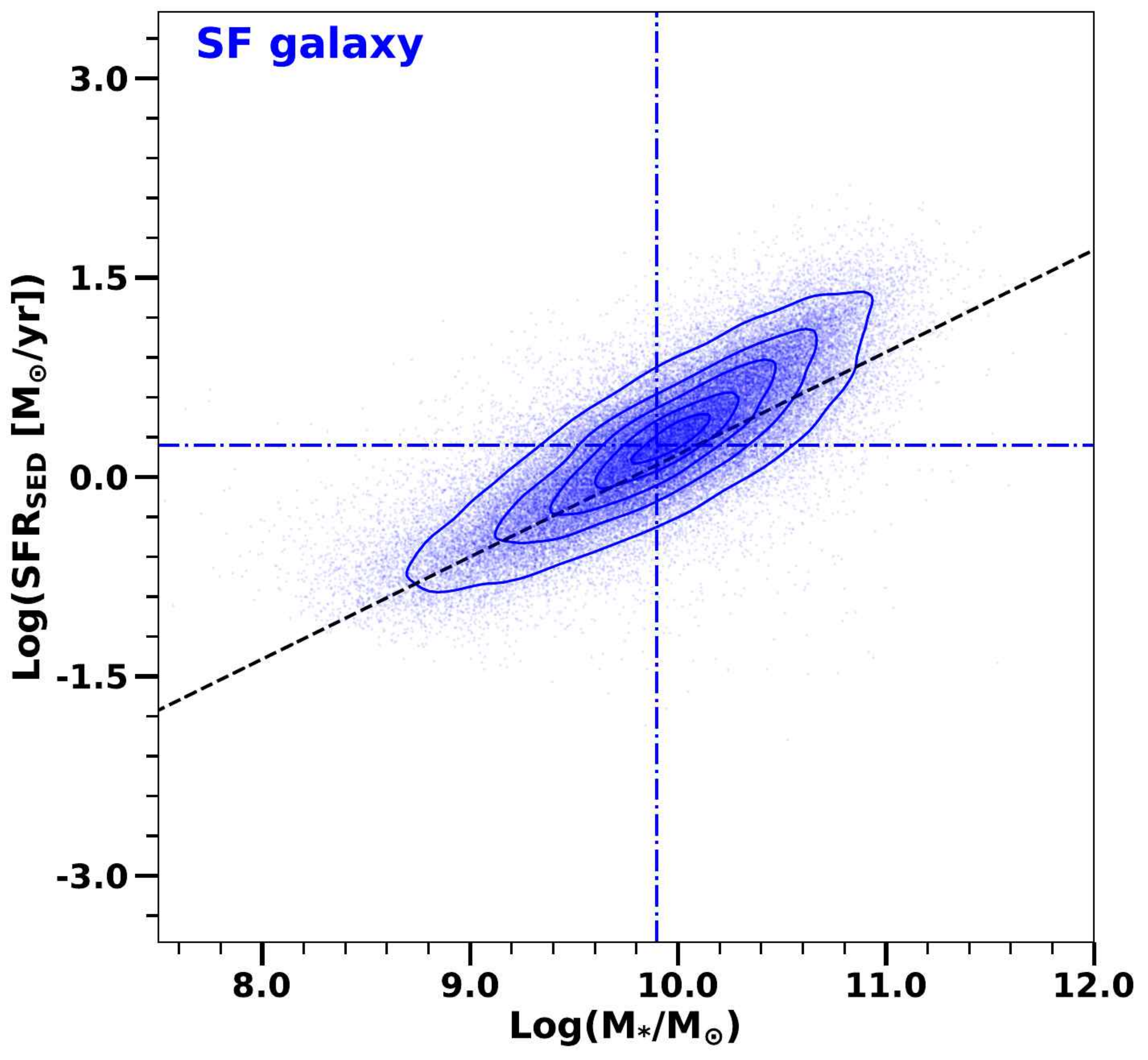}
\includegraphics[width=0.245\textwidth]{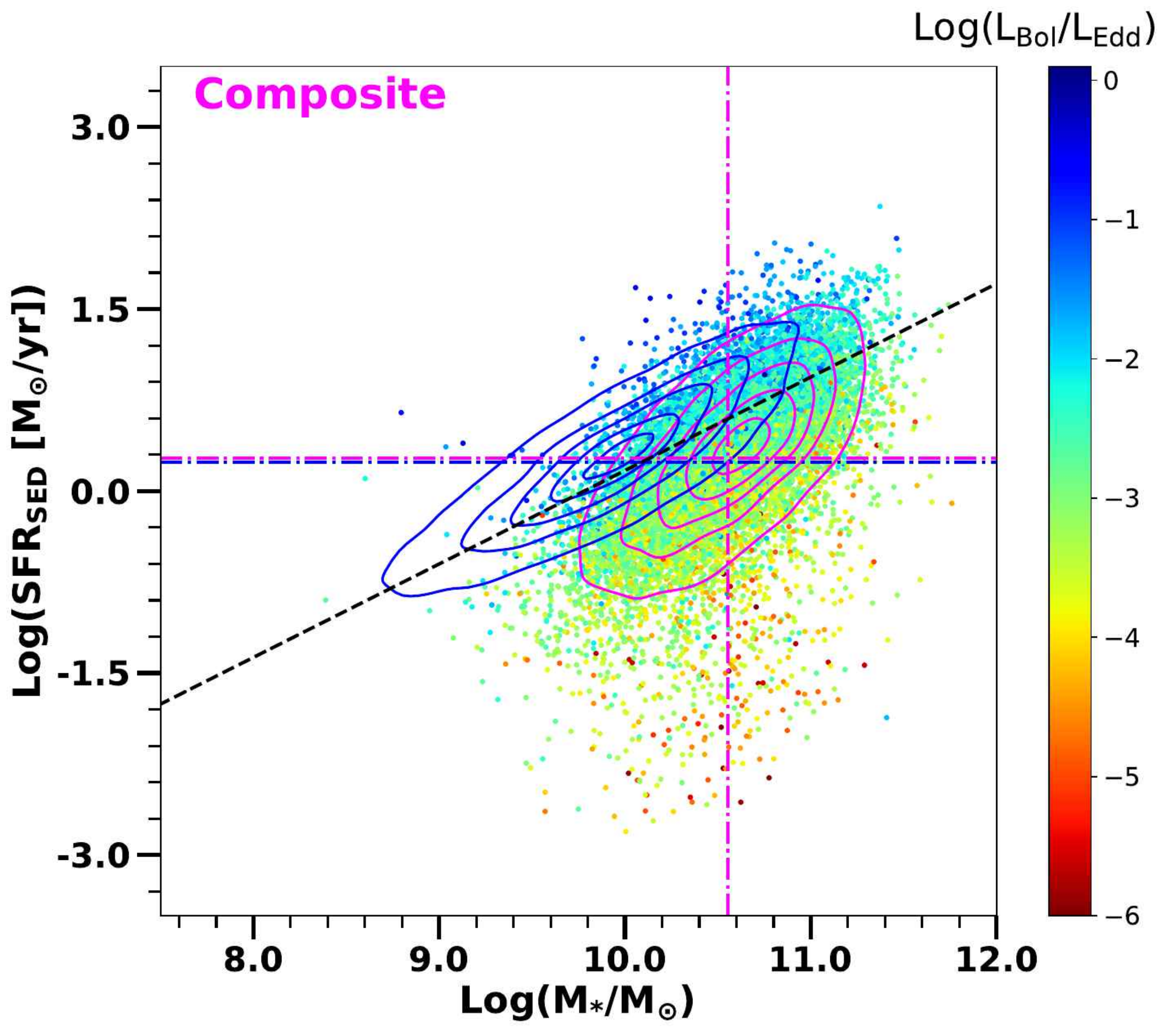}
\includegraphics[width=0.245\textwidth]{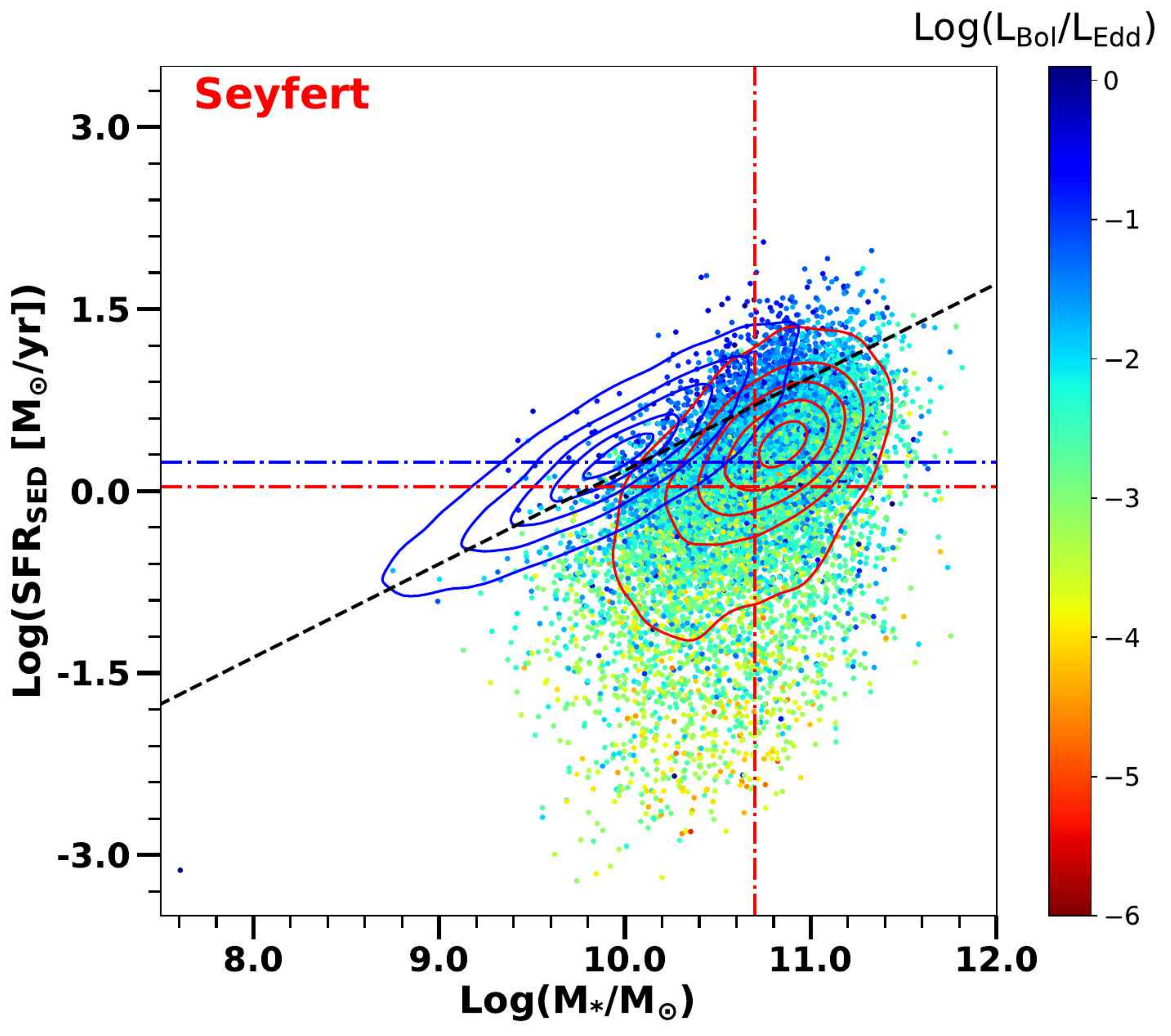}
\includegraphics[width=0.245\textwidth]{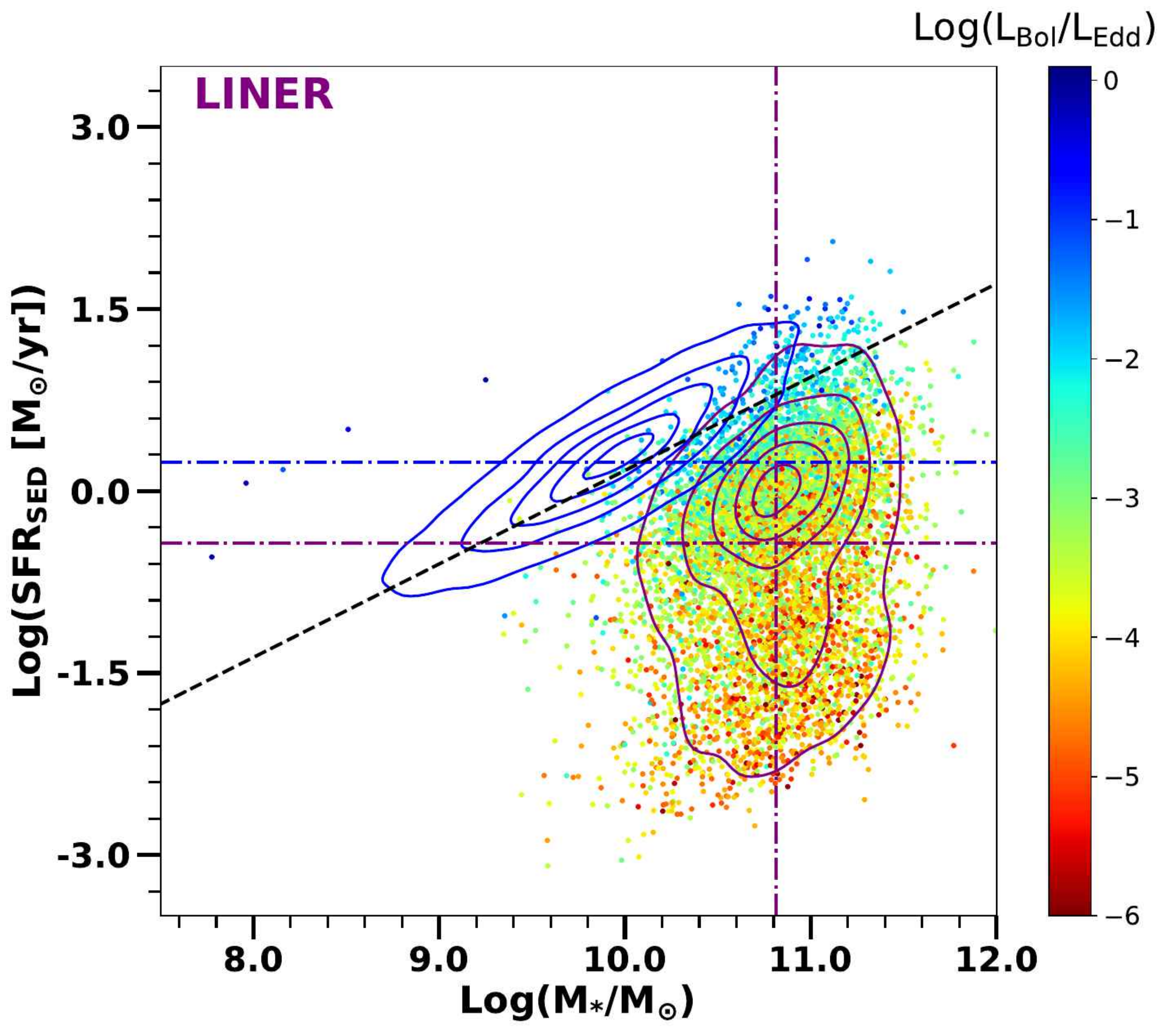}
	\caption{SFRs as a function of stellar mass. SFR values are presented by $\rm SFR_{SED}$. Each panel displays the classifications of SF, composite, Seyfert, and LINER galaxies, respectively. The dashed lines show the local MS relation for blue galaxies determined by \citet{Elbaz+07}. The $\rm SFR-M_{*}$ space is divided into a grid of 150x150 bins, and contour lines are drawn to represent 10, 30, 50, 70, and 90 percent of the maximum number density. The color scales represent Eddington ratio values. The blue contour, representing SF galaxies, is relocated to the panels of composite, Seyfert, and LINER galaxies for a proper comparison. The full set of other SFR tracers as a function of stellar mass is available in the online version.
	\label{fig:contour}}
\end{figure*}

Figure \ref{fig:contour} shows $\rm SFR_{SED}$ as a function of stellar mass. We find that SF galaxies align well with the local MS line for blue galaxies, as determined by \citet{Elbaz+07}. In contrast, composite, Seyfert, and LINER galaxies deviate from the MS line in the $\rm SFR$-$\text{M}_{*}$ relationship, with sources having lower SFRs and Eddington ratios tending to leave the MS line vertically. Interestingly, a clear variation in the distribution of Eddington ratio is observed among composite, Seyfert, and LINER galaxies. Within this distribution, Eddington ratio exhibits higher values on or above the MS line and decreases for sources located below the MS line. Notably, Seyfert galaxies exhibit significantly higher Eddington ratio values compared to both composite and LINER galaxies, suggesting that AGN strength activities are more powerful in Seyfert galaxies. Positive correlations between Eddington luminosity and SFRs have been observed in recent studies (e.g., \citealp[]{Woo+20, Zhuang+20}), highlighting the significant impact of AGN strength activities on the SFRs within their host galaxies. 

\begin{figure*}
\centering
        \includegraphics[width=0.31\textwidth]{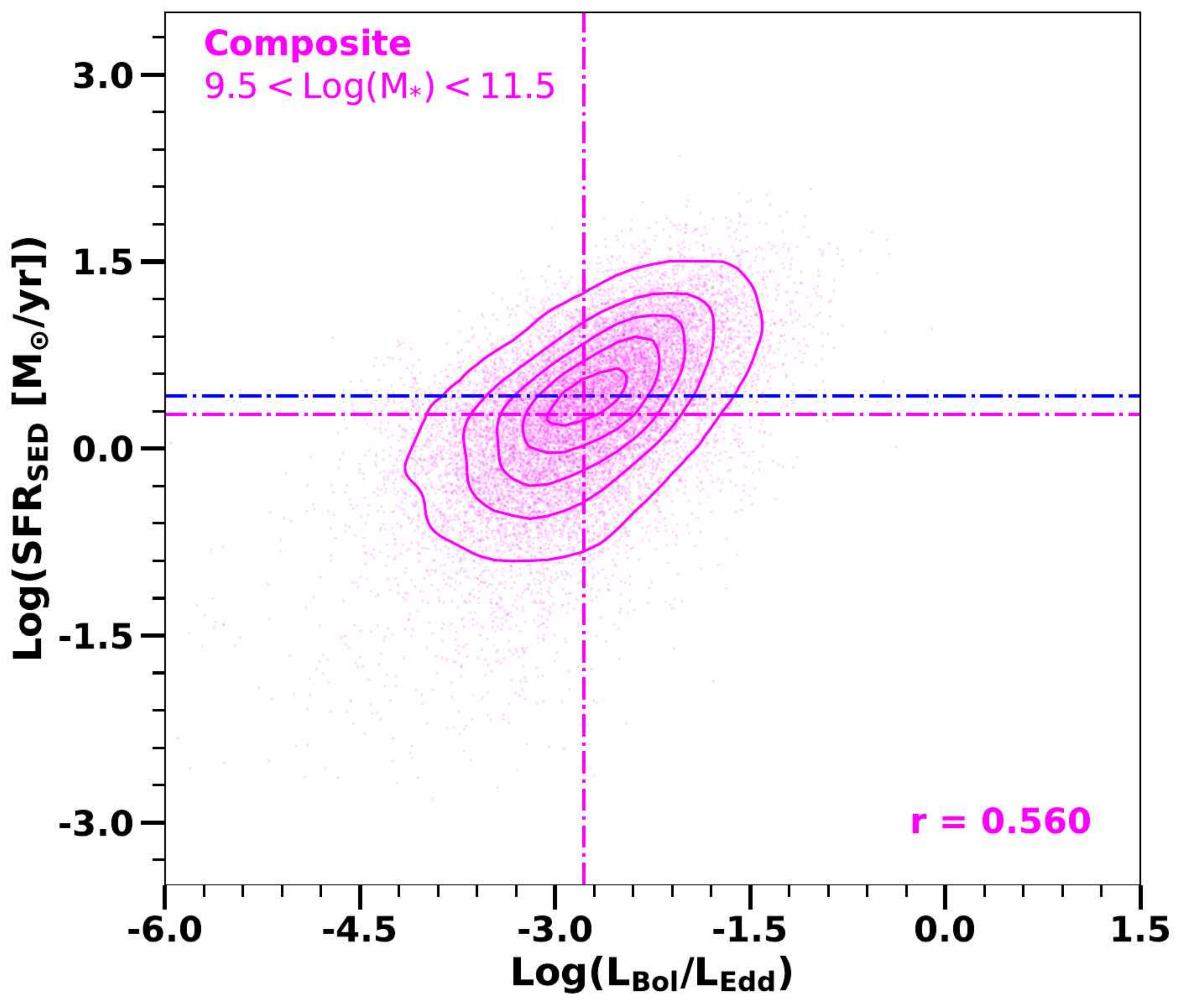}
        \includegraphics[width=0.31\textwidth]{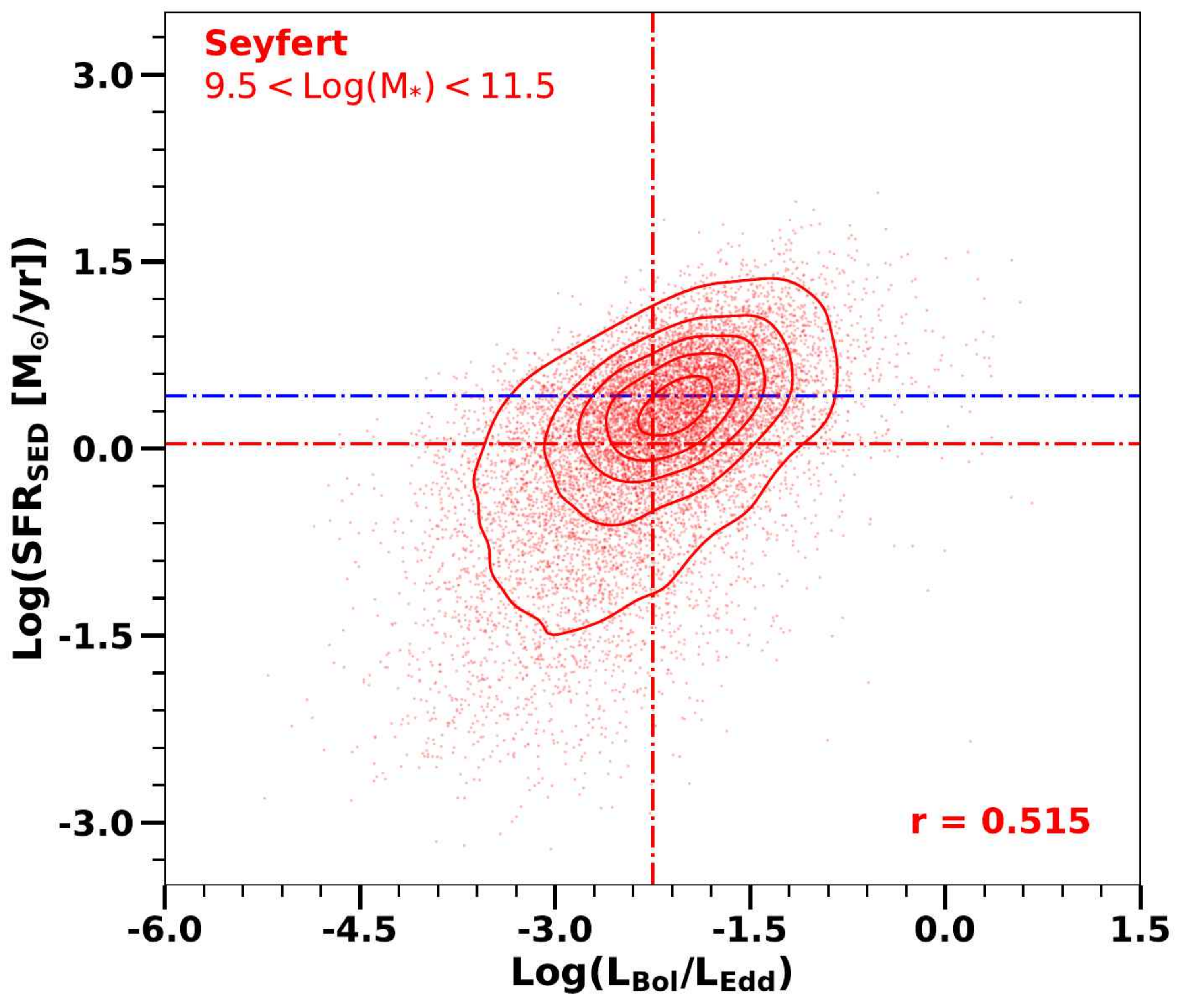}
        \includegraphics[width=0.31\textwidth]{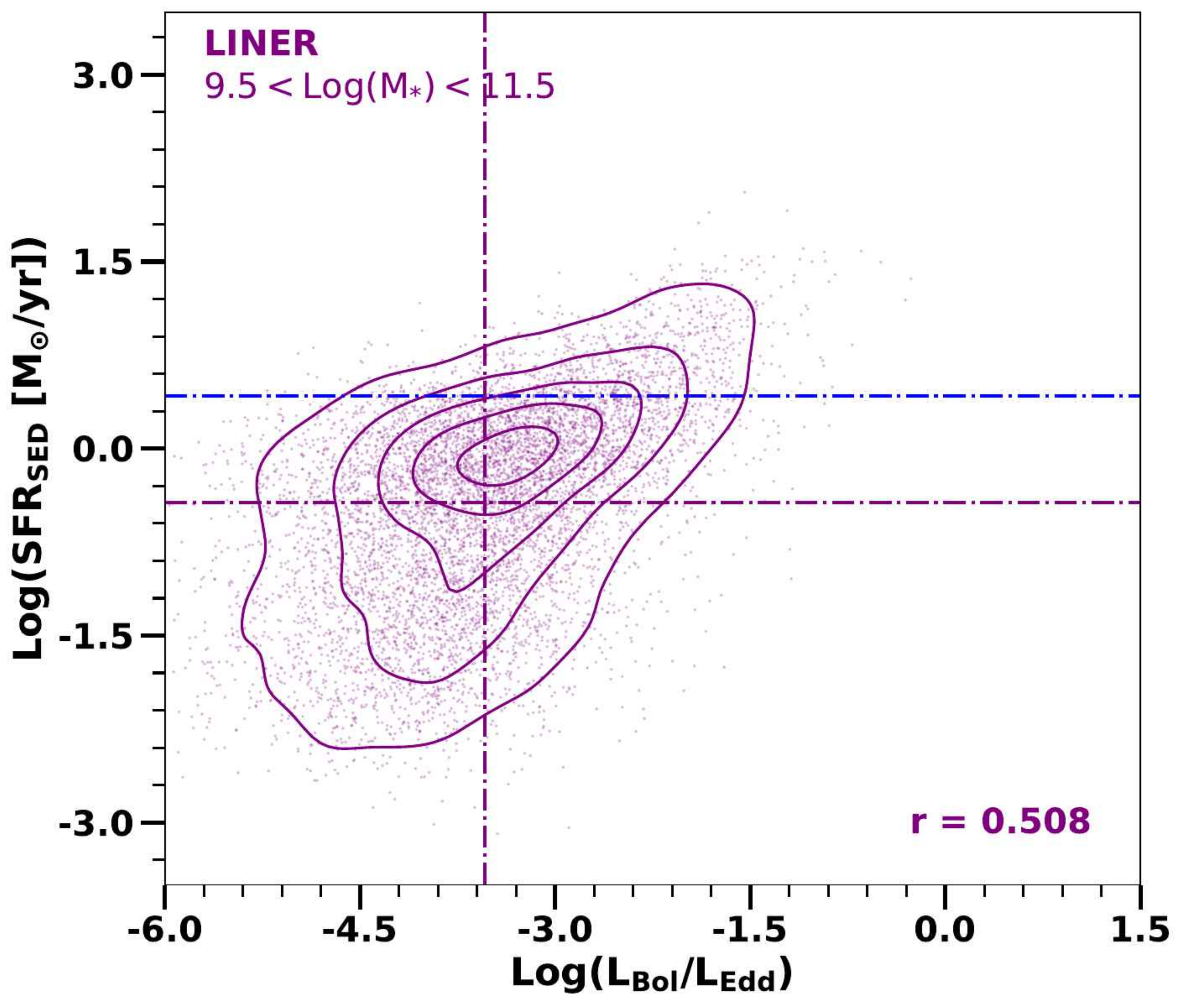}
	\caption{SFRs as a function of Eddington ratio for a stellar mass range from 9.5 $<$ $\rm \log M_{*}$ $<$ 11.5. SFR values are presented by $\rm SFR_{SED}$. Horizontal dashed lines in blue, pink, red, and cyan correspond to the median SFR values for SF, composite, Seyfert, and LINER galaxies, respectively. Vertical dashed lines indicate the median Eddington ratio values for each galaxy type.
	\label{fig:sfr_edd}}
\end{figure*}
\begin{figure*}
\centering
    \includegraphics[width=0.31\textwidth]{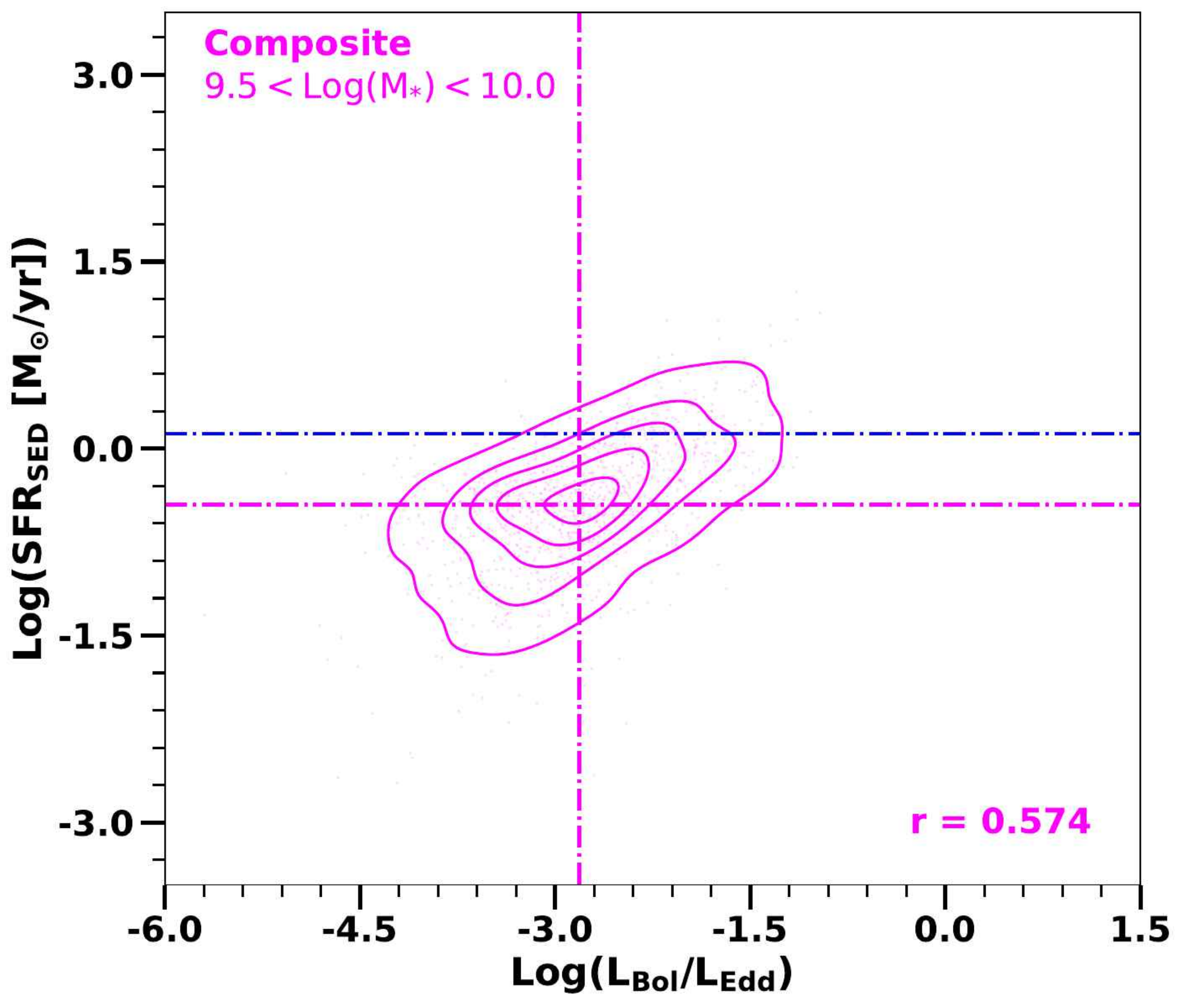}
    \includegraphics[width=0.31\textwidth]{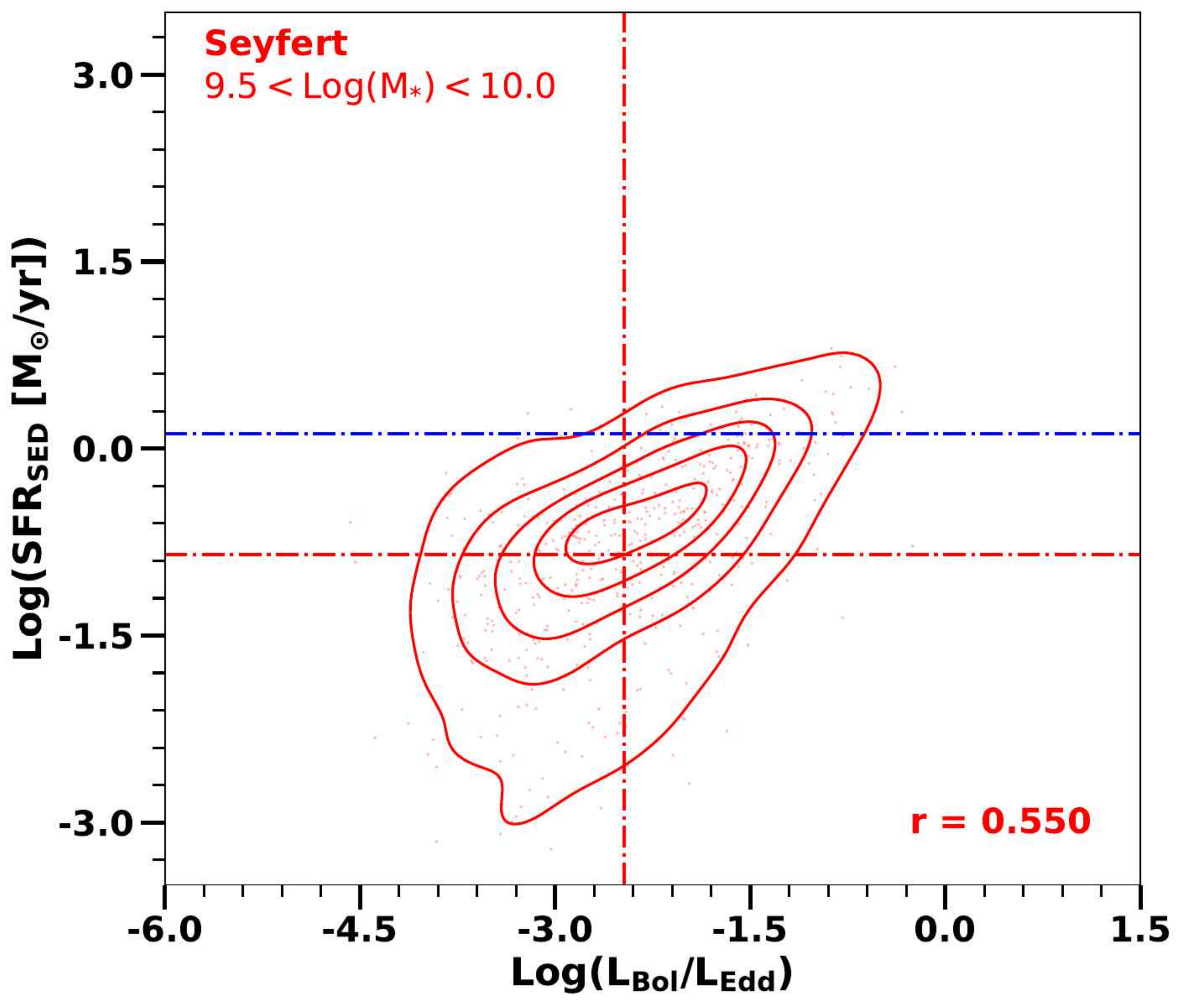}
    \includegraphics[width=0.31\textwidth]{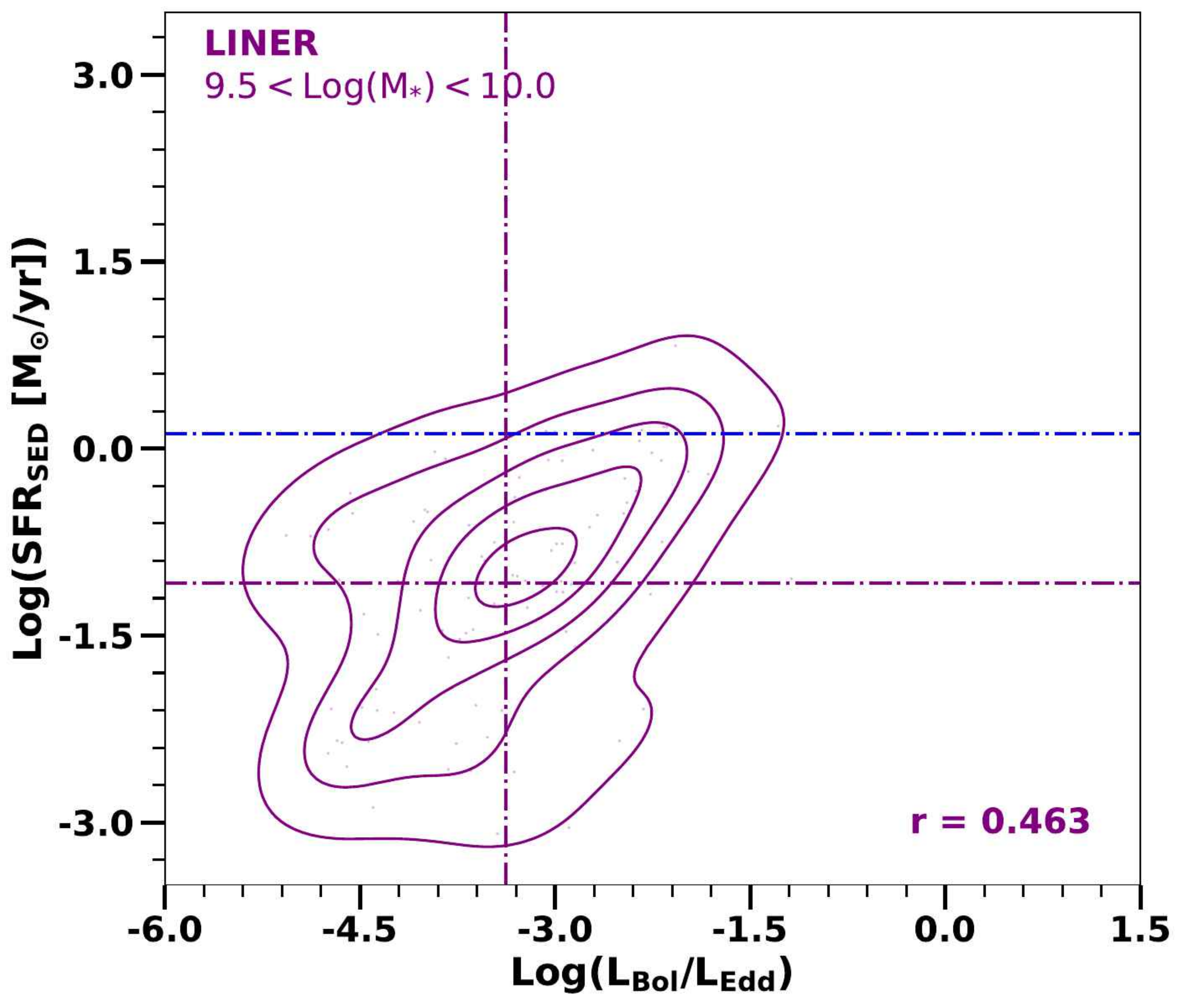}
    \includegraphics[width=0.31\textwidth]{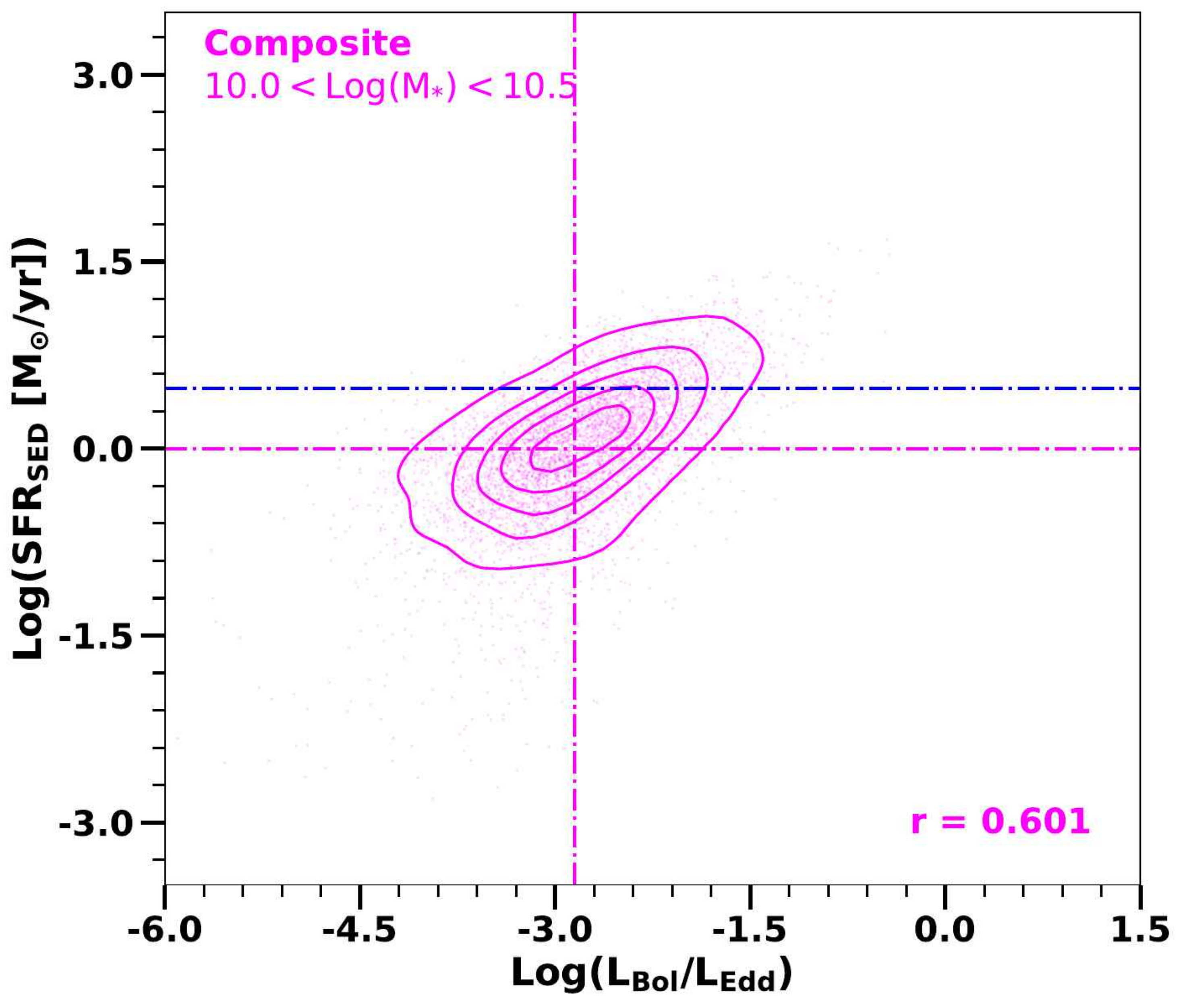}
    \includegraphics[width=0.31\textwidth]{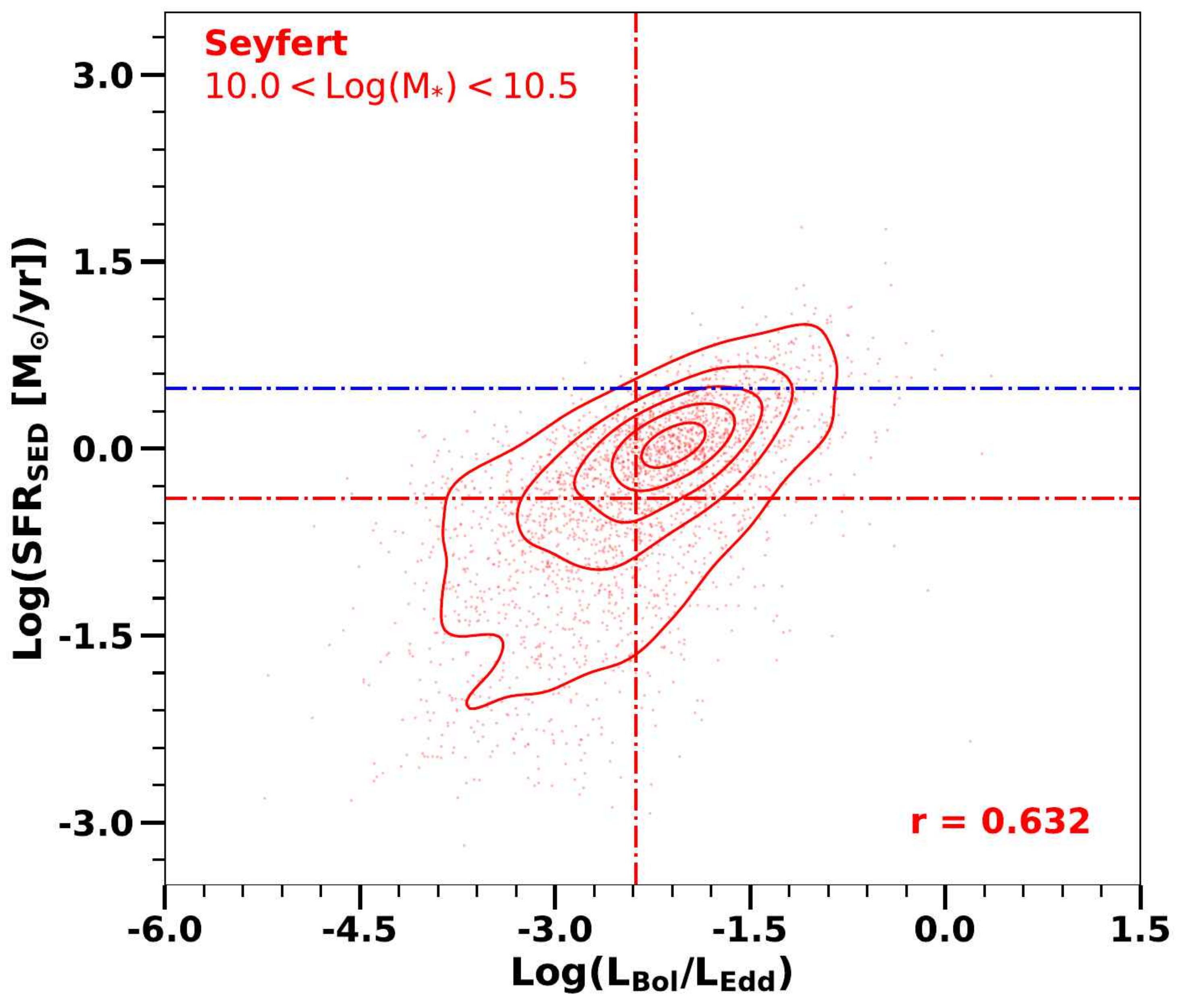}
    \includegraphics[width=0.31\textwidth]{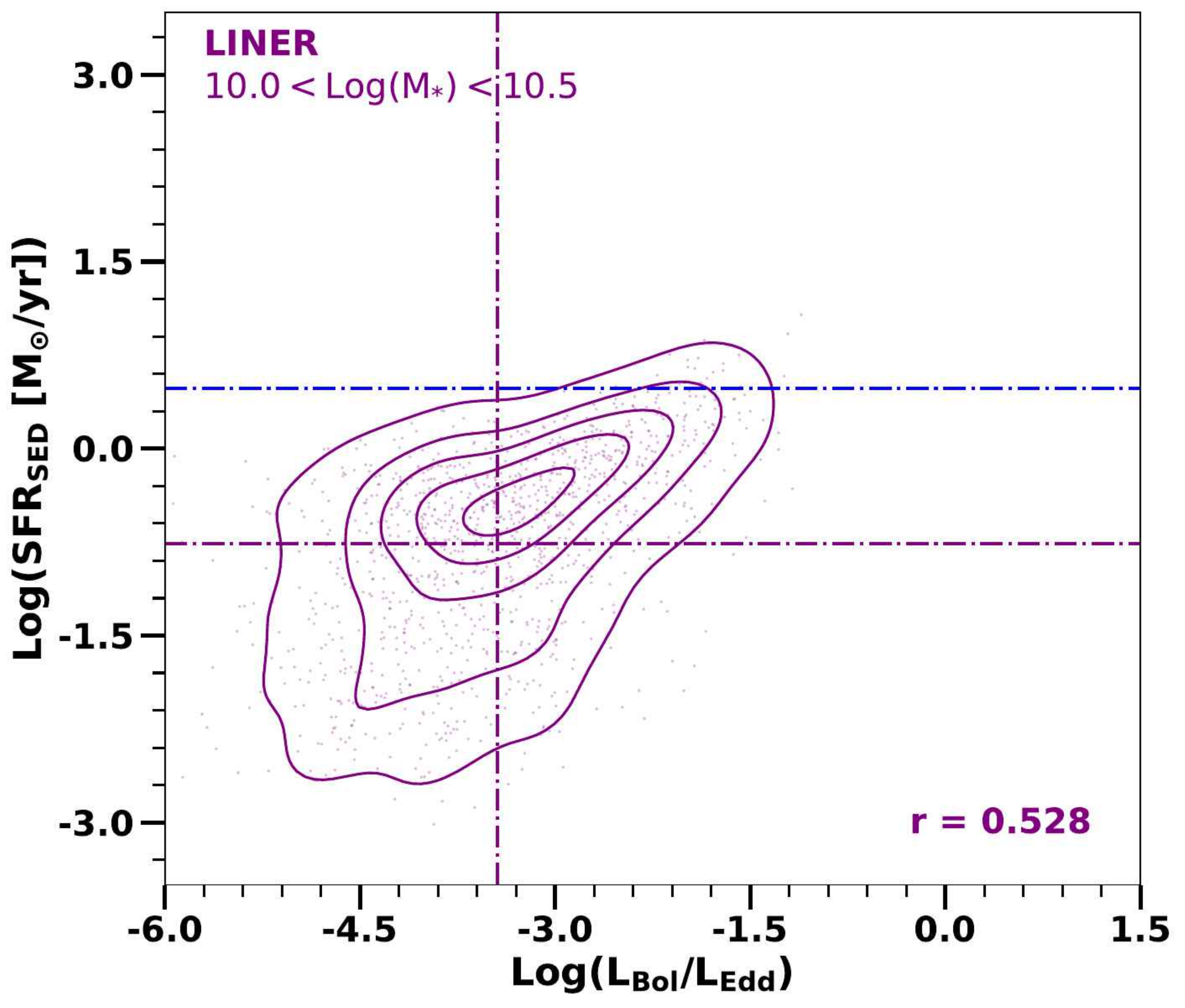}
    \includegraphics[width=0.31\textwidth]{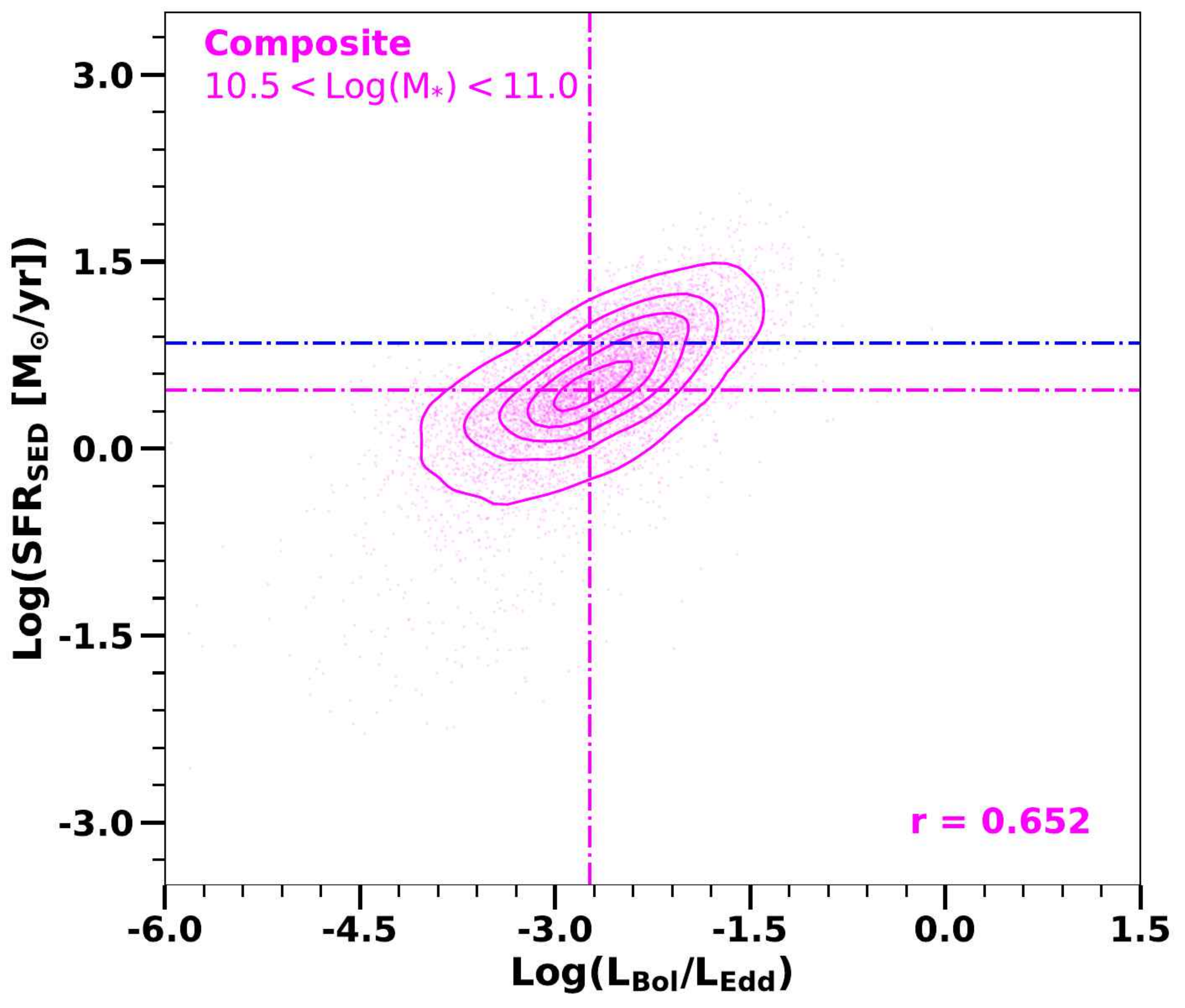}
    \includegraphics[width=0.31\textwidth]{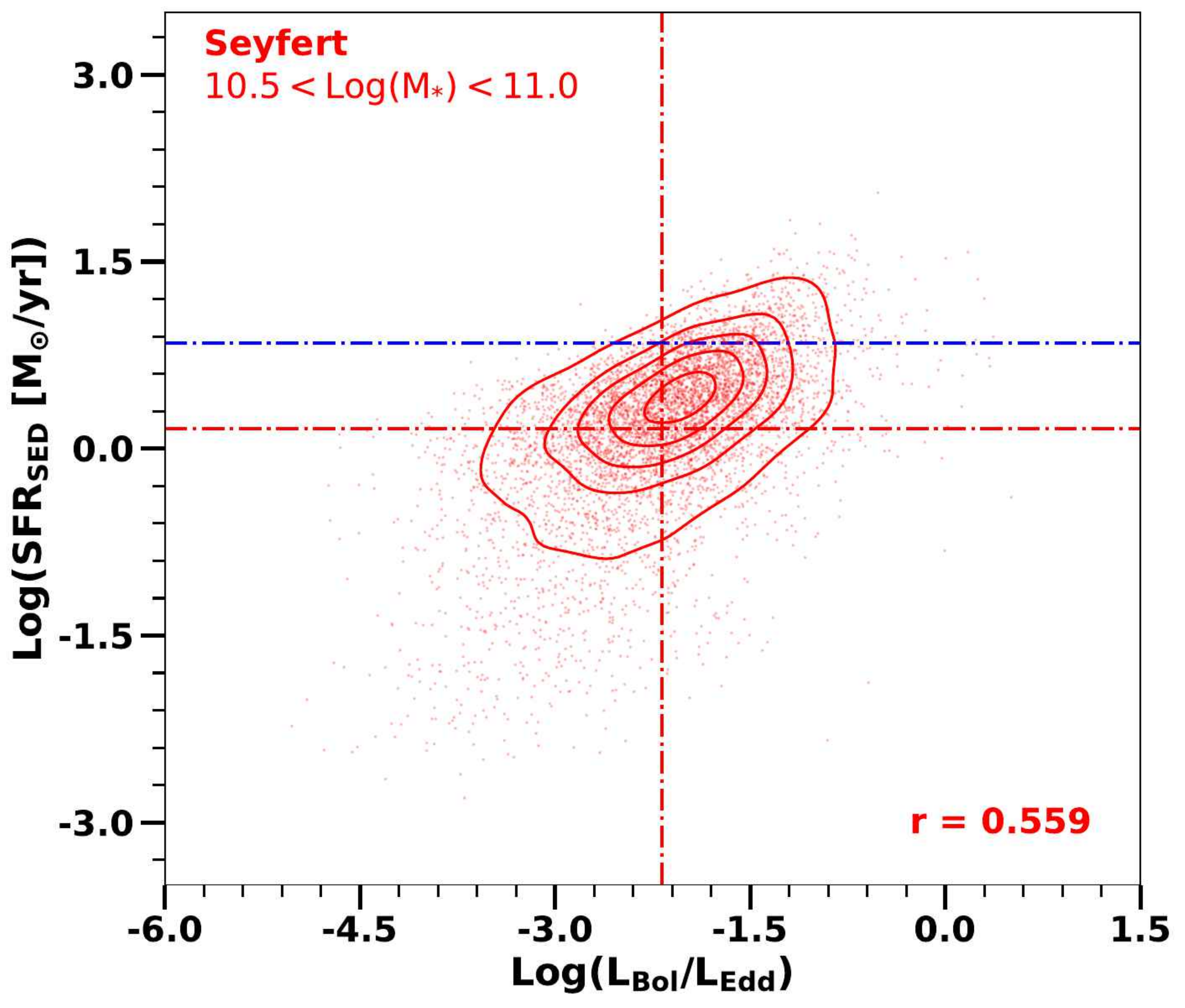}
    \includegraphics[width=0.31\textwidth]{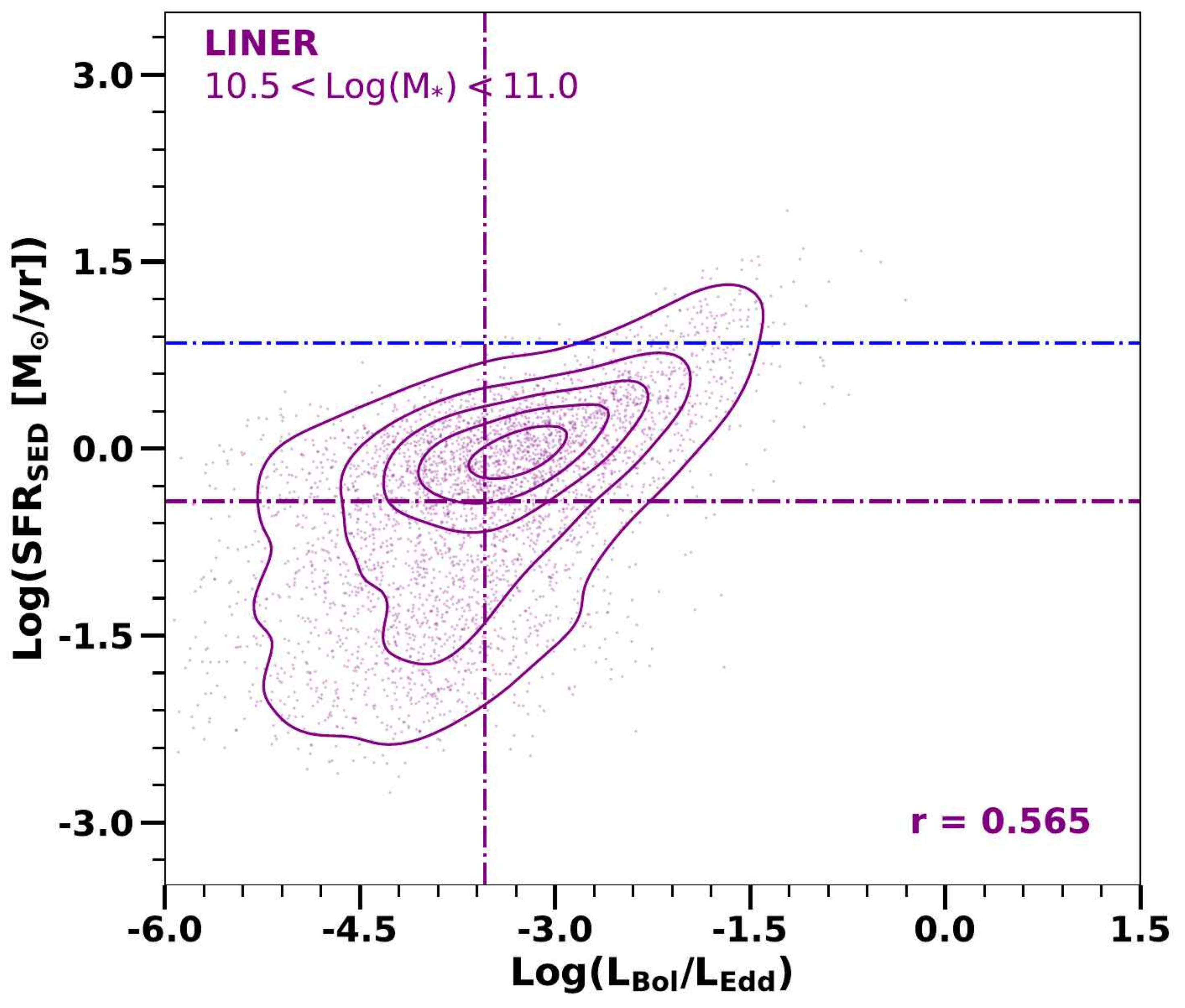}
    \includegraphics[width=0.31\textwidth]{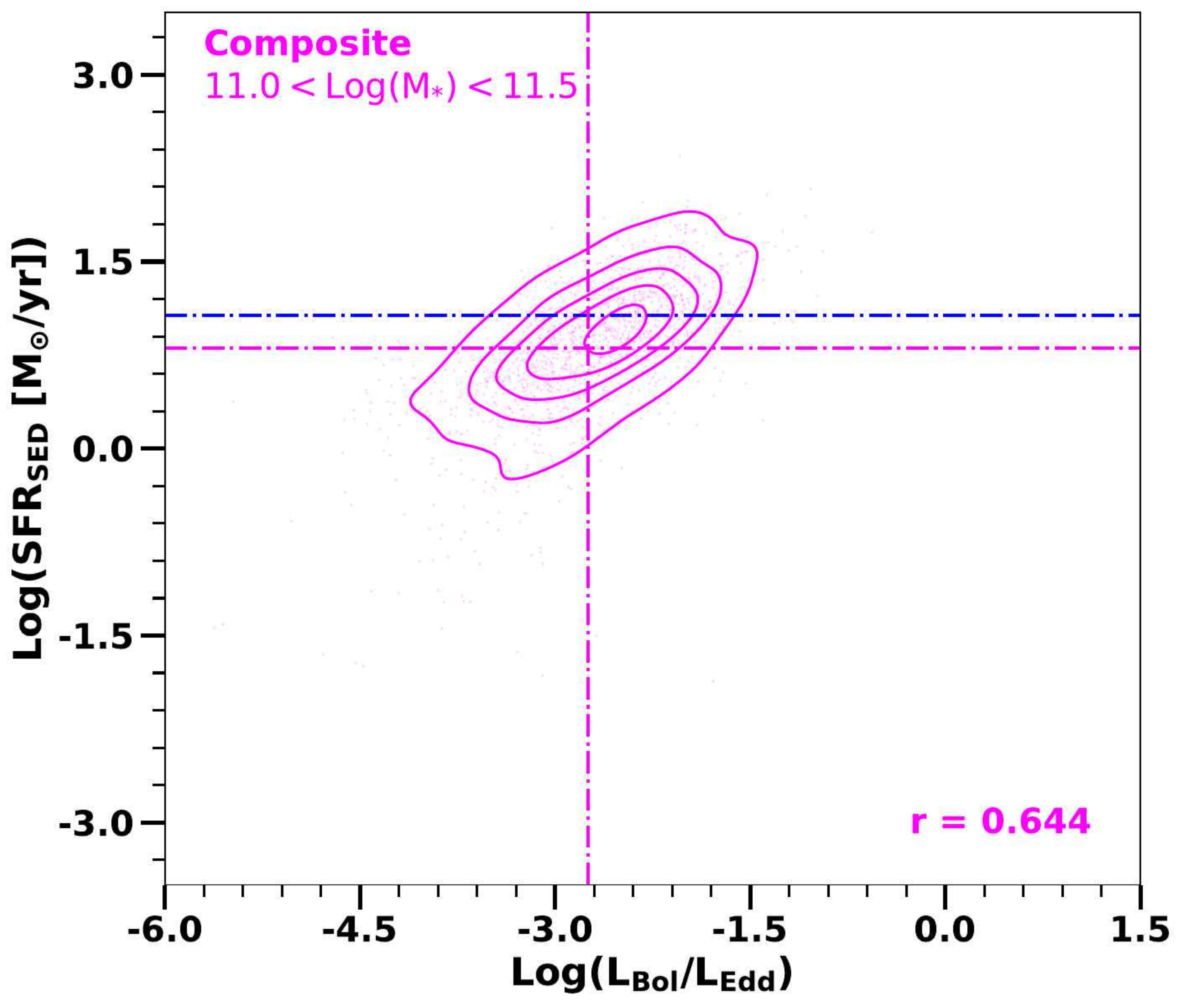}
    \includegraphics[width=0.31\textwidth]{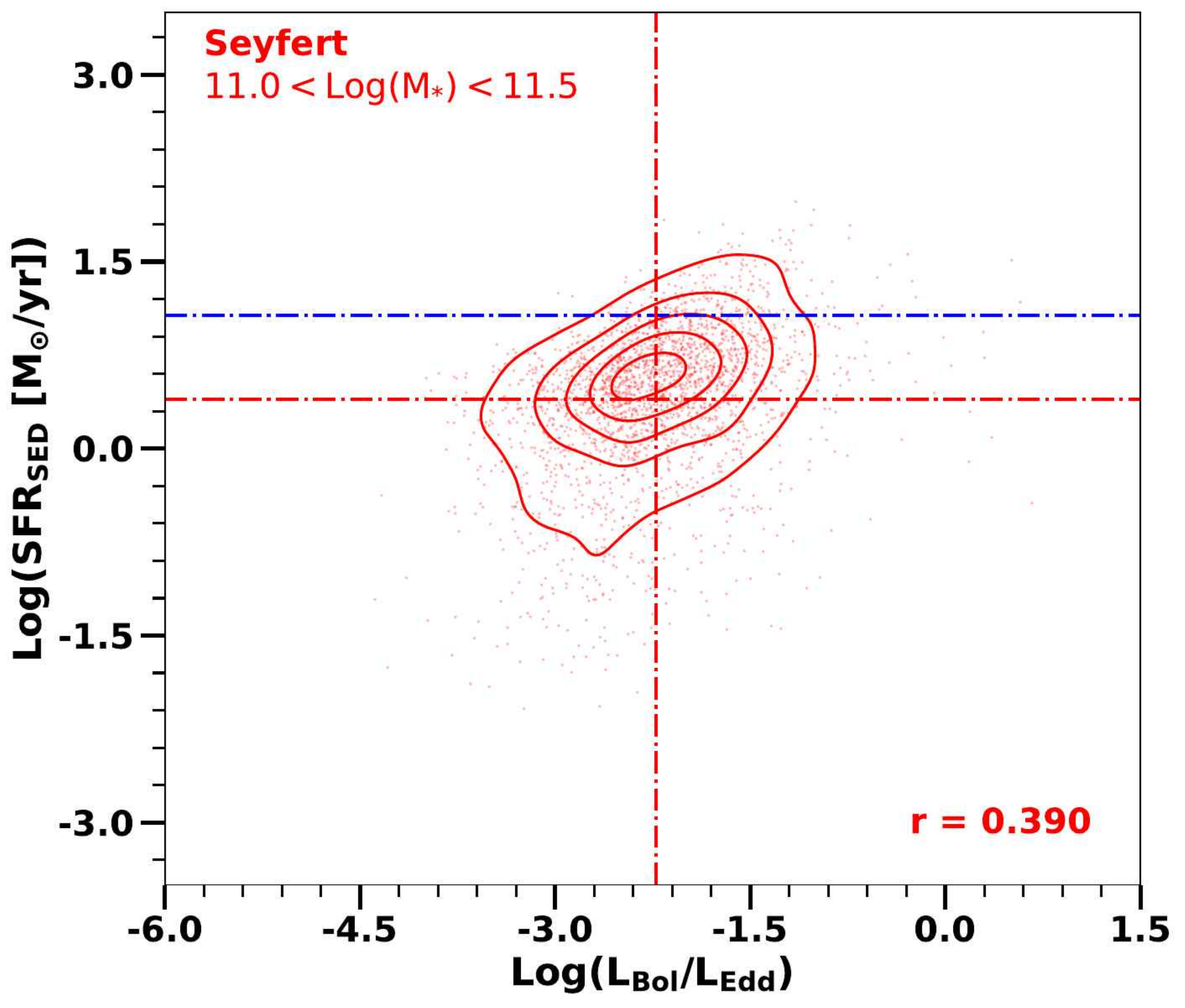}
    \includegraphics[width=0.31\textwidth]{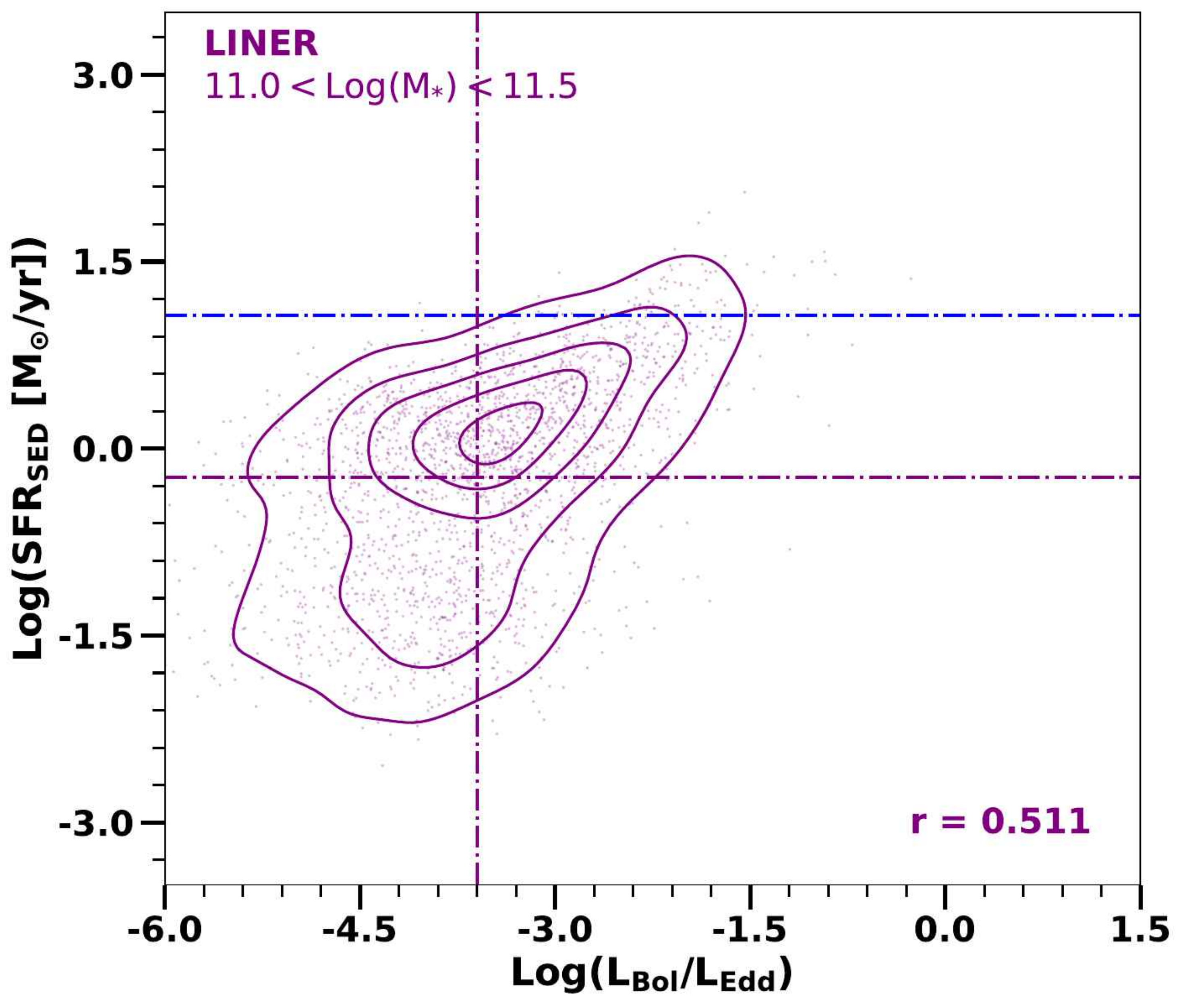}     
    \caption{Similar to Fig \ref{fig:sfr_edd}, but with the stellar mass range divided into four bins of $\rm \log M_{*}$ = 9.5$-$10.0, 10.0$-$10.5, 10.5$-$11.0, and 11.0$-$11.5, respectively. The full set of other SFR tracers as a function of Eddington ratio is available in the online version.
    \label{fig:sfr_edd_all}}
\end{figure*}

Figure \ref{fig:sfr_edd} displays the SFRs as a function of Eddington ratio for a stellar mass range of 9.5 $<$ $\rm \log M_{*}$ $<$ 11.5. Overall, all SFR tracers exhibit a strong correlation with Eddington ratio. To quantify these correlations, we calculate the Spearman correlation coefficient (r) and display it in each panel. Among galaxy types, Seyfert galaxies have the strongest Eddington ratios compared to composite and LINER galaxies. To further investigate this relationship, we divide the stellar mass into four different bins with a size of $\rm \Delta\log M_{*}$ = 0.5, enabling us to examine the correlation between SFRs and Eddington ratio for galaxies within similar stellar mass bins (Figure \ref{fig:sfr_edd_all}). Notably, there are robust correlations between SFRs and Eddington luminosity across all stellar mass bins. As the stellar mass increases, both SFRs and Eddington luminosity also increase and tend to have stronger correlations. This result may suggest that AGN activity is more pronounced in more massive galaxies, leading to a stronger correlation between SF and AGN activity.

We also present all different SFR tracers as a function of stellar mass and Eddington ratio for a proper comparison which are provided in the online version of this paper. These trends are consistent across all types of SFR indicators used in this study, albeit with variations in offsets and scatters. Median values for each SFR indicator and stellar mass are presented in Table \ref{table:median_sfr}. 

%

%
\subsection{Specific SFRs versus Eddington ratio}\label{subsec:ssfr_edd}

\begin{figure*}
\centering
	
        \includegraphics[width=0.223\textwidth]{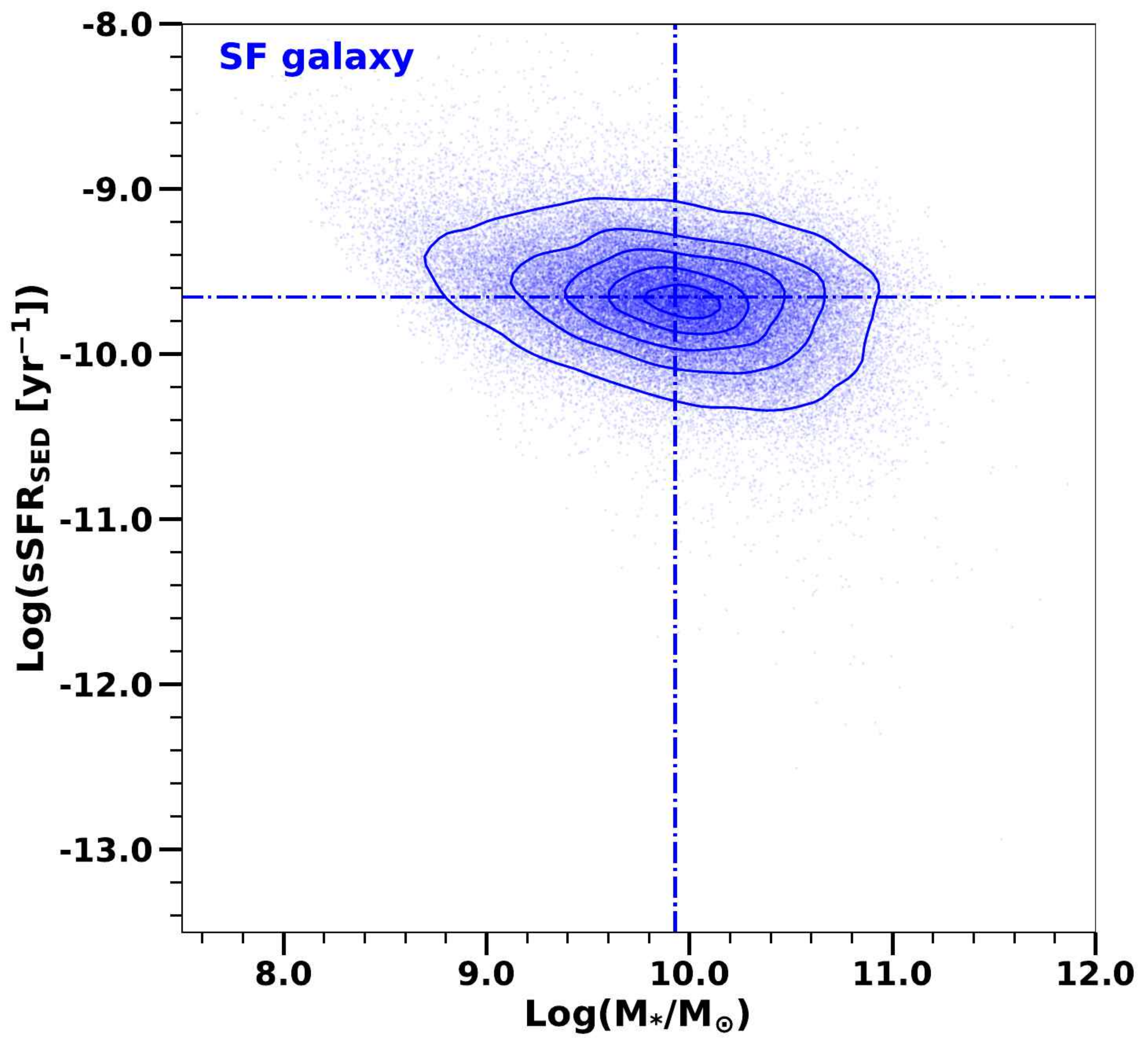}
        \includegraphics[width=0.245\textwidth]{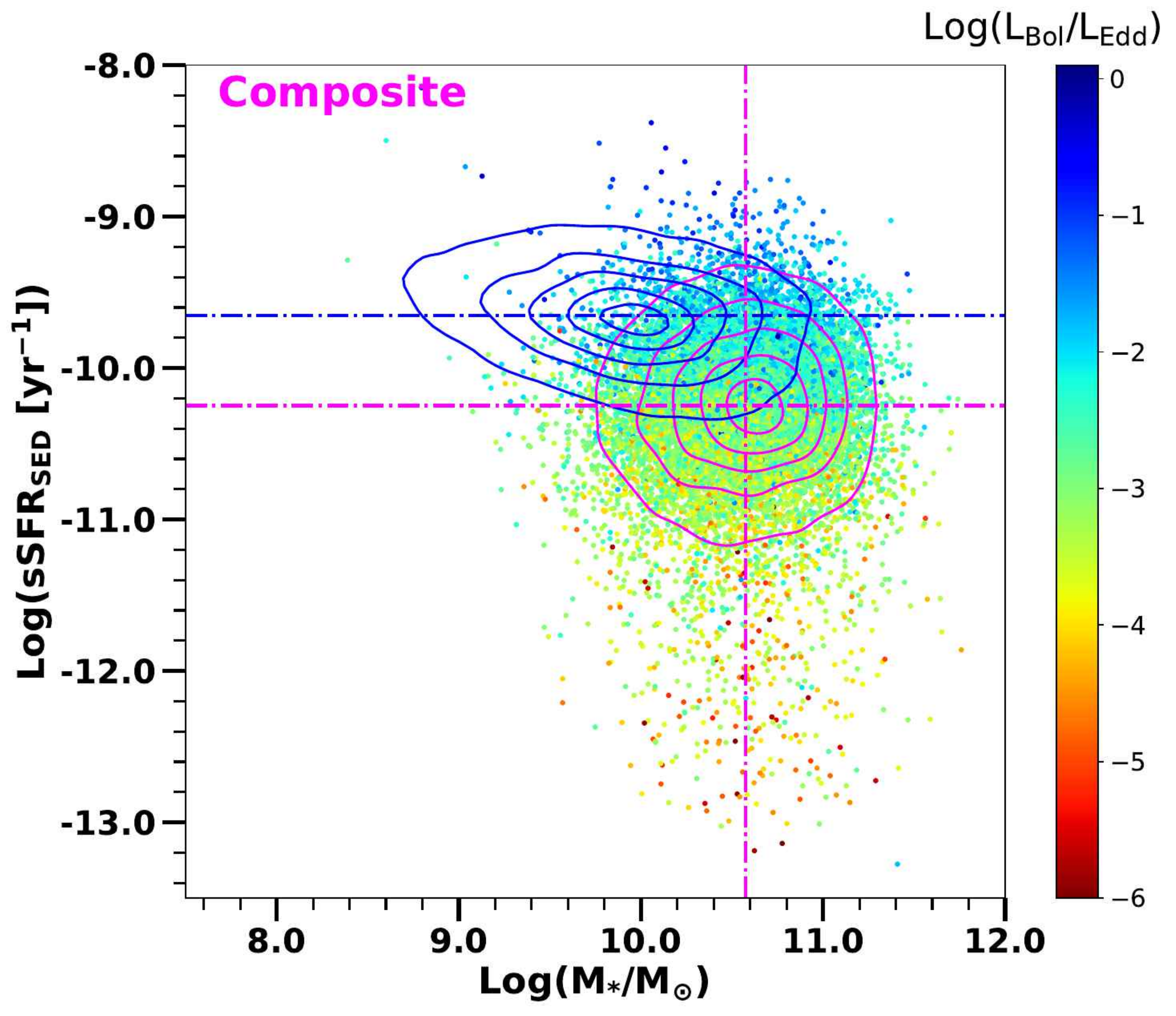}
        \includegraphics[width=0.245\textwidth]{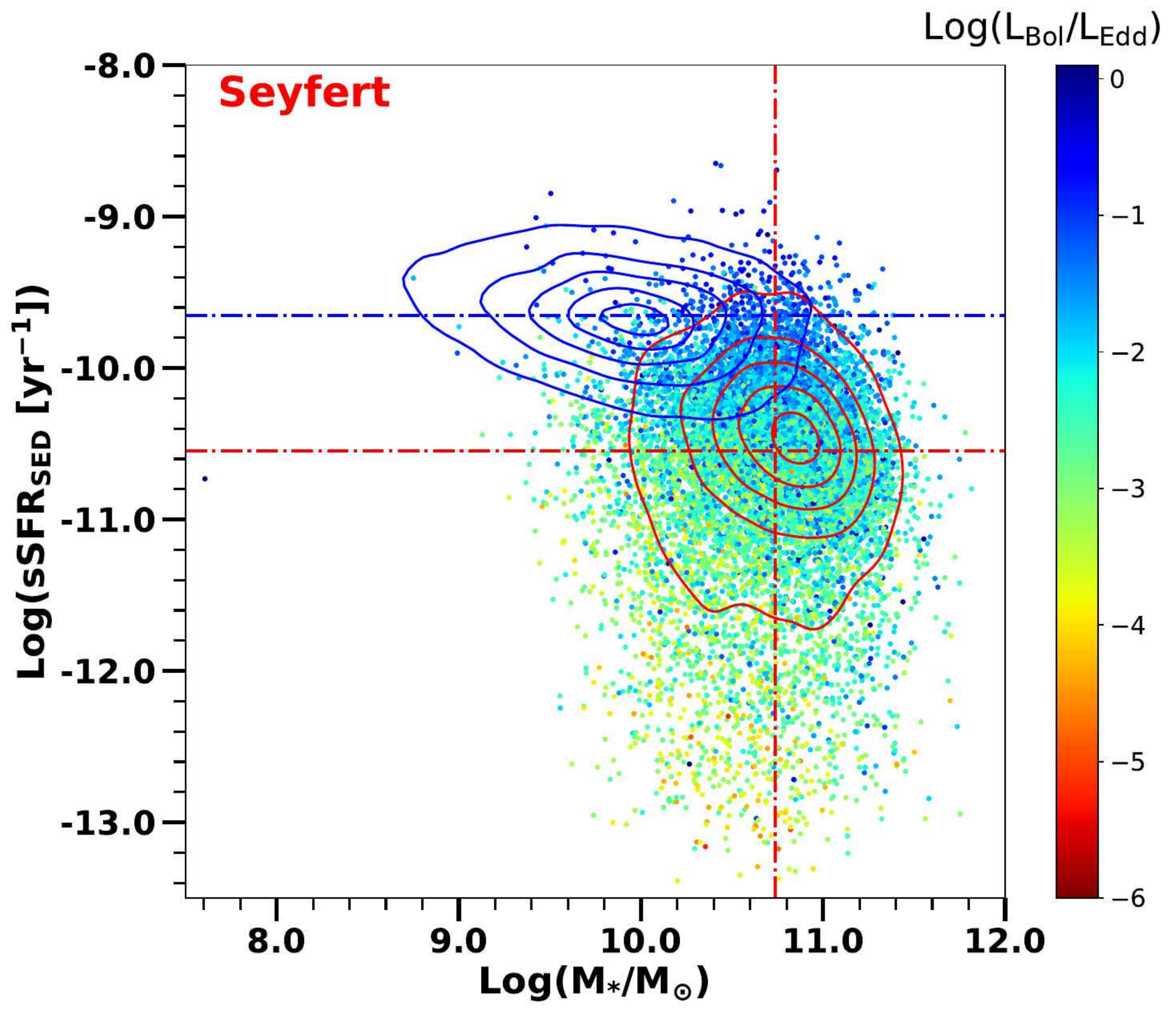}
        \includegraphics[width=0.245\textwidth]{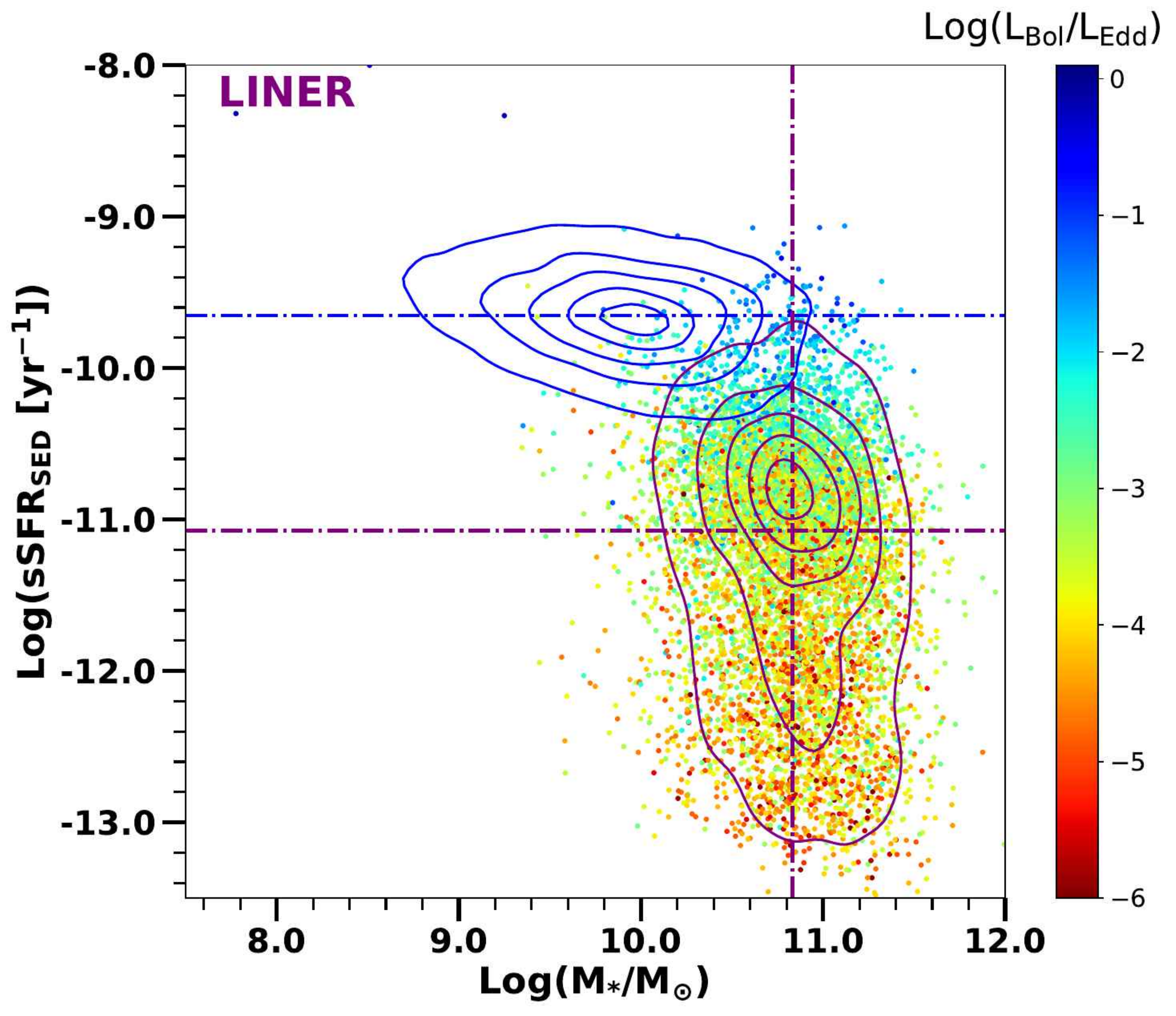}
	\caption{Specific SFRs as a function of stellar mass. sSFR values are presented by $\rm sSFR_{SED}$. Each panel displays the classifications of SF, composite, Seyfert, and LINER galaxies, respectively. The $\rm sSFR-M_{*}$ space is divided into a grid of 150x150 bins, and contour lines are drawn to represent 10, 30, 50, 70, and 90 percent of the maximum number density. The blue contour, representing SF galaxies, is relocated to the panels of composite, Seyfert, and LINER galaxies for a proper comparison. The color scales represent Eddington ratio values. Horizontal dashed lines in blue, pink, red, and cyan correspond to the median sSFR values for SF, composite, Seyfert, and LINER galaxies, respectively. Vertical dashed lines indicate the median stellar mass values for each galaxy type. The full set of other sSFR tracers as a function of stellar mass is available in the online version.
	\label{fig:ssfr_contour_edd}}
\end{figure*}

In this section, we investigate the relationship between Eddington ratio and sSFRs. Figure \ref{fig:ssfr_contour_edd} presents $\rm sSFR_{SED}$ as a function of stellar mass, with Eddington ratio values displayed as color scales in the sSFRs$-$$\rm M_{*}$ plane. Similar to the comparison of SFRs, we find smooth transitions of Eddington ratio values from high to low values corresponding to high and low sSFRs. Additionally, we observe a decrease in sSFRs from SF, composite, Seyfert, and LINER galaxies, respectively. As mentioned in \citet{Xue+10}, it is important to do analysis for mass-matched samples because this allows us to have a detailed study of AGN activity as a function of stellar mass. Also, \citet{Salim+16} argued that comparing sSFRs yields more insightful results compared to the comparison of SFRs alone, as sSFRs reflect galaxies with the same physical properties. The consistent decrease in all sSFR tracers in our result may support the argument of \citet{Salim+16} that comparing sSFRs is more informative than comparing SFRs. As shown in Figure \ref{fig:contour}, we do not observe this consistent decrease from SF, composite, Seyfert, and LINER galaxies, respectively.

\begin{figure*}
\centering
        \includegraphics[width=0.31\textwidth]{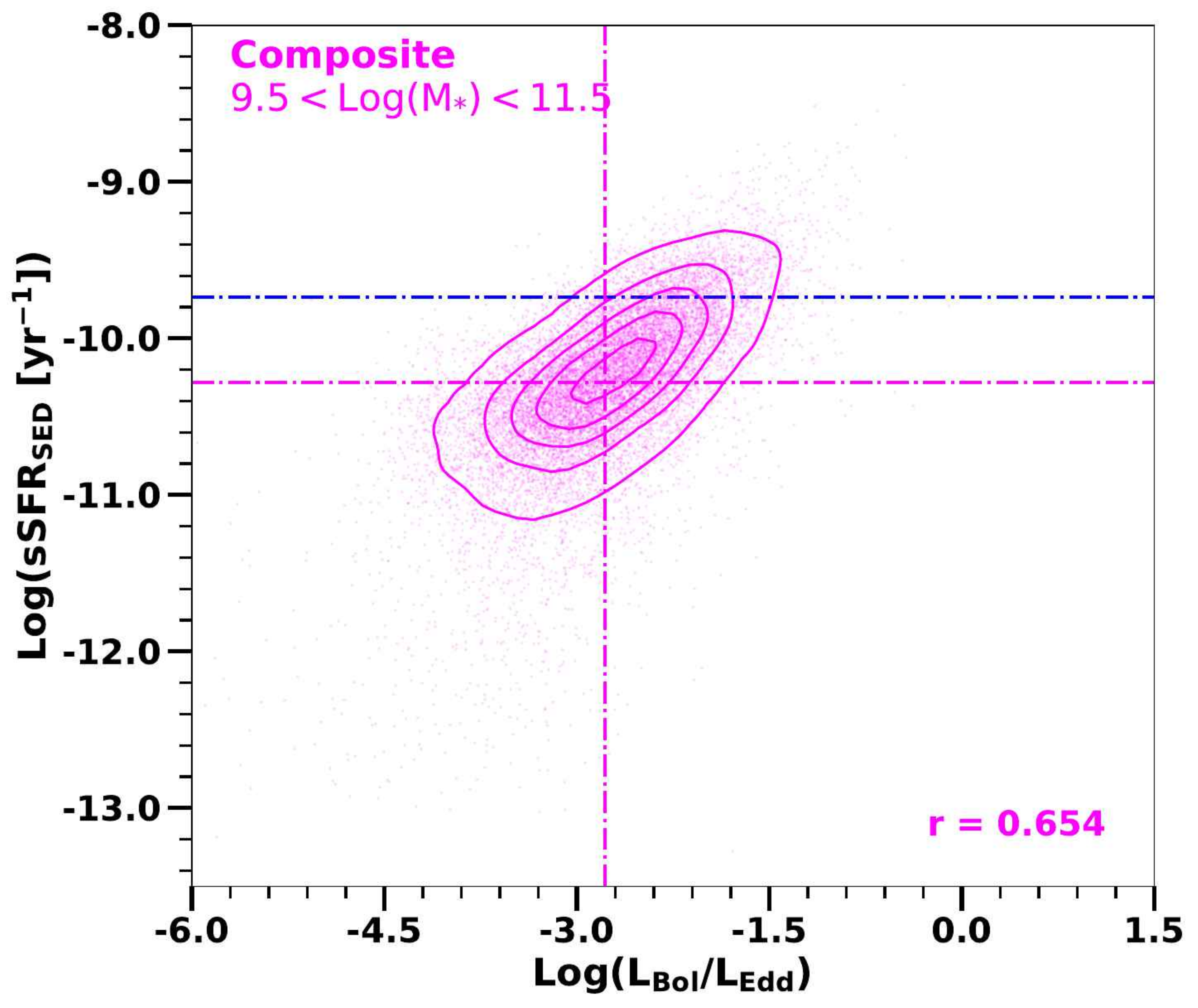}
        \includegraphics[width=0.31\textwidth]{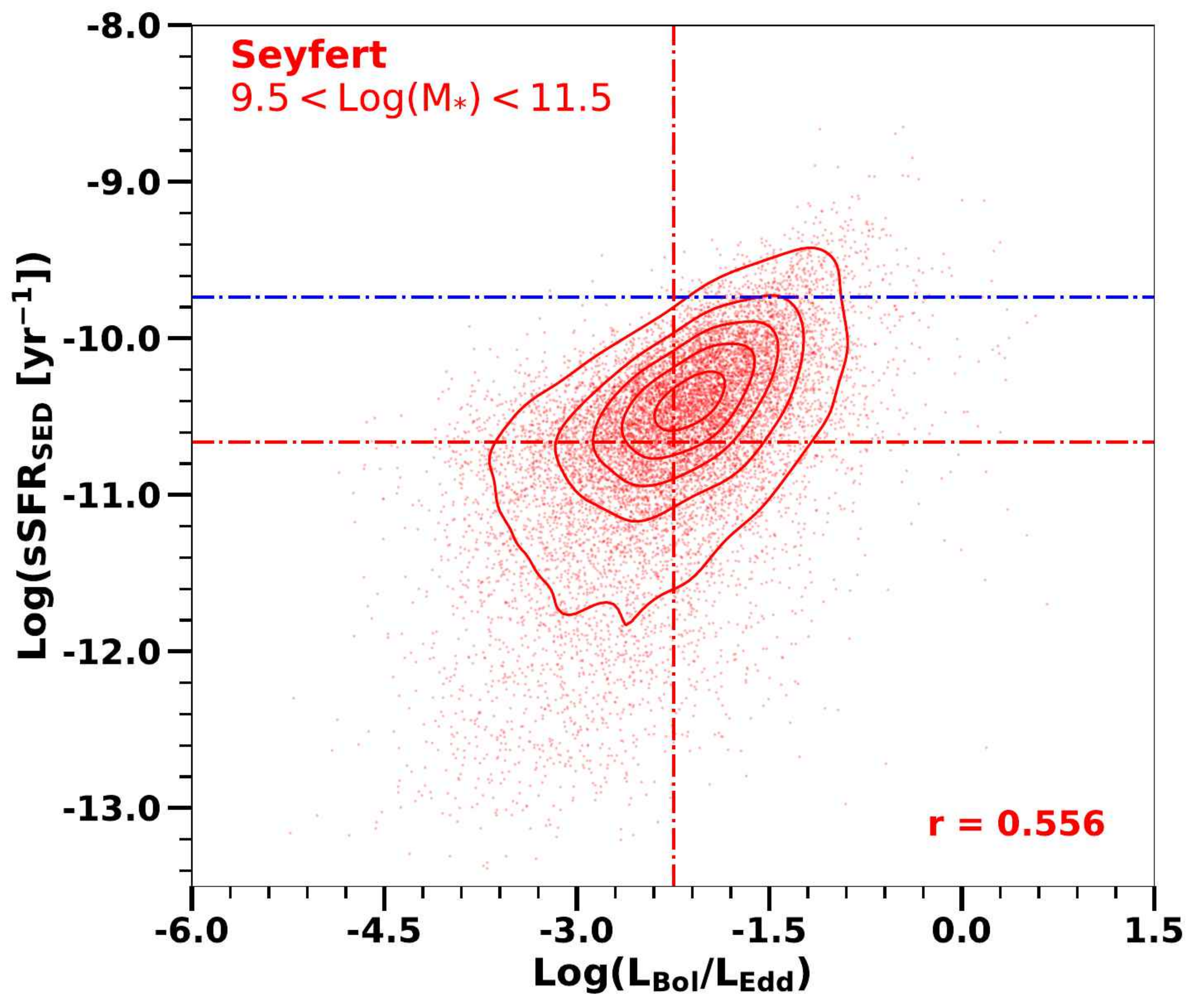}
        \includegraphics[width=0.31\textwidth]{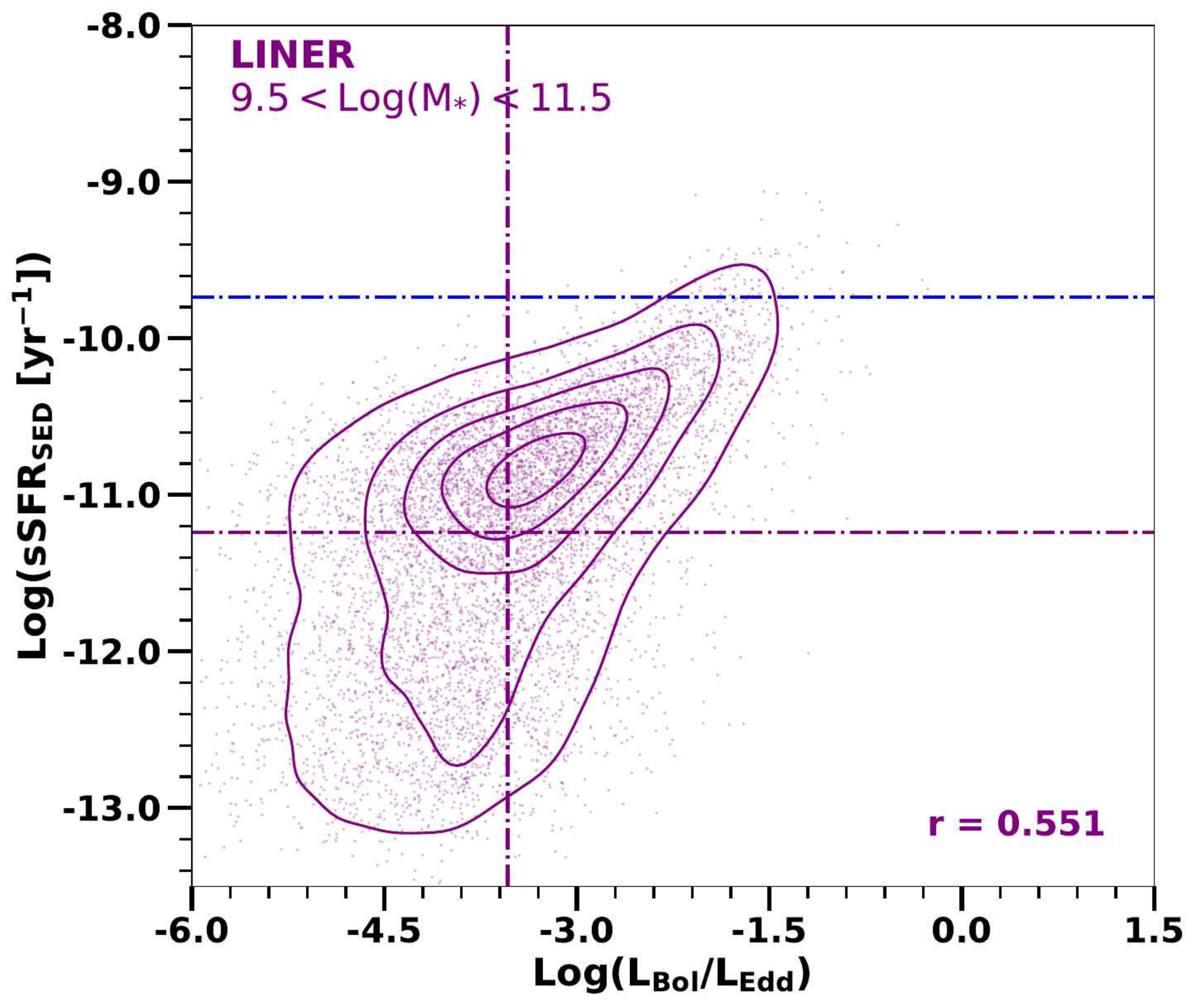}
	\caption{Specific SFRs as a function of Eddington ratio for a total stellar mass range from 9.5 $<$ $\rm \log M_{*}$ $<$ 11.5. sSFR values are presented by $\rm sSFR_{SED}$. Horizontal dashed lines in blue, pink, red, and cyan correspond to the median sSFR values for SF, composite, Seyfert, and LINER galaxies, respectively. Vertical dashed lines indicate the median Eddington ratio values for each galaxy type.
	\label{fig:ssfr_edd}}
\end{figure*}

\begin{figure*}
\centering
    \includegraphics[width=0.31\textwidth]{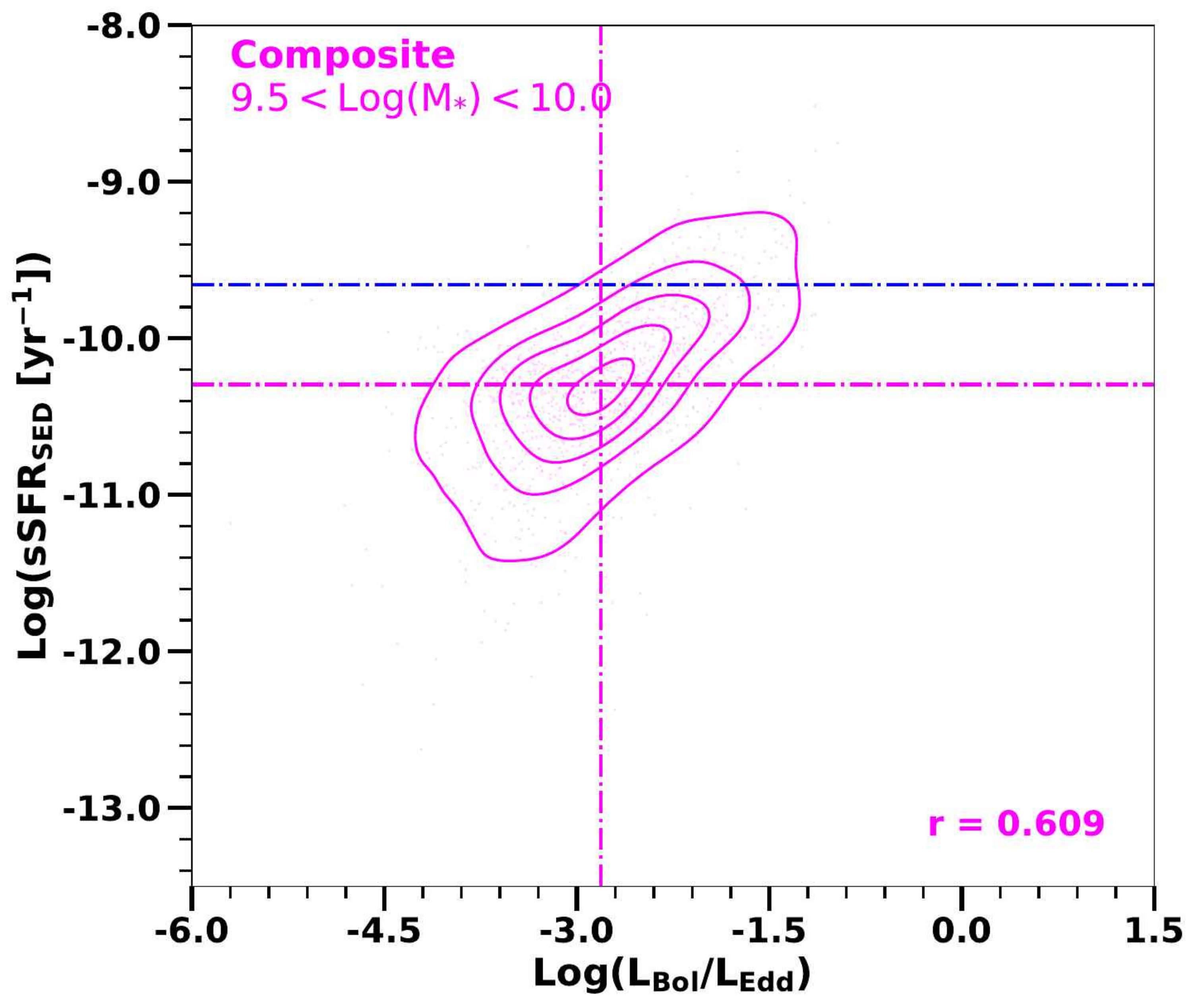}
    \includegraphics[width=0.31\textwidth]{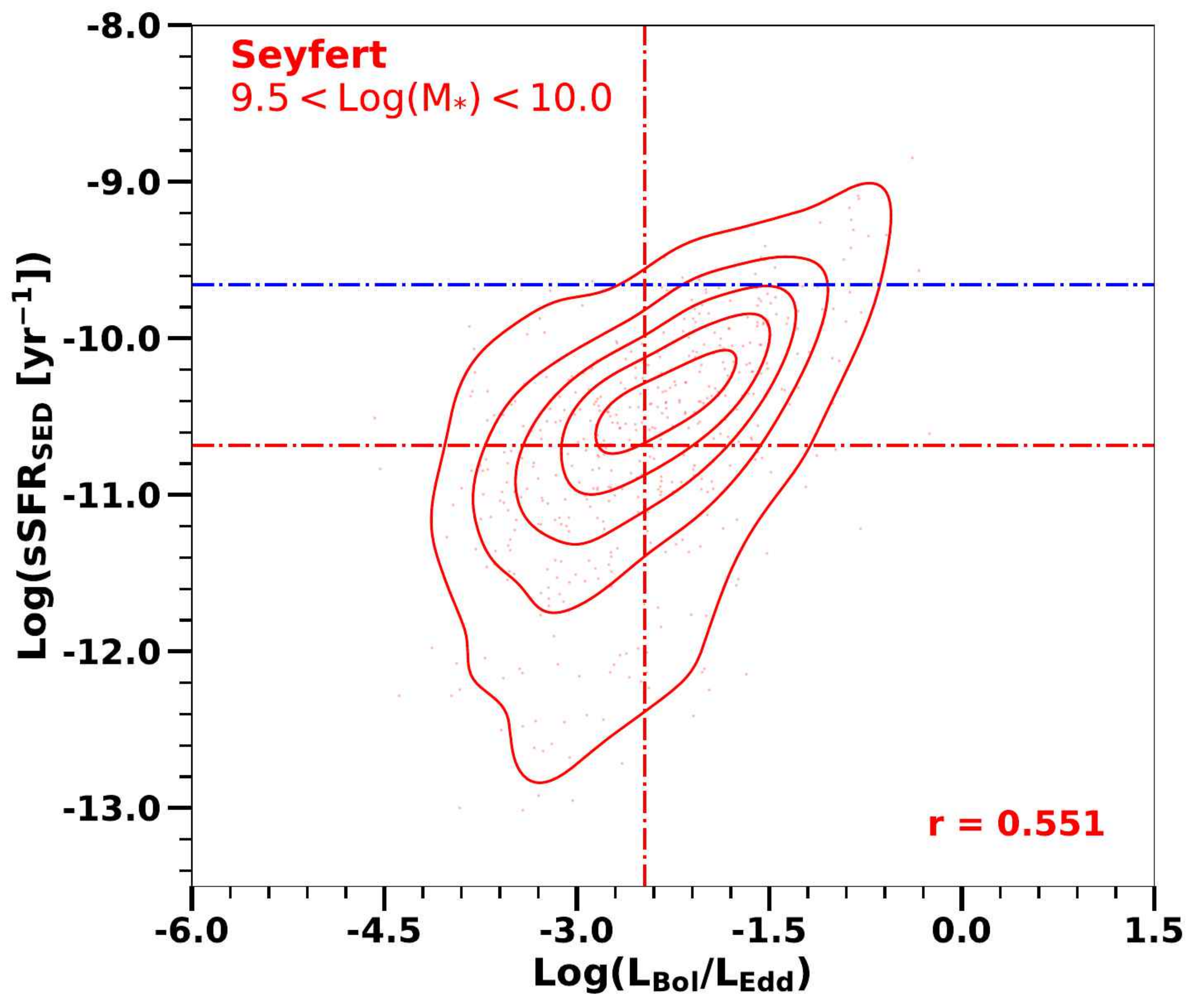}
    \includegraphics[width=0.31\textwidth]{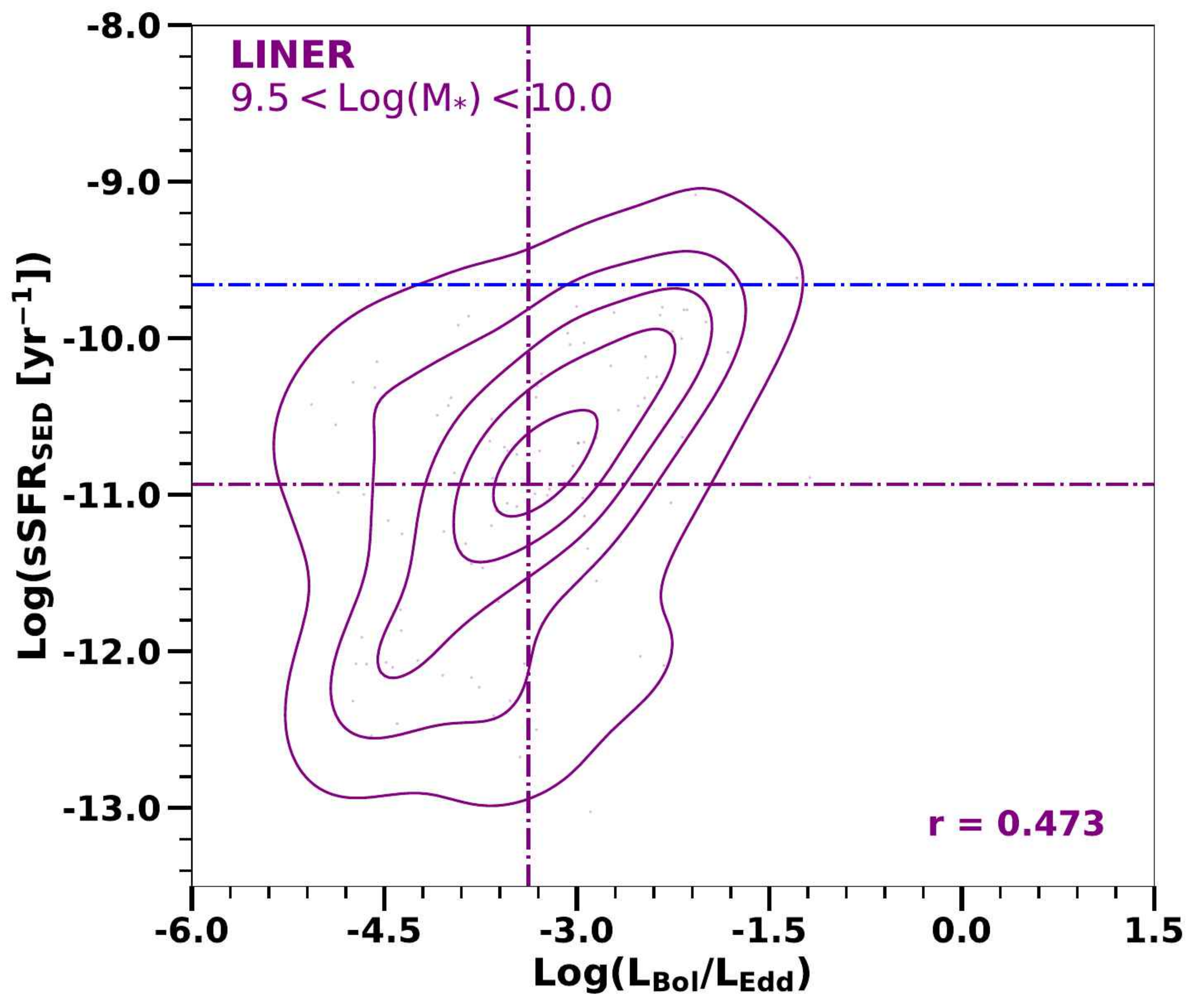}
    \includegraphics[width=0.31\textwidth]{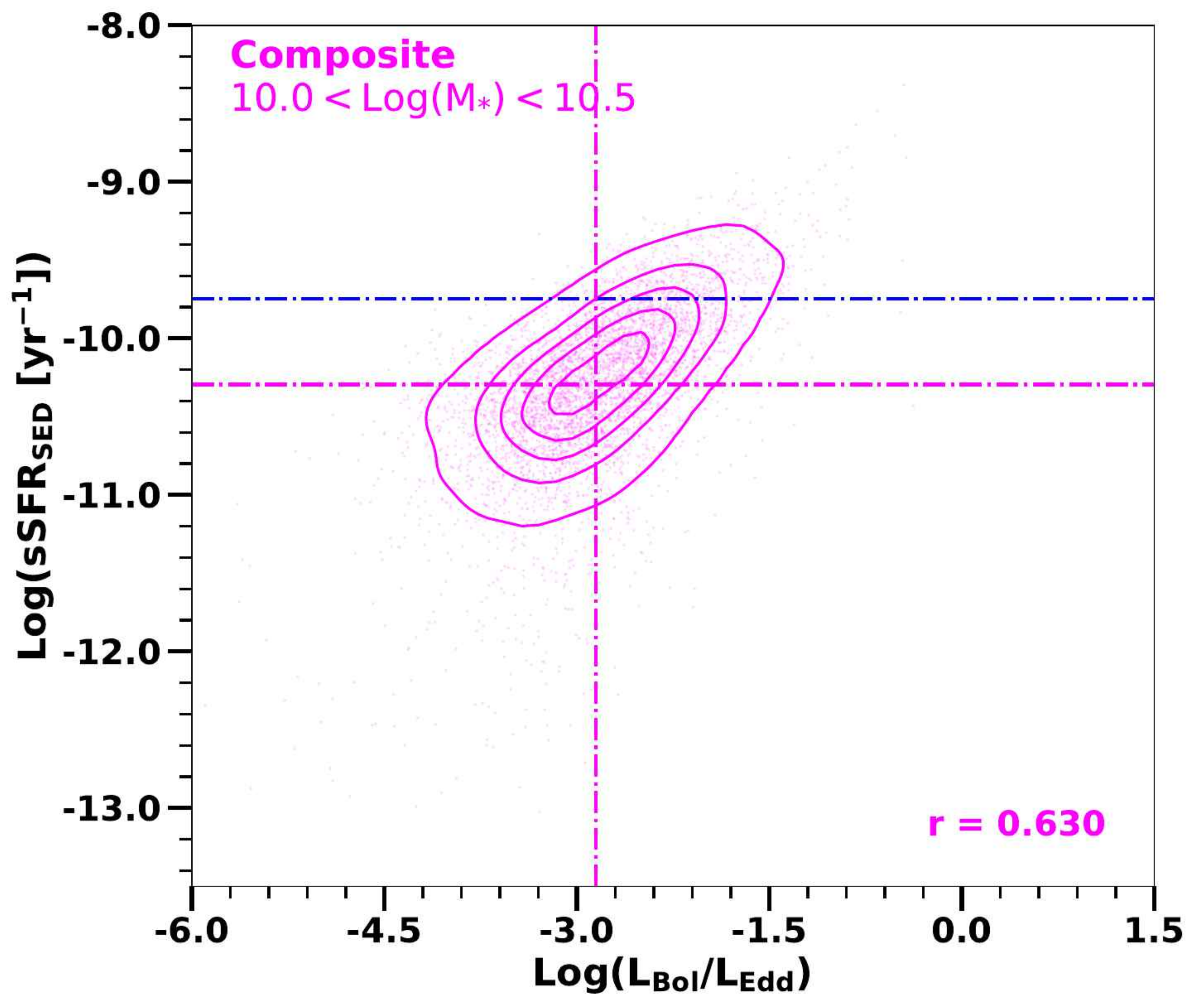}
    \includegraphics[width=0.31\textwidth]{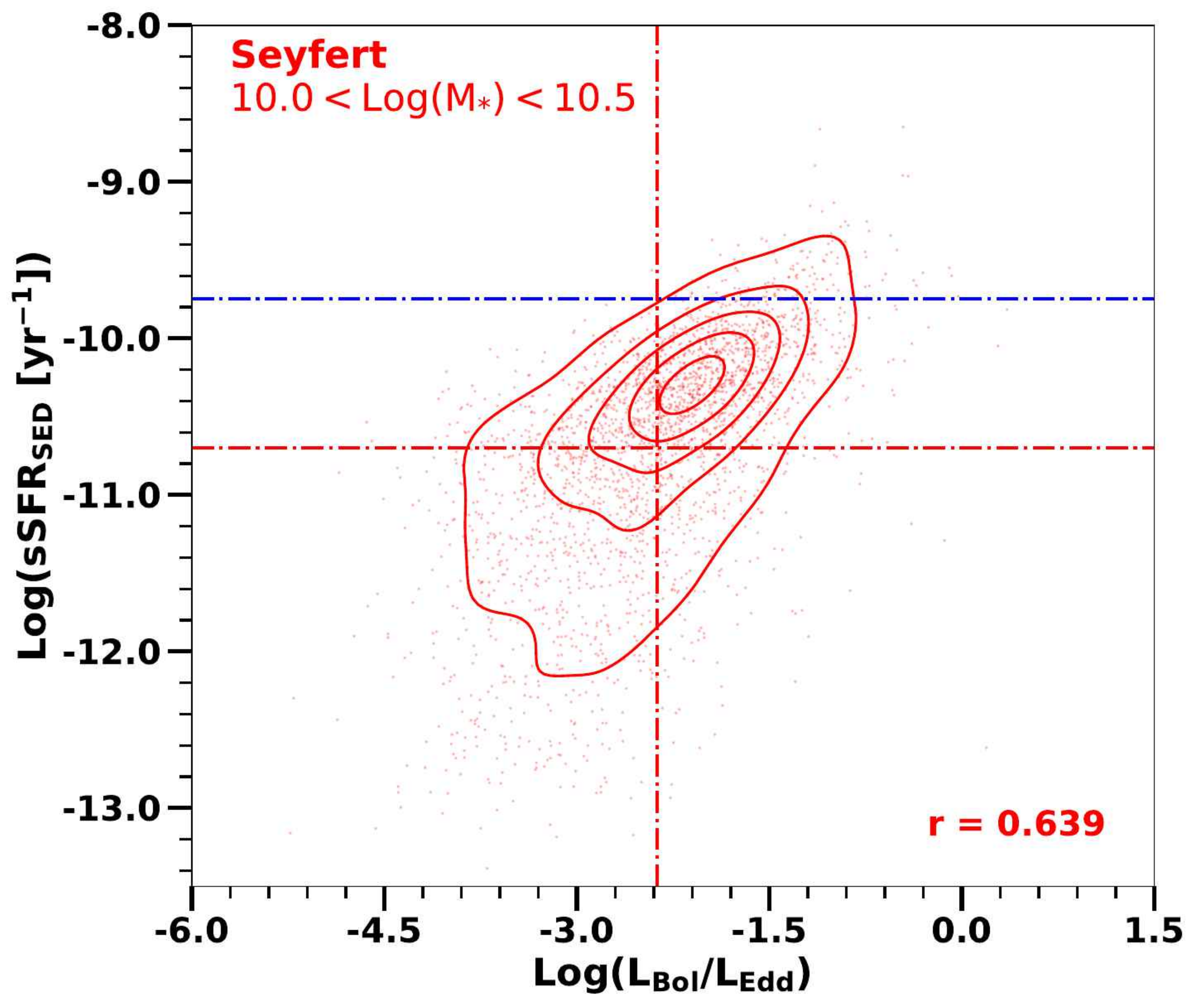}
    \includegraphics[width=0.31\textwidth]{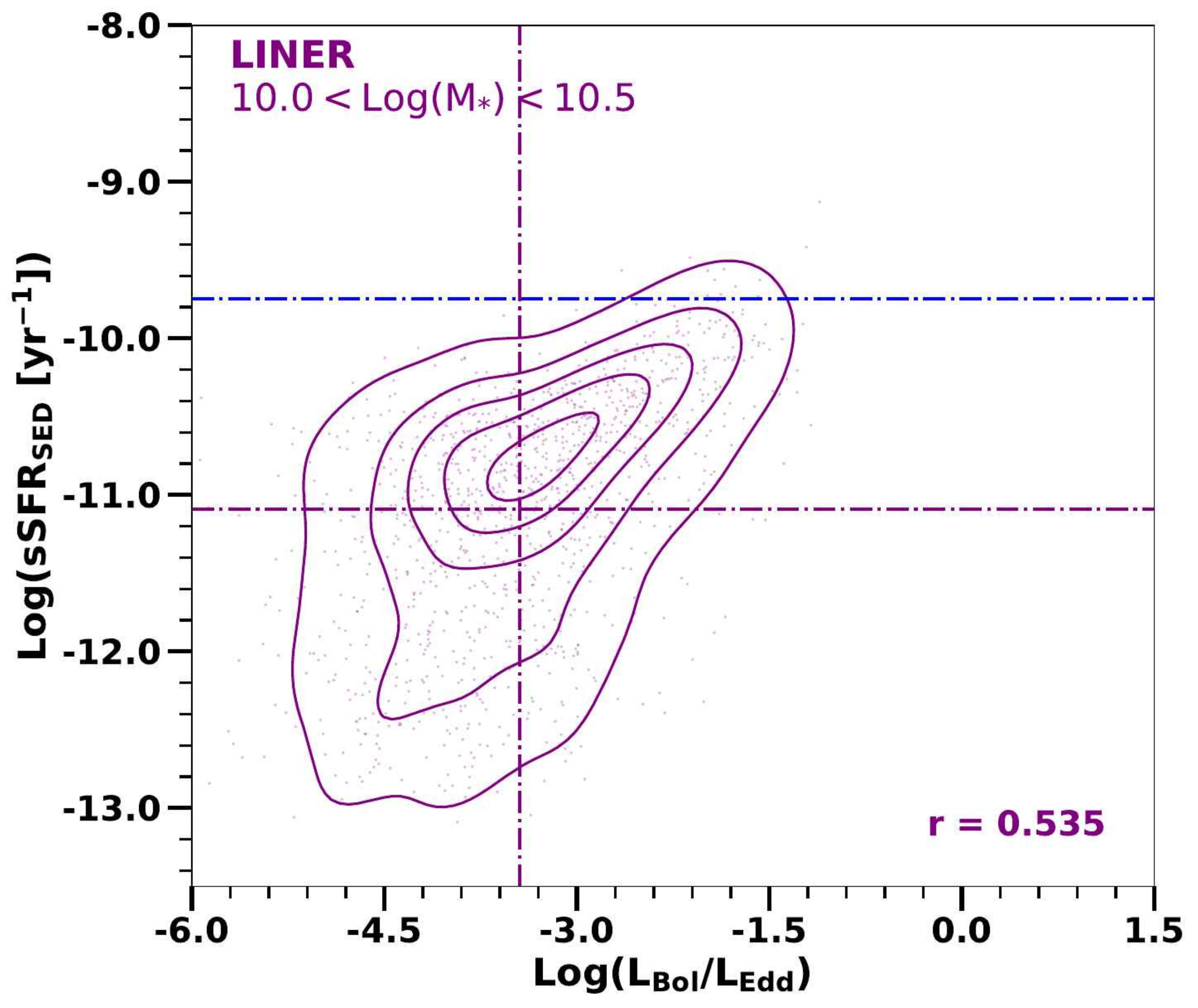}
    \includegraphics[width=0.31\textwidth]{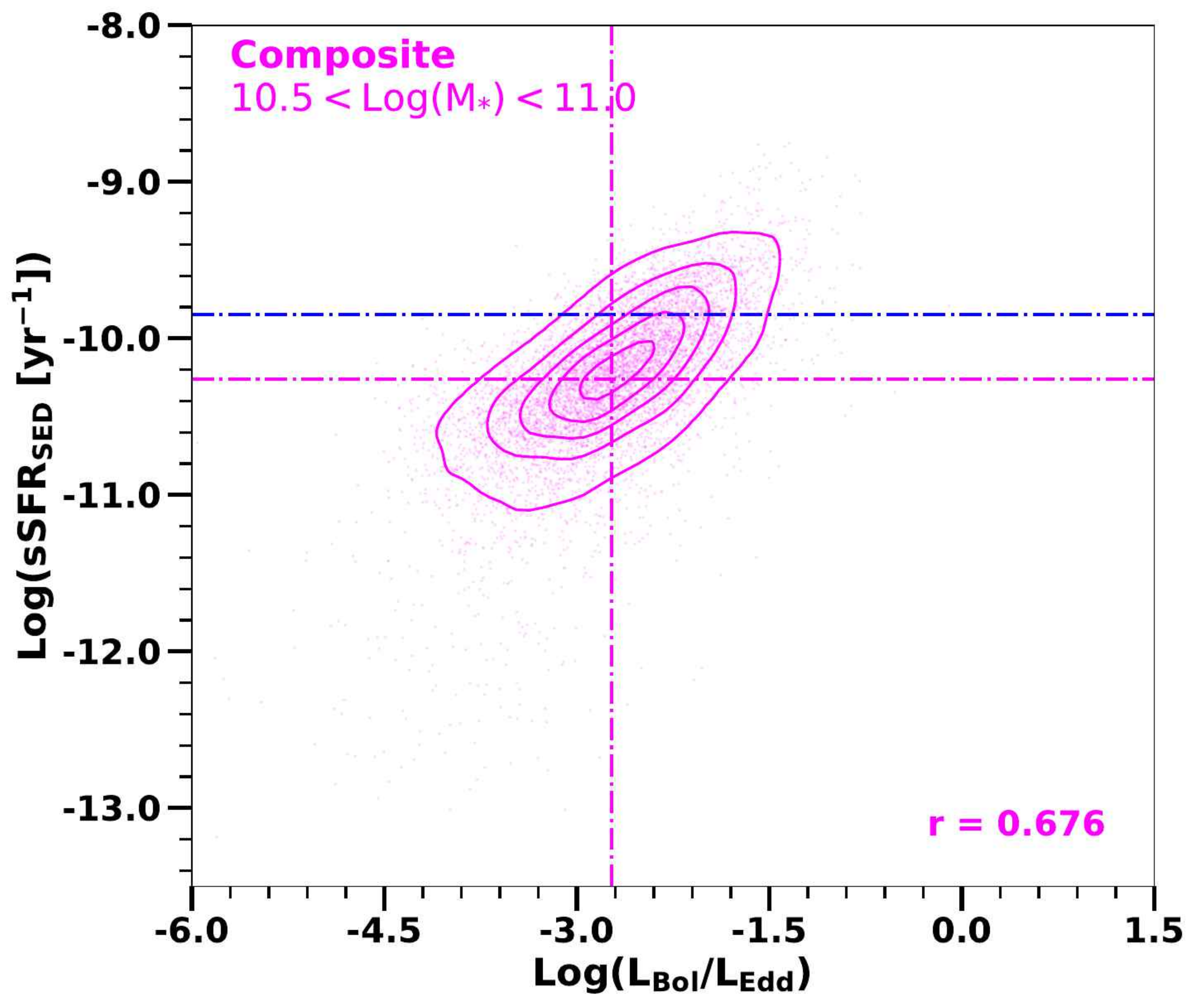}
    \includegraphics[width=0.31\textwidth]{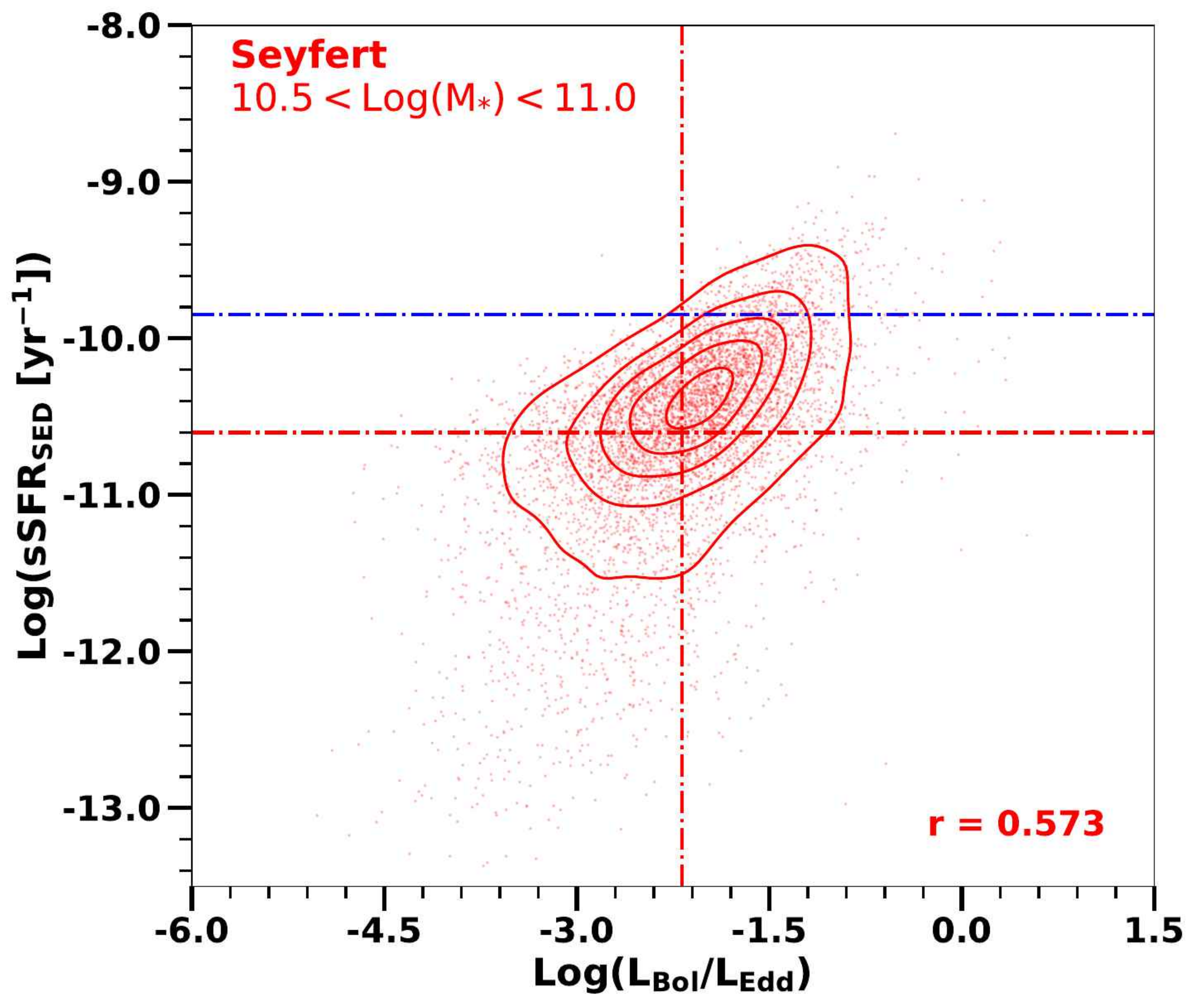}
    \includegraphics[width=0.31\textwidth]{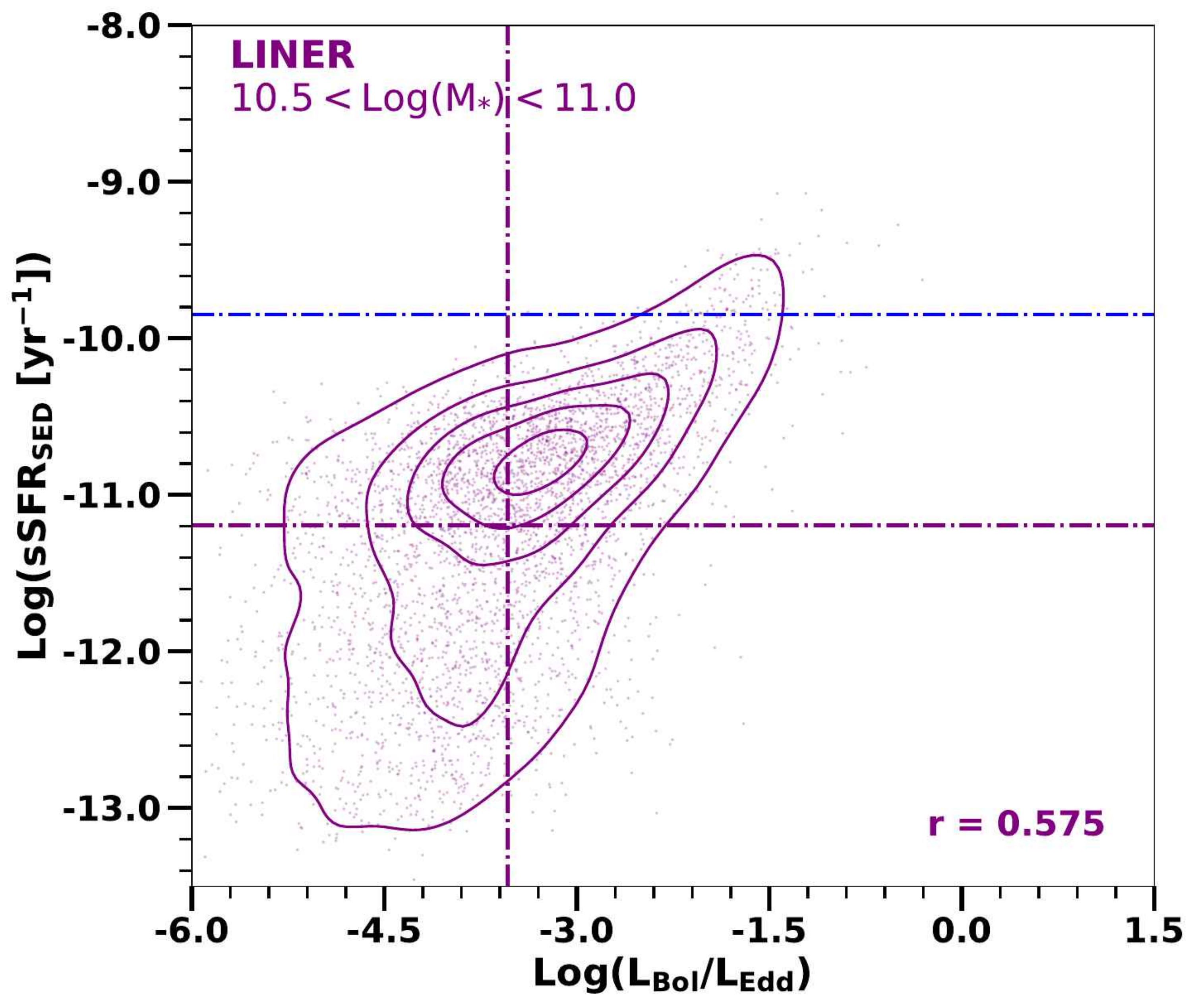}
    \includegraphics[width=0.31\textwidth]{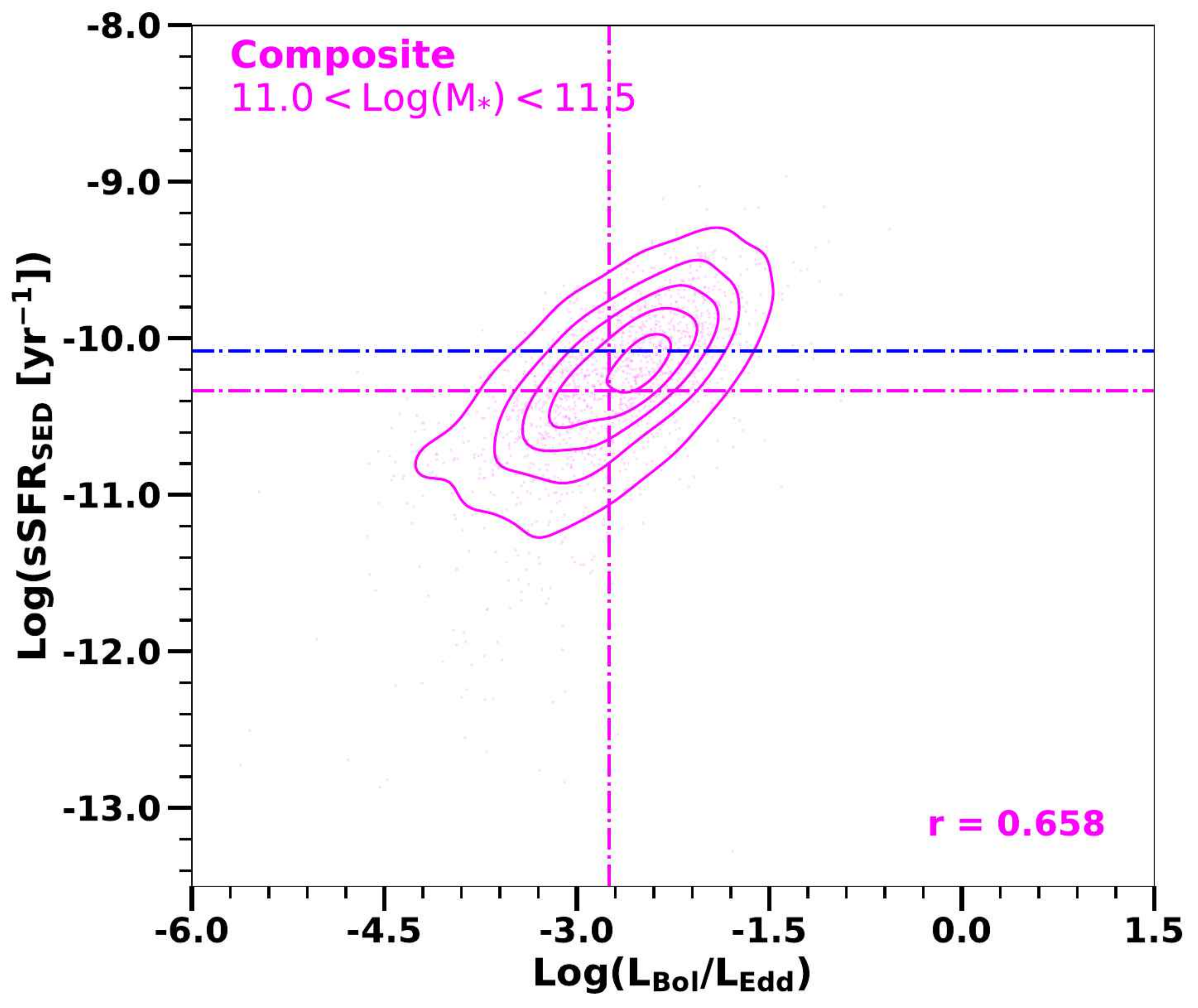}
    \includegraphics[width=0.31\textwidth]{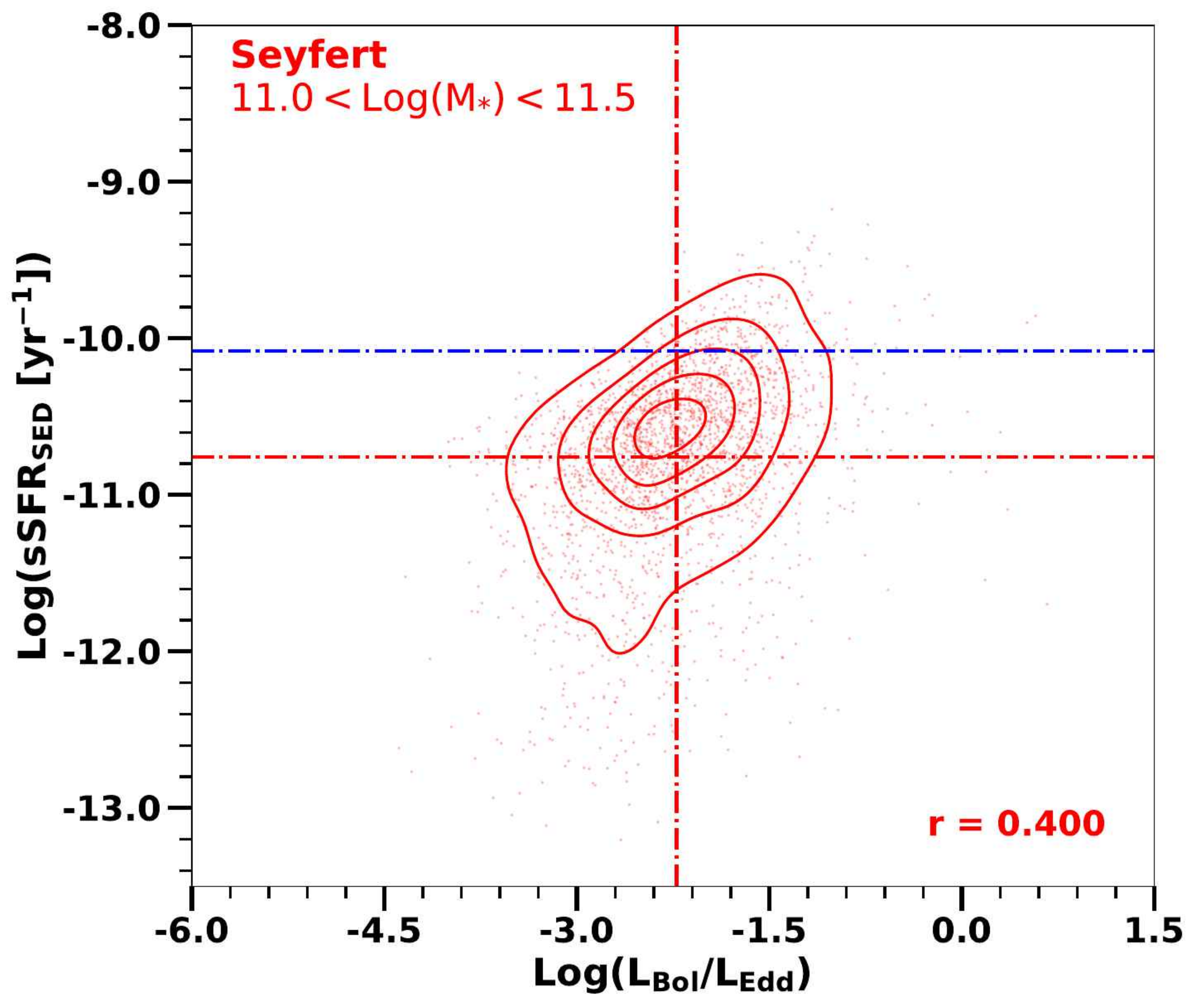}
    \includegraphics[width=0.31\textwidth]{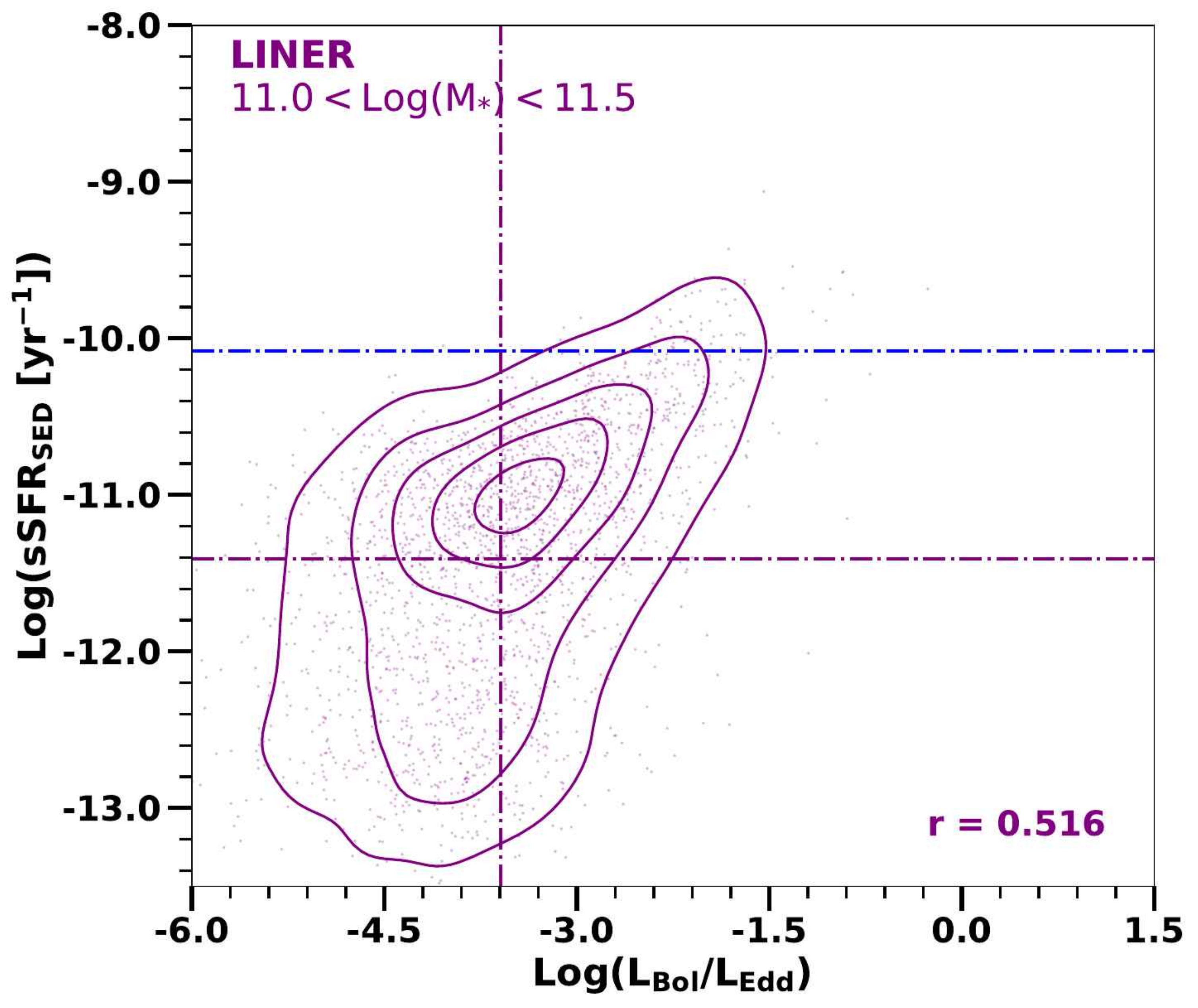}
    \caption{Similar to Fig \ref{fig:ssfr_edd}, but with the stellar mass range divided into four bins of $\rm \log M_{*}$ = 9.5$-$10.0, 10.0$-$10.5, 10.5$-$11.0, and 11.0$-$11.5, respectively. The full set of other sSFR tracers as a function of Eddington ratio is available in the online version.
    \label{fig:ssfr_edd_all}}
\end{figure*}

Figure \ref{fig:ssfr_edd} displays $\rm sSFR_{SED}$ as a function of Eddington ratio for a stellar mass range of 9.5 $<$ $\rm \log M_{*}$ $<$ 11.5. Similar to the comparison of SFRs, all sSFR tracers exhibit a strong correlation with Eddington ratio, and Seyfert galaxies have the highest Eddington ratio compared to composite and LINER galaxies. There are decreasing trends in sSFRs for SF, composite, Seyfert, and LINER galaxies, respectively. Additionally, we divide the stellar mass into four different bins with a size of $\rm \Delta\log M_{*}$ = 0.5 to examine the correlation between sSFRs and Eddington ratio for galaxies within similar stellar mass bins (Figure \ref{fig:ssfr_edd_all}). We found robust correlations between sSFRs and Eddington luminosity across all stellar mass bins. As the stellar mass increases, the correlation between sSFRs and Eddington ratio appears to strengthen. This result suggests that in more massive galaxies, AGN activity has a more pronounced impact, leading to a stronger correlation between SF and AGN activity. 

We also present various sSFR tracers for a proper comparison which are provided in the online version of this paper. The decrease in sSFRs as a function of stellar mass, along with their strong correlation with Eddington ratio, remains consistent across all SFR tracers, despite differing offsets and scatters. Table \ref{table:median_sfr} summarizes the median sSFRs and stellar mass values for all galaxy types.

%

%
%
\section{Discussions}\label{section:discuss}

\subsection{SDSS Fiber Aperture Coverage Effect}\label{section:aperture}

\begin{figure*}
\centering
        \includegraphics[width=0.90\textwidth]{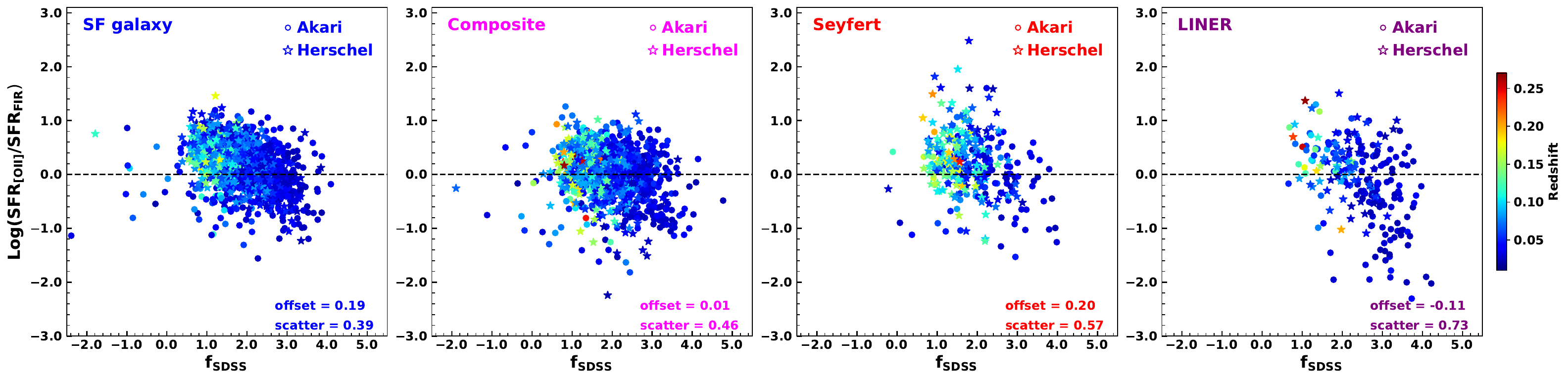}
        \includegraphics[width=0.90\textwidth]{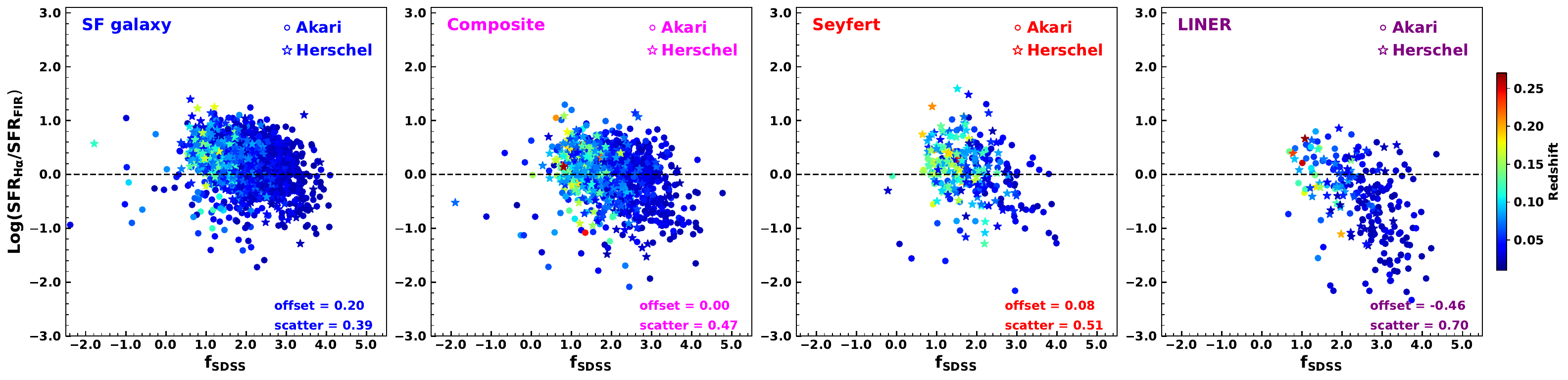}
	\caption{The ratios of $\mathrm{SFR_{[OII]}}$/$\mathrm{SFR_{FIR}}$  and $\mathrm{SFR_{H\alpha}}$/$\mathrm{SFR_{FIR}}$ as a function of fraction of u-band light captured by the SDSS fiber, defined as the difference in magnitude, $\rm f_{SDSS} = U_{Fiber}  -  U_{Total}$. Redshift is shown as the color scale. 
	\label{fig:oii_halpha_fsdss}}
\end{figure*}

We investigate the issue of SDSS fiber aperture covering different physical scales across our sample as a function of redshift. As mentioned in Section~\ref{section:sample}, the limited light captured by the 3$\arcsec$ SDSS fiber for galaxies at $z < 0.05$ may affect the comparison of SFRs derived from \OII\ and \Ha\ emission lines with those obtained from other tracers. Figure~\ref{fig:oii_halpha_fsdss} shows the ratios of $\mathrm{SFR_{[OII]}}$/$\mathrm{SFR_{FIR}}$ and $\mathrm{SFR_{H\alpha}}$/$\mathrm{SFR_{FIR}}$ as a function of the fiber light fraction, defined as $\rm f_{SDSS} = U_{Fiber} - U_{Total}$. We find that the galaxy light captured by the SDSS fiber is lower at smaller redshift, and both $\mathrm{SFR_{[OII]}}$ and $\mathrm{SFR_{H\alpha}}$ tend to be lower than $\mathrm{SFR_{FIR}}$ when compared to galaxies at higher redshift, where the fraction of galaxy light captured by the SDSS fiber is higher. At $z < 0.05$, the SDSS fiber (3$\arcsec$ $\sim$ 0.1 kpc) may only capture part of the total galaxy light (most of the central region of galaxies). The loss of light captured by the SDSS fiber for the outer parts of galaxies may lead to an underestimation of $\mathrm{SFR_{[OII]}}$ and $\mathrm{SFR_{H\alpha}}$, compared to $\mathrm{SFR_{FIR}}$ measurements obtained with AKARI/Herschel, which have significantly larger beam sizes.

\subsection{AGN Fraction}\label{section:agn_frac}

The variation in the spatial scales covered by the SDSS fiber across our sample may also affect the AGN contributions in the \OII\ and \Ha\ emission lines. In lower-redshift galaxies, where the 3$\arcsec$ SDSS fiber primarily captures the central regions, the observed \OII\ and \Ha\ lines are more likely to be dominated by AGN emission compared to higher-redshift galaxies. As shown in Figure~\ref{fig:stellar_fsdss}, the fraction of light captured by the SDSS fiber is lower at $z < 0.05$. In contrast, for galaxies at $z > 0.05$, the fraction of light captured by the SDSS fiber is higher and remains relatively constant with redshift. We thus expect that for galaxies at $z > 0.05$, the impact of aperture size on AGN contamination in emission lines is minimal. However, for galaxies at $z < 0.05$, the fiber misses the outer regions while capturing AGN-dominated nuclear emission.

\begin{figure*}
\centering
        \includegraphics[width=0.90\textwidth]{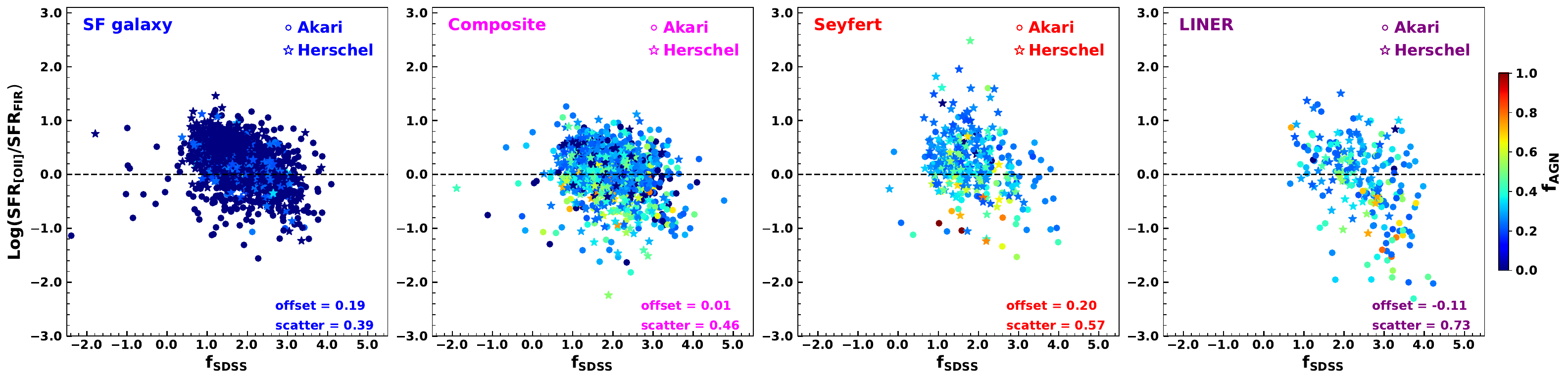}
        \includegraphics[width=0.90\textwidth]{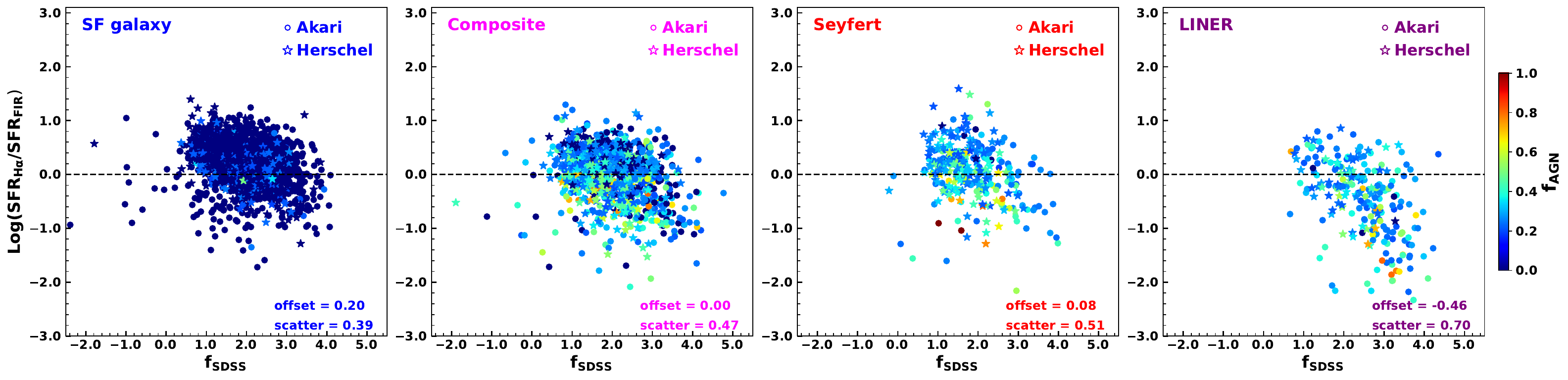}
        \includegraphics[width=0.90\textwidth]{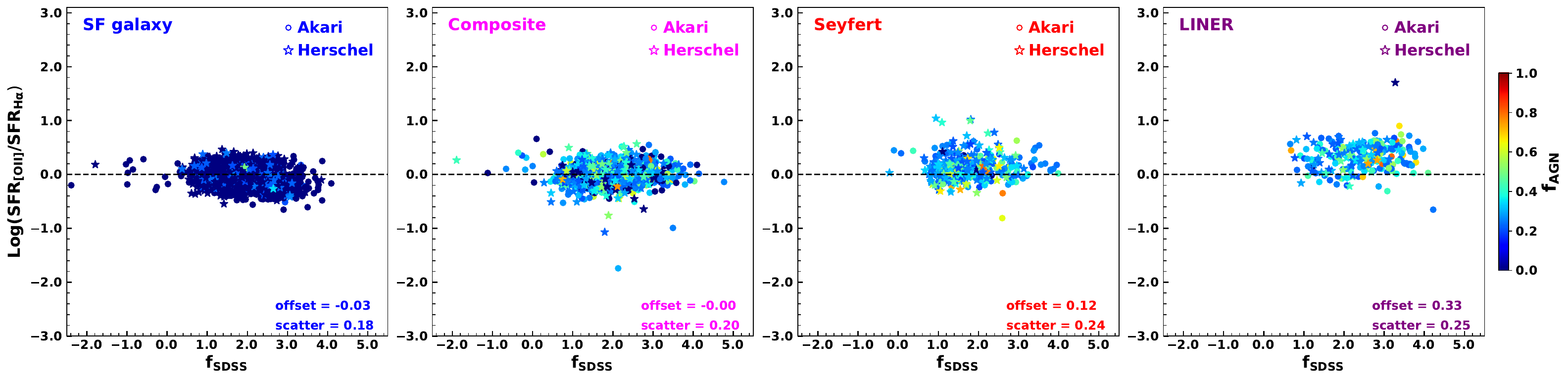}
	\caption{Top and middle panels: The ratios of $\mathrm{SFR_{[OII]}}$/$\mathrm{SFR_{FIR}}$  and $\mathrm{SFR_{H\alpha}}$/$\mathrm{SFR_{FIR}}$ as a function of fraction of u-band light captured by the SDSS fiber, defined as the difference in magnitude, $\rm f_{SDSS} = U_{Fiber}  -  U_{Total}$. Bottom panel: The ratio of $\mathrm{SFR_{[OII]}}$/$\mathrm{SFR_{H\alpha}}$ as a function of the $\rm f_{SDSS}$. AGN fraction ($\rm f_{AGN}$) is shown as the color scale. 
	\label{fig:oii_halpha_fagn}}
\end{figure*}

Figure~\ref{fig:oii_halpha_fagn} shows the ratios $\mathrm{SFR_{[OII]}}/\mathrm{SFR_{FIR}}$ and $\mathrm{SFR_{H\alpha}}/\mathrm{SFR_{FIR}}$ as a function of the fiber light fraction, $\rm f_{SDSS}$, with the AGN fraction ($\rm f_{AGN}$) indicated by the color scale. We find that sources with higher $\rm f_{AGN}$ tend to have lower $\mathrm{SFR_{[OII]}}$ and $\mathrm{SFR_{H\alpha}}$ compared to $\mathrm{SFR_{FIR}}$. In addition, for targets with lower fraction of light captured by SDSS fiber (lower-redshift sources), $\rm f_{AGN}$ is higher for non-SF sources, and both $\mathrm{SFR_{[OII]}}$ and $\mathrm{SFR_{H\alpha}}$ are lower relative to $\mathrm{SFR_{FIR}}$. We also examine the variation of $\rm f_{AGN}$ in the $\mathrm{SFR_{[OII]}}/\mathrm{SFR_{H\alpha}}$ versus $\rm f_{SDSS}$ plane. As shown in the bottom of Figure \ref{fig:oii_halpha_fagn}, non-SF sources show higher $\mathrm{SFR_{[OII]}}$ relative to $\mathrm{SFR_{H\alpha}}$. In addition, we find that for non-SF sources with higher $\rm f_{AGN}$ exhibit larger $\mathrm{SFR_{[OII]}}$ relative to $\mathrm{SFR_{H\alpha}}$. This may suggest that AGN contamination has a stronger impact on the \OII\ emission line than on the \Ha\ line. Our results are consistent with the findings from \citet{Maddox18} that \OII\ emission line could be contributed by the extended emission-line regions (EELRs) ionized by the AGN.

\begin{figure*}
\centering
        \includegraphics[width=0.95\textwidth]{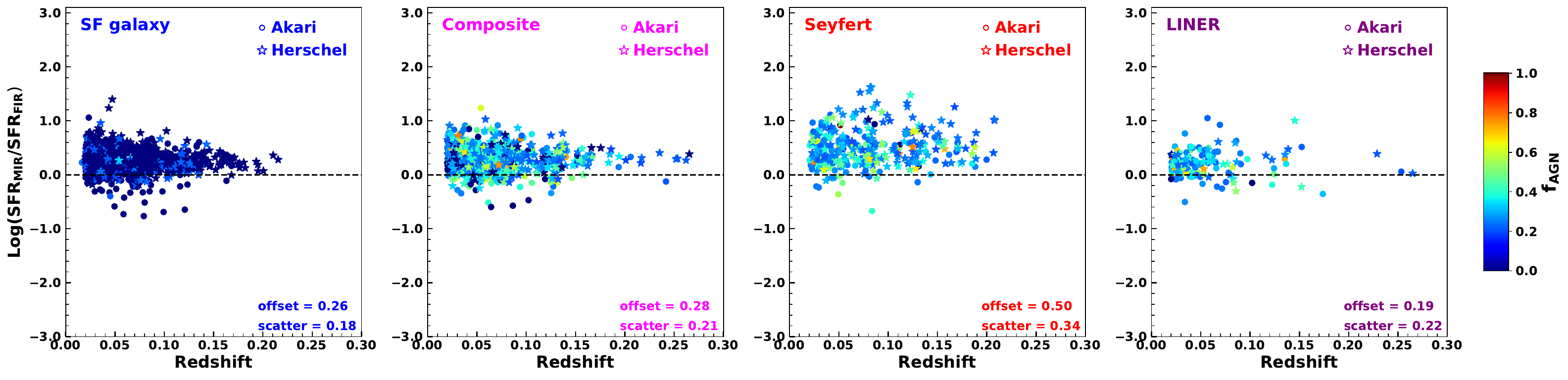}
	\caption{The ratio of $\mathrm{SFR_{MIR}}$/$\mathrm{SFR_{FIR}}$ as a function of redshift. AGN fraction ($\rm f_{AGN}$) is shown as the color scale.
	\label{fig:mir_fir_redshift}}
\end{figure*}

Figure~\ref{fig:mir_fir_redshift} shows the ratio $\mathrm{SFR_{MIR}}/\mathrm{SFR_{FIR}}$ as a function of redshift, with the AGN fraction ($\rm f_{AGN}$) indicated by the color scale. Overall, $\mathrm{SFR_{MIR}}$ is systematically higher than $\mathrm{SFR_{FIR}}$, particularly for Seyfert galaxies, which exhibit an average offset of $\sim$0.50 dex. For targets with higher AGN fraction, $\mathrm{SFR_{MIR}}$ also tends to be higher than $\mathrm{SFR_{FIR}}$. This may suggest that MIR emission may include the contribution from the hot dusty torus surrounding AGNs (e.g., \citealp{Netzer+07}). Our results are consistent with \citet{Kirkpatrick+15}, performing IR SEDs of 343 (U)LIRGs with cold and warm dust components and found that the hot-dust temperature rises with increasing AGN luminosity. Furthermore, \citet{McKinney+21} argued that AGN heating can also contribute to galaxy-scale cold dust emission in the FIR, which may explain why, despite the large $\mathrm{SFR_{MIR}}/\mathrm{SFR_{FIR}}$ offset, the two tracers remain well correlated across redshift.

\subsection{Discrepancies Among Various SFR Indicators}\label{subsec:diff_sfr}

From the comparison among various SFR tracers, we found that most SF and composite galaxies are young and exhibit blue colors (e.g., $\mathrm{D_{n}4000} < 1.8$), while Seyfert and LINER galaxies tend to be older and display redder colors (e.g., $\mathrm{D_{n}4000} > 1.8$). The UV/optical SFRs directly measure the stellar light and thus provide a straightforward measure of the ongoing SF activity in galaxies on short timescales ($<$10 Myr, e.g., \citealp{Murphy+11}). In contrast, the IR emission, which arises from dust heated by stellar light in the UV/optical bands, traces not only currently forming stars but also previously formed stars that have heated the dust over extended periods. Therefore, IR-based SFRs provide information on SF activity with a longer timescale ($<$100 Myr) compared to SFRs traced by the UV/optical bands \citep[e.g.,][]{Kennicutt+12, Calzetti+13}.

In Figures~\ref{fig:ir_compare} and \ref{fig:ann_compare}, galaxies with older stellar populations tend to have lower $\mathrm{SFR_{[OII]}}$, $\mathrm{SFR_{H\alpha}}$, $\mathrm{SFR_{Dn4000}}$, and $\mathrm{SFR_{SED}}$ relative to $\mathrm{SFR_{FIR}}$ and $\mathrm{SFR_{ANN}}$. \citet{Brinchmann+04} estimated SFRs for non-SF objects using the anti-correlation between $\mathrm{D_{n}4000}$ and specific SFRs in starburst galaxies. However, because their calibration sample was dominated by young galaxies, this approach can underestimate SFRs when applied to older and redder galaxies \citep[e.g.,][]{Matsuoka+15}. For SF galaxies, we also find that those with younger stellar populations tend to have lower $\mathrm{SFR_{FIR}}$ compared to UV/optical tracers. This may indicate that FIR emission is sensitive to probing long timescale SF activities, which could lead to an underestimation of SFRs in sources dominated by very recent star formation. In addition to the effects of SDSS aperture size and AGN fraction discussed in Sections~\ref{section:aperture} and \ref{section:agn_frac}, we find that $\mathrm{SFR_{[OII]}}$ and $\mathrm{SFR_{H\alpha}}$ are systematically lower for galaxies with older stellar populations compared to $\mathrm{SFR_{FIR}}$. This may support that SFRs may be underestimated when traced by $\mathrm{SFR_{[OII]}}$ or $\mathrm{SFR_{H\alpha}}$ for galaxies with old and red stellar populations.

%

The predicted $\mathrm{SFR_{ANN}}$ demonstrates good consistency with other SFR tracers for young and blue SF galaxies. However, for older and redder galaxies, $\mathrm{SFR_{ANN}}$ tends to be higher compared to other SFR tracers, except $\mathrm{SFR_{MIR}}$. This discrepancy may arise from the limitations of the training sample of the ANN, which predominantly includes galaxies with $\mathrm{SFR} > 0.3~\mathrm{M_\odot\ yr^{-1}}$. Consequently, $\mathrm{SFR_{ANN}}$ could be overestimated for redder galaxies with lower SFRs (e.g., \citealp[]{Ellison+16b}). In addition, for higher redshift galaxies ($z > 0.15$), $\mathrm{SFR_{ANN}}$ is systematically higher than other tracers. This may result from the lack of high-$z$ galaxies in the training sample, as \citet{Ellison+16b} cross-matched sample of Herschel and SDSS only for sources of $0.04 < z < 0.15$.

%
%
%
Discrepancies among various SFR tracers vary significantly, making the choice of tracer crucial when studying the interplay between AGN activity and SFRs in host galaxies. The selected tracer can introduce varying degrees of scatter and uncertainty compared to other SFR tracers.

\subsection{The Relation between AGN Strength and Star Formation Activities}

Through the results of this study, we find that there is a tight correlation between SFRs and the Eddington ratio. What does this tight correlation tell us? For both AGN strength and SF activities, there is a fundamental connection between them that they share the same \emph{gas supply} within a galaxy system. It is a natural expectation that because both AGN strength and SF activities share the same gas supply, they are inherently interconnected and mutually related, leading to a strong correlation between them. If the gas supply is abundant, we expect both SFRs and Eddington luminosity to be strong. Conversely, when the gas supply diminishes, both SFRs and Eddington luminosity weaken. Indeed, we find a smooth transition in Eddington ratio, transitioning from high to lower values in tandem with the transition from high to lower values of both SFRs and sSFRs, as shown in Figures \ref{fig:contour} and \ref{fig:ssfr_contour_edd}. Additionally, \citet{Woo+20} interpreted this tight correlation trend between SFRs and Eddington ratio as \emph{the delayed AGN feedback}. In this scenario, both AGNs and SF are triggered by the same gas supply. AGNs with high Eddington ratios generate powerful outflowing gas, while SF remains active in the ISM. It may take a dynamical timescale for the outflowing gas to impact the SF activity within the ISM, where the outflowing gas pushes and clears out the gas supply, resulting in the observation of low SFRs and a weak outflow gas with a low Eddington ratio. 

Our results are in agreement with some previous studies based on X-ray, NIR, FIR, and optically selected samples (e.g., \citealp[]{Netzer+07, Netzer+09, Matsuoka+15, Zhuang+20, Mountrichas+22}). For example, \citet{Netzer+09} employed a large sample that included polycyclic aromatic hydrocarbon (PAH) features in the MIR and cold dust emission in the FIR to estimate SF luminosity. They found a strong correlation between SF luminosity and AGN bolometric luminosity. Similarly, based on a sample of 492 type-2 AGNs observed with the AKARI/Herschel, \citet{Matsuoka+15} discovered a strong correlation between FIR and AGN luminosities. Additionally, utilizing a large sample from the SDSS, \citet{Zhuang+20} identified a strong correlation between black hole accretion rate and SFRs. Also, in the study of approximately 1800 X-ray AGNs in the eROSITA Final Equatorial-Depth Survey (eFEDS), \citet{Mountrichas+22} found that AGNs with higher X-ray luminosity sources ($ \rm L_{X, 2-10keV} > 10^{44.2}$ $\rm erg\ s^{-1}$) tend to exhibit higher SFRs compared to SF sources.

Our results are consistent with recent findings by \citet{Mountrichas24}, that different galaxy types may not originate from the same AGN activity source. In the early-stage, gas-rich environment, a high accretion black hole is expected, such as in Seyfert 2 galaxies. In contrast, in the later-stage and gas-poor environment, there will be a low accretion black hole, such as in LINERS. Indeed, in our results, Seyfert galaxies show high SFRs and Eddington ratios (strong AGN activities). While LINERS show low SFRs and low Eddington ratios (weak AGN activities). Additionally, in our study, we primarily focus on the relationship between Eddington ratios and (s)SFRs. In recent years, by using 122 X-ray AGNs in the XMM-XXL field and 3371 galaxies from the VIPERS survey, \citet{Mountrichas23} found that black hole mass may be the main driving factor of the SF activities in the host galaxies instead of other physical properties, e.g., X-ray luminosity. We plan to further investigate this in our sample in future studies.

Both SFRs and sSFRs exhibit good correlations with Eddington ratios. Notably, we find more significant correlations with Eddington ratios (as indicated by the Spearman correlation factors) in the cases of $\rm SFR_{[OII]}$ and $\rm SFR_{H\alpha}$ compared to other SFR tracers (Figures \ref{fig:sfr_edd_all} and \ref{fig:ssfr_edd_all}). As we calculated AGN bolometric luminosity from \OIII\ luminosity, this may explain the stronger correlation of Eddington ratios with $\rm SFR_{[OII]}$ and $\rm SFR_{H\alpha}$ compared to other SFR tracers. 

Additionally, in the comparison between sSFRs and SFRs with Eddington ratio, we confirm that comparing sSFRs is more informative than comparing SFRs because sSFRs reflect the comparison for galaxies with the same physical properties \citep[e.g.,][]{Xue+10, Salim+16}. As we can see in Figure~\ref{fig:ssfr_contour_edd}, when sSFR is shown as a function of stellar mass for different galaxy types, a clear decreasing trend is seen from SF to composite to Seyfert and to LINER galaxies. In contrast, such a variance in SFR is not shown in Figure~\ref{fig:contour}. Similarly, we find a more significant decrease in sSFRs than in SFRs when compared with Eddington ratios in Figures~\ref{fig:ssfr_edd} and \ref{fig:sfr_edd}, respectively. We found that the decreasing trends in sSFRs are consistent across all SFR tracers, from composite to Seyfert and LINER galaxies, this consistency is not observed in the comparison of SFRs (Table \ref{table:median_sfr}). 


In addition, it is essential to mention that our sample was selected based on the BPT diagram. There is a possibility that we may misclassify weak AGNs with low Eddington ratios as SF galaxies with high SFRs because their observed emission lines can be strongly influenced by SF processes. To mitigate this limitation of our sample selection using the BPT diagram, conducting a similar study with X-ray data (e.g., \citealp{Xue+17}) is necessary. Furthermore, the observed emission lines in composite galaxies may not only originate from SF or AGNs but also from shock excitations. Similarly, narrow emission lines in LINER galaxies could be caused not only by AGN but also by non-nuclear sources, e.g., shocks or hot/evolved stars (e.g., \citealp{Ho08}).

\section{Summary}\label{section:sum}

We conducted a detailed analysis with a large sample of $\sim$113,000 galaxies ($z < 0.3$) characterized by high-quality data. Our study compares SFRs within host galaxies using various diagnostic methods. We categorized the sample into four types of galaxies: SF, composite, Seyfert, and LINER galaxies. We discuss the disparities in SFR estimates obtained through different tracers, including observed infrared flux from AKARI/Herschel ($\sim$4,100 sources), the MPA-JHU, ANN, GSWLC catalogs, as well as \OII, and \Ha\ emission lines. From the estimated SFR tracers, we investigated the interplay between AGN strength and SF activities in our sample. We summarize our main results as follows: 

\smallskip 
(1) For SF galaxies, SFR measurements showed offsets and scatters below 0.26 dex and 0.29 dex, respectively. In contrast, non-SF galaxies, especially LINERs, exhibited larger discrepancies, with offsets up to 0.85 dex and a scatter of 0.57 dex. Determining SFRs in non-SF galaxies is challenging, and the choice of tracer is critical, as each will have different levels of scatter and uncertainty.

\smallskip
(2) The predicted $\mathrm{SFR_{ANN}}$ shows good agreement with other SFR tracers for SF galaxies. However, for non-SF galaxies, $\mathrm{SFR_{ANN}}$ tends to exceed the one-to-one line compared to other tracers. Notably, $\mathrm{SFR_{ANN}}$ and $\mathrm{SFR_{MIR}}$ are comparable, with minimal offsets and scatters.

\smallskip
(3) We found that SF and composite galaxies are young and exhibit blue colors, while Seyfert and LINER galaxies tend to be older and display redder colors with higher stellar mass, which is consistent with previous studies.

\smallskip
(4) We observed that both SFRs and sSFRs exhibit strong correlations with Eddington ratio. We also found that comparing sSFRs is more informative than comparing SFRs because they account for galaxies with the same physical properties.

\smallskip
(5) Eddington ratio displays smooth distribution transitions in the SFRs/sSFRs-stellar mass diagrams. Galaxies with a high Eddington ratio exhibit high SF activity, similar to that of blue SF galaxies. We also found decreasing trends in sSFRs from SF galaxies to composite, Seyfert, and LINER galaxies.

\smallskip
(6) Gas supply may be crucial in the strong correlations between SF and AGN strength activities. When the gas supply is high, we observe a high Eddington ratio and high SF activity. Conversely, SF and AGN strength activities decrease when the gas supply decreases. 

\bigskip
We thank the referee for valuable suggestions and comments that significantly improved the presentation and clarity of this paper. This work has been supported by the National Key R\&D Program of China No. 2022YFF0503401, the National Natural Science Foundation of China (12473014, 12203047, 12025303, and 12003031), and the Basic Science Research Program through the National Research Foundation (NRF) of the Korean Government (2021R1A2C3008486). H. A. N. Le acknowledges the support from the Fundamental Research Funds for the Central Universities (WK2030000057). 

\begin{deluxetable*}{ccccc}
\tablecaption{List of SFR tracers \label{table:sfr_summary}}
\tablewidth{\textwidth}
\tabletypesize{\footnotesize}
\tablehead{
\colhead{SFR Tracer} & \colhead{References} & \colhead{IMF} & \colhead{Wavelength} & \colhead{Method}
}
\startdata
(1) & (2) & (3) & (4) & (5) \\
\hline
$\rm SFR_{FIR}$       & \citet{Kennicutt+98}    & Salpeter & FIR: 90$\rm \micron$, 100$\rm \micron$                  & $\rm L_{FIR}$: $\rm L_{90\micron}$, $\rm L_{100\micron}$ \\
$\rm SFR_{ANN}$       & \citet{Ellison+16a}     & Chabrier & IR: 8$-$1000 $\rm \micron$                   & $\rm L_{ANN}$: ANN technique \\
$\rm SFR_{MIR}$       & \citet{Salim+18}        & Chabrier & MIR: 22 $\rm \micron$                      & MIR photometry from WISE catalog \\
$\rm SFR_{Dn4000}$    & \citet{Brinchmann+04}   & Kroupa   & UV: 4000 \AA\ break             & Dn4000 vs. $\rm SFR_{H\alpha}/M_*$\\
$\rm SFR_{SED}$       & \citet{Salim+18}        & Chabrier & UV/optical/MIR                  &  SED fitting  \\
$\rm SFR_{[OII]}$     & \citet{Zhuang+19}       & Kroupa   & Optical: [OII] 3727 \AA, [OIII] 5007 \AA & $\rm L_{[OII]}$ + $\rm L_{[OIII]}$ (subtracted $\rm f_{AGN}$) \\
$\rm SFR_{H\alpha}$   & \citet{Murphy+11}       & Kroupa   & Optical: H$\alpha$ 6563 \AA     & $\rm L_{H\alpha}$ (subtracted $\rm f_{AGN}$) \\
\enddata
\tablecomments{Col. (1): SFR tracers. Col. (2): References for derivation. Col. (3): IMF assumed in original calibration (all SFRs rescaled to \citealp{Kroupa01} IMF in this work). Col. (4): Wavelength bands. Col. (5): Determined methods.}
\end{deluxetable*}

\begin{table*}
\caption{Target properties}\label{table:fits}
\centering
\scalebox{1.0}{
\begin{tabular}{llll}
\hline\hline
Parameter & Format & Units & Description \\
(1) & (2) & (3) & (4) \\
\hline
OBJID & DOUBLE & \nodata & Object ID \\
RA & DOUBLE & degree & J2000 Right ascension \\
DEC & DOUBLE & degree &  J2000 Declination \\
PLATE & LONG & \nodata & Plate number of the SDSS spectrum \\
FIBER & LONG & \nodata & Fiber ID of the SDSS spectrum \\
MJD & LONG & day &  Modified Julian Date of the SDSS spectrum \\
$\rm SN_{spec}$ & DOUBLE & \nodata &  Optical continuum 5100\AA\ signal-to-noise of the spectra \\ 
OCLASS & DOUBLE & \nodata &  Classification of targets: 1: Composite, 2: SF galaxy, 3: Seyfert, 4: LINER\\ 
$\rm Z_{PPXF}$ & DOUBLE & \nodata &  Measured redshift from the PPXF\\ 
$\rm f_{AGN}$ & DOUBLE & \nodata &  AGN fraction calculated from the rainbow package \\ 
$\rm U_{Fiber}$ & DOUBLE & mag &  U-band FIBERMAG from the SDSS \\ 
$\rm U_{Total}$ & DOUBLE & mag &  U-band MODELMAG from the SDSS \\ 
$\rm \log(M_{*})$ & DOUBLE & $\rm M_{\odot}$ &  Measured stellar mass from the MPA-JHU catalog\\ 
$\rm D_{n}(4000)$ & DOUBLE & \nodata &  Measured $\rm D_{n}(4000)$ flux from the MPA-JHU catalog\\
$\rm Err\ D_{n}(4000)$ & DOUBLE & \nodata &  Error measured $\rm D_{n}(4000)$ flux from the MPA-JHU catalog\\ 
$\rm \log(SFR_{[OII]})$ & DOUBLE & $\rm M_{\odot}\ yr^{-1}$ &  Measured SFRs from \OII\ emisison line\\
$\rm Err\ \log(SFR_{[OII]})$ & DOUBLE & $\rm M_{\odot}\ yr^{-1}$ &  Error measured SFRs from \OII\ emisison line\\
$\rm \log(SFR_{H\alpha})$ & DOUBLE & $\rm M_{\odot}\ yr^{-1}$ &  Measured SFRs from \Ha\ emisison line\\
$\rm Err\ \log(SFR_{H\alpha})$ & DOUBLE & $\rm M_{\odot}\ yr^{-1}$ &  Error measured SFRs from \Ha\ emisison line\\
\hline\\
\end{tabular}
}
\label{table:catalog}
\raggedright
\tablecomments{Col. (1): Parameter types; Col. (2): Format of parameters; Col. (3): Units of measured parameters; Col. (4): Parameter descriptions. The full table is available in FITS format in the online version of the paper.}
\end{table*}


\begin{deluxetable*}{ccccccccc}
\tablecolumns{9}
\tablewidth{\textwidth}
\tabletypesize{\footnotesize}
\tablecaption{Comparison between different star formation rate tracers}
\tablehead{
\colhead{SFR Ratio} &
\multicolumn{2}{c}{Star-forming} &
\multicolumn{2}{c}{Composite} &
\multicolumn{2}{c}{Seyfert} &
\multicolumn{2}{c}{LINER} \\
\cmidrule(lr){2-3}
\cmidrule(lr){4-5}
\cmidrule(lr){6-7}
\cmidrule(lr){8-9}
& \colhead{Offset(Scatter)} & \colhead{Sample} &
\colhead{Offset(Scatter)} & \colhead{Sample} &
\colhead{Offset(Scatter)} & \colhead{Sample} &
\colhead{Offset(Scatter)} & \colhead{Sample}
}
\startdata
(1) & (2) & (3) & (4) & (5) & (6) & (7) & (8) & (9)  \\
\tableline
$\rm SFR_{ANN} / SFR_{FIR}$ & $0.183(0.270)$ & 1691 & $0.190(0.257)$ & 986 & $0.416(0.381)$ & 276 & $0.159(0.363)$ & 148 \\
$\rm SFR_{H\alpha} / SFR_{FIR}$ & $0.184(0.401)$ & 2211 & $0.003(0.477)$ & 1222 & $0.086(0.499)$ & 376 & $-0.460(0.699)$ & 252 \\
$\rm SFR_{Dn4000} / SFR_{FIR}$ & $0.140(0.349)$ & 2211 & $0.118(0.357)$ & 1222 & $0.002(0.395)$ & 376 & $-0.247(0.481)$ & 252 \\
$\rm SFR_{[OII]} / SFR_{FIR}$ & $0.178(0.396)$ & 1956 & $0.012(0.469)$ & 1158 & $0.213(0.561)$ & 370 & $-0.114(0.726)$ & 239 \\
$\rm SFR_{SED} / SFR_{FIR}$ & $0.261(0.287)$ & 1880 & $0.039(0.370)$ & 1038 & $0.035(0.454)$ & 281 & $-0.083(0.514)$ & 219 \\
$\rm SFR_{MIR} / SFR_{FIR}$ & $0.257(0.179)$ & 1857 & $0.284(0.206)$ & 1070 & $0.501(0.337)$ & 314 & $0.187(0.218)$ & 184 \\
\tableline
$\rm SFR_{Dn4000} / SFR_{ANN}$ & $-0.017(0.188)$ & 63048 & $-0.186(0.268)$ & 19118 & $-0.594(0.409)$ & 12672 & $-0.703(0.407)$ & 7192 \\
$\rm SFR_{[OII]} / SFR_{ANN}$ & $-0.022(0.311)$ & 62134 & $-0.360(0.421)$ & 18638 & $-0.294(0.556)$ & 12089 & $-0.516(0.652)$ & 6072 \\
$\rm SFR_{H\alpha} / SFR_{ANN}$ & $-0.106(0.326)$ & 62902 & $-0.390(0.403)$ & 18793 & $-0.515(0.460)$ & 12194 & $-0.857(0.567)$ & 6128 \\
$\rm SFR_{SED} / SFR_{ANN}$ & $0.064(0.189)$ & 54956 & $-0.214(0.320)$ & 16595 & $-0.524(0.494)$ & 10545 & $-0.650(0.617)$ & 6235 \\
$\rm SFR_{MIR} / SFR_{ANN}$ & $-0.007(0.176)$ & 42544 & $-0.037(0.220)$ & 12936 & $-0.011(0.312)$ & 7327 & $-0.132(0.254)$ & 2413 \\
\tableline
$\rm SFR_{SED} / SFR_{Dn4000}$ & $0.088(0.188)$ & 59936 & $-0.017(0.283)$ & 17591 & $0.074(0.423)$ & 11082 & $0.091(0.528)$ & 6318 \\
$\rm SFR_{H\alpha} / SFR_{Dn4000}$ & $-0.094(0.372)$ & 69031 & $-0.205(0.413)$ & 20329 & $0.086(0.539)$ & 13489 & $-0.194(0.582)$ & 7334 \\
$\rm SFR_{[OII]} / SFR_{Dn4000}$ & $0.002(0.352)$ & 65682 & $-0.175(0.461)$ & 19931 & $0.308(0.673)$ & 13283 & $0.150(0.698)$ & 7177 \\
$\rm SFR_{[OII]} / SFR_{H\alpha}$ & $0.080(0.168)$ & 65690 & $0.026(0.194)$ & 19936 & $0.219(0.276)$ & 13284 & $0.336(0.246)$ & 7179 \\
$\rm SFR_{H\alpha} / SFR_{SED}$ & $-0.183(0.361)$ & 59940 & $-0.190(0.479)$ & 17596 & $-0.012(0.602)$ & 11083 & $-0.290(0.773)$ & 6320 \\
$\rm SFR_{OII} / SFR_{SED}$ & $-0.090(0.338)$ & 57061 & $-0.159(0.522)$ & 17252 & $0.215(0.734)$ & 10911 & $0.057(0.861)$ & 6184 \\
\enddata
\label{table:result}
\tablecomments{Col. (1): Ratio of $\rm SFR_{B}$ to $\rm SFR_{A}$. Cols. (2,3): Mean offset and scatter (in parentheses) and number of sources for SF galaxies. Cols. (4,5): Similar to cols. (2,3) but for composite galaxies. Cols. (6,7): Similar to cols. (2,3) but for Seyfert galaxies. Cols. (8,9): Similar to cols. (2,3) but for LINER galaxies. 
}
\end{deluxetable*}

\begin{deluxetable*}{ccccccccc}
\tablecolumns{9}
\tablewidth{\textwidth}
\tabletypesize{\footnotesize}
\tablecaption{Median star formation rate and specific star formation rate for different tracers}
\tablehead{
\colhead{SFR Tracer}&
\colhead{$\rm log(SFR)$}&
\colhead{$\rm log(sSFR)$}&
\colhead{$\rm log(SFR)$}&
\colhead{$\rm log(sSFR)$}&
\colhead{$\rm log(SFR)$}&
\colhead{$\rm log(sSFR)$}&
\colhead{$\rm log(SFR)$}&
\colhead{$\rm log(sSFR)$}
}
\startdata
(1) & (2) & (3) & (4) & (5) & (6) & (7) & (8)  & (9)  \\
\hline
&  \multicolumn{2}{c}{Star-forming} & \multicolumn{2}{c}{Composite} & \multicolumn{2}{c}{Seyfert} &  \multicolumn{2}{c}{LINER} \\
\cmidrule(lr){1-1}
\cmidrule(lr){2-3}
\cmidrule(lr){4-5}
\cmidrule(lr){6-7}
\cmidrule(lr){8-9}
\\[-5pt]
$\rm SFR_{FIR}$ & $0.349(0.469)$ & $-9.911(0.337)$ & $0.569(0.498)$ & $-10.093(0.466)$ & $0.535(0.555)$ & $-10.250(0.520)$ & $0.310(0.610)$ & $-10.581(0.647)$ \\
$\rm SFR_{ANN}$ & $0.146(0.505)$ & $-9.751(0.309)$ & $0.496(0.571)$ & $-10.061(0.401)$ & $0.586(0.615)$ & $-10.141(0.481)$ & $0.156(0.537)$ & $-10.622(0.509)$ \\
$\rm SFR_{MIR}$ & $0.306(0.512)$ & $-9.728(0.342)$ & $0.612(0.521)$ & $-9.992(0.435)$ & $0.781(0.546)$ & $-9.985(0.470)$ & $0.283(0.590)$ & $-10.545(0.569)$ \\
$\rm SFR_{Dn4000}$ & $0.158(0.503)$ & $-9.738(0.326)$ & $0.351(0.631)$ & $-10.187(0.494)$ & $0.057(0.696)$ & $-10.645(0.614)$ & $-0.641(0.70)$ & $-11.439(0.705)$ \\
$\rm SFR_{SED}$ & $0.247(0.520)$ & $-9.653(0.334)$ & $0.320(0.630)$ & $-10.248(0.516)$ & $0.179(0.754)$ & $-10.548(0.668)$ & $-0.274(0.829)$ & $-11.074(0.823)$ \\
$\rm SFR_{[OII]}$ & $0.152(0.668)$ & $-9.733(0.509)$ & $0.149(0.931)$ & $-10.414(0.836)$ & $0.363(1.116)$ & $-10.366(1.045)$ & $-0.31(1.134)$ & $-11.148(1.116)$ \\
$\rm SFR_{H\alpha}$ & $0.074(0.728)$ & $-9.816(0.476)$ & $0.13(0.804)$ & $-10.441(0.664)$ & $0.166(0.993)$ & $-10.578(0.888)$ & $-0.688(0.886)$ & $-11.528(0.852)$ \\
\enddata
\label{table:median_sfr}
\tablecomments{Col. (1): SFR tracers. Cols. (2) and (3) : Median $\rm \log(SFR)$ and $\rm log(sSFR)$ of SF galaxy. Cols. (4) and (5): Median $\rm \log(SFR)$ and $\rm log(sSFR)$ of composite galaxy. Cols. (6) and (7): Median $\rm \log(SFR)$ and $\rm log(sSFR)$ of Seyfert galaxy. Cols. (8) and 9: Median $\rm \log(SFR)$ and $\rm log(sSFR)$ of LINER galaxy. The standard deviation values are shown in parentheses.}
\end{deluxetable*}



\end{CJK*}

\end{document}